\documentclass[longauth]{aa}  

\usepackage{graphicx}
\usepackage{txfonts}
\usepackage{xcolor}
\usepackage{scalerel}
\usepackage{multirow}
\usepackage{colortbl}
\usepackage{color}
\usepackage{caption}
\usepackage{pdflscape}
\usepackage{placeins}
\usepackage{hyperref}  
\hypersetup{colorlinks=true,linkcolor=[rgb]{1.,0.2,0.2},citecolor=[rgb]{0.1,0.4,1.},filecolor=[rgb]{0.7,0.2,0.2},urlcolor=[rgb]{0.7,0.2,0.2}}
\usepackage{orcidlink}
\newcommand{\orcid}[1]{\href{https://orcid.org/#1}{\textcolor[HTML]{A6CE39}{\aiOrcid}}}

\newcommand\Hb{$\rm H\beta$}

\newcommand\OIII{O\protect\scaleto{$III$}{1.2ex}}
\newcommand\OIIIa{[O\protect\scaleto{$III$}{1.2ex}]$\lambda5007$}
\newcommand\OIIIb{[O\protect\scaleto{$III$}{1.2ex}]$\lambda4959$}

\newcommand\HII{H\protect\scaleto{$II$}{1.2ex}}
\newcommand\HI{H\protect\scaleto{$I$}{1.2ex}}

\newcommand\lya{Ly$\rm \alpha$}
\newcommand{\JWST}{\textrm{JWST}}
\newcommand{\lenstool}{\texttt{Lenstool}}
\newcommand{\glafic}{\texttt{glafic}}
\newcommand{\bagpipes}{\texttt{Bagpipes}}
\newcommand{\EVCG}{V25}
\defcitealias{Vanzella2025}{\EVCG}
\def\kms{km\,s$^{-1}$}

\graphicspath{{figures/}} 

\begin{document} 

   \title{JWST spectroscopic confirmation of the Cosmic Gems arc at z=9.625}
   \subtitle{Insights into the small-scale structure of a post-burst system}

\authorrunning{Matteo Messa et al.}
\author{M.~Messa\inst{\ref{inafbo}}\fnmsep\thanks{\email{matteo.messa@inaf.it}}$^{\orcidlink{0000-0003-1427-2456}}$
\and E.~Vanzella\inst{\ref{inafbo}}$^{\orcidlink{0000-0002-5057-135X}}$
\and F.~Loiacono\inst{\ref{inafbo}}$^{\orcidlink{0000-0002-8858-6784}}$
\and A.~Adamo\inst{\ref{univstock}}$^{\orcidlink{0000-0002-8192-8091}}$  
\and M.~Oguri\inst{\ref{chibaUNI},\ref{chibaGSS}}$^{\orcidlink{0000-0003-3484-399X}}$ 
\and K.~Sharon\inst{\ref{um}}$^{\orcidlink{0000-0002-7559-0864}}$      
\and L.~D.~Bradley\inst{\ref{stsci}}$^{\orcidlink{0000-0002-7908-9284}}$ 
\and L.~Christensen\inst{\ref{nbi},\ref{dawn}}$^{\orcidlink{0000-0001-8415-7547}}$ 
\and A.~Claeyssens\inst{\ref{CRAL}}$^{\orcidlink{0000-0001-7940-1816}}$    
\and J.~Richard\inst{\ref{CRAL}}$^{\orcidlink{0000-0001-5492-1049}}$
\and Abdurro'uf\inst{\ref{JHU},\ref{stsci}}$^{\orcidlink{0000-0002-5258-8761}}$ 
\and F.~E.~Bauer\inst{\ref{Tarapaca}}$^{\orcidlink{0000-0002-8686-8737}}$  
\and P.~Bergamini\inst{\ref{unimi},\ref{inafbo}}$^{\orcidlink{0000-0003-1383-9414}}$ 
\and A.~Bolamperti\inst{\ref{mpa},\ref{inafiasf}}$^{\orcidlink{0000-0001-5976-9728}}$
\and M.~Brada\v{c} \inst{\ref{uniLjubljana}}$^{\orcidlink{0000-0001-5984-0395}}$  
\and F.~Calura\inst{\ref{inafbo}}$^{\orcidlink{0000-0002-6175-0871}}$
\and D.~Coe\inst{\ref{stsci}}$^{\orcidlink{0000-0001-7410-7669}}$ 
\and J.~M.~Diego\inst{\ref{ifca}}$^{\orcidlink{0000-0001-9065-3926}}$ 
\and C.~Grillo \inst{\ref{unimi},\ref{inafiasf}}$^{\orcidlink{0000-0002-5926-7143}}$
\and T.~Y-Y.~Hsiao\inst{\ref{CfA},\ref{JHU},\ref{stsci}}$^{\orcidlink{0000-0003-4512-8705}}$
\and A.~K.~Inoue\inst{\ref{WasedaPhys},\ref{WISE}}$^{\orcidlink{0000-0002-7779-8677}}$
\and S.~Fujimoto\inst{\ref{uoft},\ref{dunlap}}$^{\orcidlink{0000-0001-7201-5066}}$ 
\and M.~Lombardi\inst{\ref{unimi}}$^{\orcidlink{0000-0002-3336-4965}}$
\and M.~Meneghetti\inst{\ref{inafbo}}$^{\orcidlink{0000-0003-1225-7084}}$ 
\and T.~Resseguier\inst{\ref{stsci},\ref{JHU}}$^{\orcidlink{0009-0007-0522-7326}}$
\and M.~Ricotti\inst{\ref{univmaryland}}$^{\orcidlink{0000-0003-4223-7324}}$
\and P.~Rosati \inst{\ref{unife},\ref{inafbo}}$^{\orcidlink{0000-0002-6813-0632}}$
\and B.~Welch\inst{\ref{Goddard},\ref{univmaryland}}$^{\orcidlink{0000-0003-1815-0114}}$
\and R.~A.~Windhorst\inst{\ref{ArizonaSU}}$^{\orcidlink{0000-0001-8156-6281}}$
\and X.~Xu\inst{\ref{CIREA}}$^{\orcidlink{0000-0000-0000-0001}}$
\and E.~Zackrisson\inst{\ref{UU}}$^{\orcidlink{0000-0003-1096-2636}}$
\and A.~Zanella\inst{\ref{inafpd}}$^{\orcidlink{0000-0001-8600-7008}}$
\and A.~Zitrin\inst{\ref{BGU}}$^{\orcidlink{0000-0002-0350-4488}}$
}

\institute{
INAF -- OAS, Osservatorio di Astrofisica e Scienza dello Spazio di Bologna, via Gobetti 93/3, I-40129 Bologna, Italy \label{inafbo} 
\and Department of Astronomy, Oskar Klein Centre, Stockholm University, AlbaNova University Centre, SE-106 91, Sweden\label{univstock}
\and Center for Frontier Science, Chiba University, 1-33 Yayoi-cho, Inage-ku, Chiba 263-8522, Japan \label{chibaUNI} 
\and Department of Physics, Graduate School of Science, Chiba University, 1-33 Yayoi-Cho, Inage-Ku, Chiba 263-8522, Japan \label{chibaGSS} 
\and Department of Astronomy, University of Michigan 1085 South University Avenue Ann Arbor, MI 48109, USA\label{um}
\and Space Telescope Science Institute, 3700 San Martin Drive, Baltimore, MD 21218, USA \label{stsci}
\and Niels Bohr Institute, University of Copenhagen, Jagtvej 128, 2200-N Copenhagen, Denmark \label{nbi}
\and{Cosmic Dawn Center (DAWN), Denmark \label{dawn}}
\and Univ Lyon, Univ Lyon1, ENS de Lyon, CNRS, Centre de Recherche Astrophysique de Lyon UMR5574, Saint-Genis-Laval, France\label{CRAL}
\and Center for Astrophysical Sciences, Department of Physics and Astronomy, The Johns Hopkins University, 3400 N Charles St. Baltimore, MD 21218, USA \label{JHU}
\and Instituto de Alta Investigaci{\'{o}}n, Universidad de Tarapac{\'{a}}, Casilla 7D, Arica, Chile \label{Tarapaca}
\and Dipartimento di Fisica, Università degli Studi di Milano, Via Celoria 16, I-20133 Milano, Italy\label{unimi}
\and Max-Planck-Institut f\"ur Astrophysik, Karl-Schwarzschild-Str. 1, D-85748 Garching, Germany \label{mpa}
\and INAF -- IASF Milano, via A. Corti 12, I-20133 Milano, Italy\label{inafiasf}
\and University of Ljubljana, Faculty of Mathematics and Physics, Jadranska ulica 19, SI-1000 Ljubljana, Slovenia\label{uniLjubljana}
\and  Instituto de F\'isica de Cantabria (CSIC-UC). Avda. Los Castros s/n. 39005 Santander, Spain \label{ifca}
\and Center for Astrophysics \text{\textbar} Harvard \& Smithsonian, 60 Garden Street, Cambridge, MA 02138, USA \label{CfA}
\and Department of Physics, School of Advanced Science and Engineering, Faculty of Science and Engineering, Waseda University, 3-4-1 Okubo, Shinjuku, Tokyo 169-8555, Japan \label{WasedaPhys}
\and Waseda Research Institute for Science and Engineering, Faculty of Science and Engineering, Waseda University, 3-4-1 Okubo, Shinjuku, Tokyo 169-8555, Japan \label{WISE}
\and David A. Dunlap Department of Astronomy and Astrophysics, University of Toronto, 50 St. George Street, Toronto, Ontario, M5S 3H4, Canada \label{uoft}
\and Dunlap Institute for Astronomy and Astrophysics, 50 St. George Street, Toronto, Ontario, M5S 3H4, Canada  \label{dunlap}
\and Department of Astronomy, University of Maryland, College Park, 20742, USA\label{univmaryland}
\and Dipartimento di Fisica e Scienze della Terra, Università degli Studi di Ferrara, Via Saragat 1, I-44122 Ferrara, Italy\label{unife}
\and Astrophysics Science Division, Code 660, NASA Goddard Space Flight Center, 8800 Greenbelt Rd., Greenbelt, MD, 20771, USA \label{Goddard}
\and School of Earth and Space Exploration, Arizona State University, Tempe, AZ 85287-1404, USA \label{ArizonaSU} 
\and Center for Interdisciplinary Exploration and Research in Astrophysics (CIERA), 1800 Sherman Avenue, Evanston, IL, 60201, USA\label{CIREA}
\and Observational Astrophysics, Department of Physics and Astronomy, Uppsala University, Box 516, SE-751 20 Uppsala, Sweden \label{UU}
\and INAF Osservatorio Astronomico di Padova, vicolo dell'Osservatorio 5, 35122 Padova, Italy\label{inafpd}
\and Department of Physics, Ben-Gurion University of the Negev, P.O. Box 653, Be'er-Sheva 84105, Israel\label{BGU}
}
   \date{Received 25 July 2025 / Accepted 31 October 2025}

 
\abstract
{We present JWST/NIRSpec integral field spectroscopy of the Cosmic Gems arc, strongly magnified by the galaxy cluster SPT-CL\, J0615$-$5746.
Six-hour integration using NIRSpec prism spectroscopy (resolution R~$\simeq 30-300$), covering the spectral range $0.8-5.3~\mu m$, reveals a pronounced \lya-continuum break at $\lambda \simeq 1.3 \mu m$, as well as weak optical \Hb\ and \OIIIb\ emission lines at $z=9.625\pm0.002$, located in the reddest part of the spectrum ($\lambda > 5.1~\mu m$). No additional ultraviolet or optical emission lines are reliably detected. A weak Balmer break is measured alongside a very blue ultraviolet slope ($\beta \leq-2.5$, F$_{\lambda} \sim \lambda^{\beta}$). 
Spectral fitting with \texttt{Bagpipes} suggests that the Cosmic Gems galaxy is in a post-starburst phase, making it the highest-redshift system currently observed in a mini-quenched state. Spatially resolved spectroscopy at tens of parsecs shows relatively uniform features across subcomponents of the arc. 
These findings align well with the physical properties previously derived from JWST/NIRCam photometry of the stellar clusters, now corroborated by spectroscopic evidence. In particular, five observed star clusters exhibit ages of $7-30$~Myr. An updated lens model constrains the intrinsic sizes and masses of these clusters, confirming they are extremely compact and denser than typical star clusters in local star-forming galaxies ($\rm \Sigma_{M_\star}=10^5-10^6~M_\odot$). Additionally, four compact stellar systems consistent with star clusters ($\lesssim10$ pc) are identified along the extended tail of the arc. A sub-parsec line-emitting H\protect\scaleto{$II$}{1.2ex} region straddling the critical line, lacking a NIRCam counterpart, is also serendipitously detected. The Cosmic Gems arc thus offers a rare opportunity to investigate, at parsec scales, the aftermath of a star formation burst in the early Universe.} 
%
\maketitle
%
\section{Introduction}
\label{sec:intro}   
One of the major surprises of surpassing the first half-gigayear threshold in cosmic time (corresponding to $z>9$) is the significant slow evolution of the space density of bright galaxies \citep[e.g.,][]{Finkelstein2024, Napolitano2025}. The discovery and subsequent confirmation of a substantial population of bright $z>9$ galaxies (typically with $M_{\rm UV} < -20$) by JWST observations have now well established this trend \citep[e.g.,][]{Naidu2022, Arrabal23, Hsiao24, Wang23, Fujimoto24, Atek23, Curtis23, Robertson23, Bunker23, Tacchella23, Arrabal23, Finkelstein2024, Castellano24, Zavala25, Helton25, Carniani24a, Robertson24, Tang2025, Donnan2025}.
Several physical mechanisms have been discussed
to explain this phenomenon, including a top-heavy initial mass function \citep[e.g.,][]{Trinca24, Hutter_2025}, extremely bursty star formation histories \citep[SFHs,][]{Mason23,Pallottini23,Garcia25,Carvajal_Bohorquez2025}, reduced feedback resulting in higher star formation efficiency \citep[e.g.,][]{Dekel23,Li24,Somerville25}, radiation-driven feedback processes \citep[][]{Ferrara_2023_monsters, Ferrara2025,SugimuraR2024}, or a combination thereof.

Efforts are now underway to extend these studies to fainter galaxies at similar redshifts (e.g., \citealp{atek2024, mowla2024, Tang2025, Whitler25}, and upcoming observations, e.g., the cycle 4 Vast Exploration for Nascent, Unexplored Sources (VENUS) large program, GO-6882, PI Fujimoto). 
The confirmation of very high-redshift galaxies primarily relies on the detection of either strong emission lines or a continuum break, the latter being feasible only for intrinsically bright sources. These techniques have led to the confirmation of the most distant known galaxies to date at $z=14.2$ \citep{Carniani24a} and $z=14.44$ \citep{naidu2025_arxiv}. 

Accessing the population of faint and weakly (or non-) emitting galaxies at $z>9$ is therefore the next frontier. These include galaxies in a post-burst or off-mode phase of star formation (SF), sometimes referred to as mini-quenched galaxies \citep[e.g.,][]{gelli2023,dome2024,Endsley25,endsley2024,topping2024,looser2024,looser2025,baker2025, CoveloPaz2025}, observed a few tens of megayears after a starburst. If SFHs in the early Universe are generally bursty, especially at low masses, then we may expect many high-z galaxies to be in a temporary "dormant" post-burst phase \citep{tacchella2016,tacchella2020,fauchergiguere2018,sun2023a,sun2023b,Mason23,GarciaR2023}.
In such systems, the most massive O-type stars have already evolved, reducing the ionizing output and leading to weak or absent emission lines \citep[e.g.,][]{stasinska1996,choi2017}, which makes detailed spectroscopic characterization particularly difficult \citep[e.g.,][]{looser2025}. 
As the equivalent width (EW) of nebular emission lines decreases, only the continuum break constrains the redshift. This requires very long or impractical exposure times, even with instruments such as JWST/NIRSpec.
One of the most effective strategies to overcome this difficulty involves leveraging gravitational lensing, which enables the spectroscopic confirmation of $z>9$ galaxies with $M_{\rm UV} >- 20$ \citep[e.g.,][]{robertsborsani2023}.

The Cosmic Gems arc \citep[][]{salmon2018,strait2020,welch2023,adamo2024a,Bradley25} provides a unique case at $z \simeq 10$, where gravitational lensing magnifies a relatively faint source (intrinsic M$_{\rm UV} \simeq -18.4$, apparent delensed magnitude m$_{\rm UV} \simeq 29.1$) into a bright arc with an observed magnitude F200W~=~24.5 \citep{Bradley25}. 
In addition to the enhanced signal-to-noise ratio provided by lensing, the strong tangential stretching has resolved internal substructures, allowing the identification of five massive, dense star clusters \citep{adamo2024a}, along with fainter components extending along the arc's tail.
While a detailed analysis of the star cluster mass function and its connection to the counterimage is presented in a companion paper \citet[hereafter \EVCG]{Vanzella2025}, here we report on JWST/NIRSpec integral field units (IFU) spectroscopy of the bright arc. These observations provide an exceptionally high S/N redshift confirmation, despite the presence of relatively weak spectral features.
This fortunate configuration enables us to spatially resolve and analyze both the star clusters and the host galaxy, offering a rare window into a post-burst system at the epoch of reionization.

This paper is structured as follows: Observations and data reduction are outlined in Section~\ref{sec:data}. The updated lens models used in this work are described in Section~\ref{sec:lens_model}. The spectral analysis of the Cosmic Gems and the study of its star cluster population are presented in Section~\ref{sec:nirspec_analysis} and \ref{sec:ysc_main}, respectively. Finally, the main results are discussed in Section~\ref{sec:discussion} and summarized in Section~\ref{sec:conclusions}.

Throughout this paper, we adopt a flat Lambda cold dark matter ($\rm \Lambda$-CDM) cosmology with $H_0=70$ km s$^{-1}$ Mpc$^{-1}$ and $\rm \Omega_M = 0.3$, a \citet{kroupa2001} initial mass function, and a solar metallicity $\rm Z_\odot=0.02$. All quoted magnitudes are in the AB system. All quoted EWs are rest-frame values. 

\section{Observations and data reduction}
\label{sec:data}

\subsection{JWST-NIRSpec}\label{sec:data:jwst}

\begin{figure*}
    \centering    \includegraphics[width=0.99\textwidth]{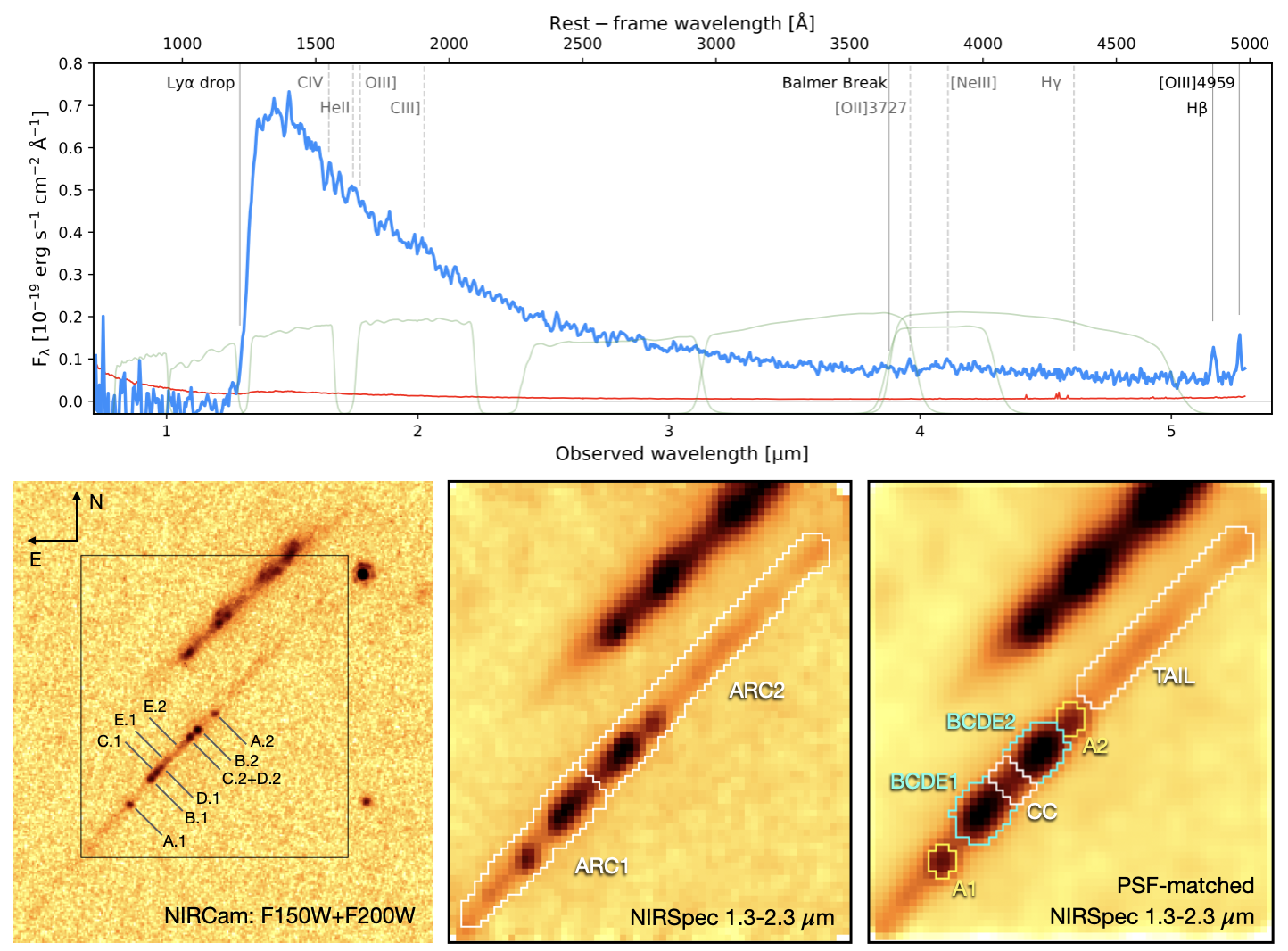}
    \caption{\JWST/NIRSpec spectrum, NIRCam, and pseudo-NIRCam images (NIRSpec-IFU based) of the Cosmic Gems arc. {\it Top:} 1D NIRSpec prism spectrum (from the PSF-matched cube) extracted from the full arc (with its aperture mask shown in the bottom-central panel), showing the large \lya-break and damping at $\simeq 1.3~\mu m$, along with two weak emission lines, \Hb\ and \OIIIb\,, on the rightmost side of the spectral range. A weak Balmer break is also detected (Section~\ref{sec:nirspec_analysis}). The 1$\sigma$ uncertainty spectrum is shown in red. The wavelengths corresponding to the main undetected UV and optical lines are shown with dashed lines. The throughputs of the NIRCam filters covering the field are shown in green, for reference.
    {\it Bottom:} From left to right, \JWST/NIRCam stacked F150W+F200W image with the NIRSpec-IFU field of view outlined; collapsed NIRSpec IFU data cube in the rest-UV (covering the observed wavelength range of F150W and F200W), before (middle) and after (right) PSF-matching. The last panel also includes the outlines of the masks used to extract the 1D spectra for the regions discussed in the main text.}
    \label{fig:nircam_nirspec}
\end{figure*}

\JWST/NIRSpec IFU observations (GO 5917, PI: Vanzella) targeting the strongly lensed Cosmic Gems arc, magnified by the galaxy cluster SPT-CL\,J0615$-$5746 (hereafter SPT0615) were acquired on February 4, 2025. Figure~\ref{fig:nircam_nirspec} shows the layout of the field of view, which includes the major emitting mirrored regions A, B, C, D, E groups 1 and 2, along with the tail of the arc oriented toward the north. A total integration time of $\simeq 5.8$ hours on target were acquired. The dataset consists of PRISM spectroscopy covering the spectral range $\rm 0.8-5.3\ \mu m$, with spectral resolution varying from $\rm R\simeq30$ (around $\rm \sim1\ \mu m$) to $\rm R\simeq300$ (at the longest wavelengths). 

The data were reduced following the same procedures described in \citet{Messa2025} (see also \citealt{Vanzella2024}). Briefly, we used the STScI pipeline 
\citep[v1.17.0 and 1322.pmap,][]{bushouse23} 
and post-processed the intermediate products of stage 2 with customized procedures that combine the eight partial cubes into the final cleaned cube. The latter included background subtraction, removal of detector defects, and the computation of the error spectrum.  
Additionally, we performed the 1/f noise subtraction in stage 1 using the \texttt{clean\_flicker\_noise} step.
In contrast with the faint targets discussed in \citet{Messa2025}, the Cosmic Gems arc was well detected in the continuum at each wavelength slice; therefore, we estimated the background at each wavelength by computing a 2D polynomial fitting in regions of the field of view free from clearly detected sources (see Appendix~\ref{sec:app:reduction}).
The zero, first, and second-order polynomial fittings produce similar results. We used the first-order fit as fiducial background subtraction. 
As in \citet{Messa2025}, the error cube was derived by storing the median deviation of each pixel from the combination of the partial cubes.

We produced data cubes at 100 and 50 milli-arcsec (mas) per pixel. By comparing 1D spectra coming from the same regions of the arc, we found no notable differences (Appendix~\ref{sec:app:reduction}); therefore, we decided to keep the 50 mas reduction as our reference, as it better samples the NIRSpec PSF.
Flux calibration was cross-checked with \JWST/NIRCam photometry extracted from the main target Cosmic Gems arc \citep[][]{Bradley25}. The detected sources (the arc and the nearby $z=2.5$ galaxy) in the reduced, post-processed, and collapsed data cubes were aligned with their \JWST/NIRCam counterparts. 

We checked the consistency of the above products with an independent reduction carried out from stage 3 of the official pipeline. Overall, the results are fully consistent, with the above post-processing producing slightly cleaner spectra (reduced spikes and defects; see Appendix~\ref{sec:app:reduction}). 

Finally, we PSF-matched the data in the spectral wavelength using as reference the PSF models produced by the \texttt{STPSF} tool\footnote{\url{https://stpsf.readthedocs.io/en/latest/index.html}} (assuming the same instrumental configurations as our data). 
The modeled PSF as a function of wavelength shows good agreement with the PSF recovered by fitting the shape of one of the bright sources in our field-of-view (see Appendix~\ref{sec:app:reduction}). The final data cube was matched to the worst PSF at 5.3 micron, applying a kernel smoothing at each wavelength slice\footnote{The kernels are produced assuming Gaussian profiles for the PSF. The PSF-correction overall is performed using the \texttt{DPUSER} interactive language, \url{https://www.mpe.mpg.de/~ott/dpuser/}.}, following the aforementioned PSF variation trend. The bottom panels in Figure~\ref{fig:nircam_nirspec} show a collapsed 1.3--2.3~$\mu$m image of the rest-frame UV part of the IFU cube before and after the PSF-matching of the cube. This correction was implemented in order to avoid a wavelength-dependent loss of flux when extracting 1D spectra from the IFU masks (Section~\ref{sec:nirspec_analysis}). The 1D spectrum (and relative uncertainty) coming from the final reduction is shown in Figure~\ref{fig:nircam_nirspec}.

Throughout the paper we refer to the image and cluster notation introduced in \citet{adamo2024a} and \citet{Bradley25}, i.e., the galaxy's image in the southeast is numbered "1" while the one in the northwest "2," and clusters are named "A" to "E" (see bottom-left panel in Fig.~\ref{fig:nircam_nirspec}).
The NIRCam imaging used in this work (Fig.~\ref{fig:nircam_nirspec},~\ref{fig:map_redshift},~\ref{fig:new_ysc} and \ref{fig:forward_mod}) is presented in \citet{Bradley25}. 

\subsection{VLT-MUSE}\label{sec:data:muse}
SPTJ0615 was observed with the MUSE instrument on the VLT between January and March 2024 (period P112, PI: F. Bauer) in the wide field mode (WFM) and using the adaptive optics (WFM-AO-N). The observations consist of 6.2 hours of observing time in total (24 individual exposures of 930 seconds each), divided into two different pointings covering the core of the cluster. The MUSE data reduction was performed using the same procedure described in \citet{richard2021}, largely following the prescription described in \citet{Weilbacher2020}, with some specific improvements for crowded fields. The final output is given as an FITS data cube with two extensions containing the flux and the associated variance over a regular 3D grid at a spatial pixel scale of $0\farcs2$ and a wavelength step of 1.25 $\AA$ between 4750 and 9350 $\AA$. The use of the AO mode results in a gap in the 5800-5980 $\AA$ wavelength range due to the AO notch filter. Individual exposures were aligned against each other and with respect to four star positions in J2000 selected from the Gaia data release 2 (\citealt{gaia2018}). The relative astrometry between the MUSE cube and the HST images was cross-checked by matching bright sources present in the two datasets. We measure a alignment rms of 0.1'', corresponding to half a MUSE spaxel. 
Following the same method as \citet{richard2021}, a line-detected sources catalog was produced directly from the  MUSE data cube, performed by running the muselet software, which is part of the MPDAF Python package (\citealt{Piqueras2019}). 

\section{Updated lens model}\label{sec:lens_model}

\begin{figure}
\centering
\includegraphics[width=\columnwidth]{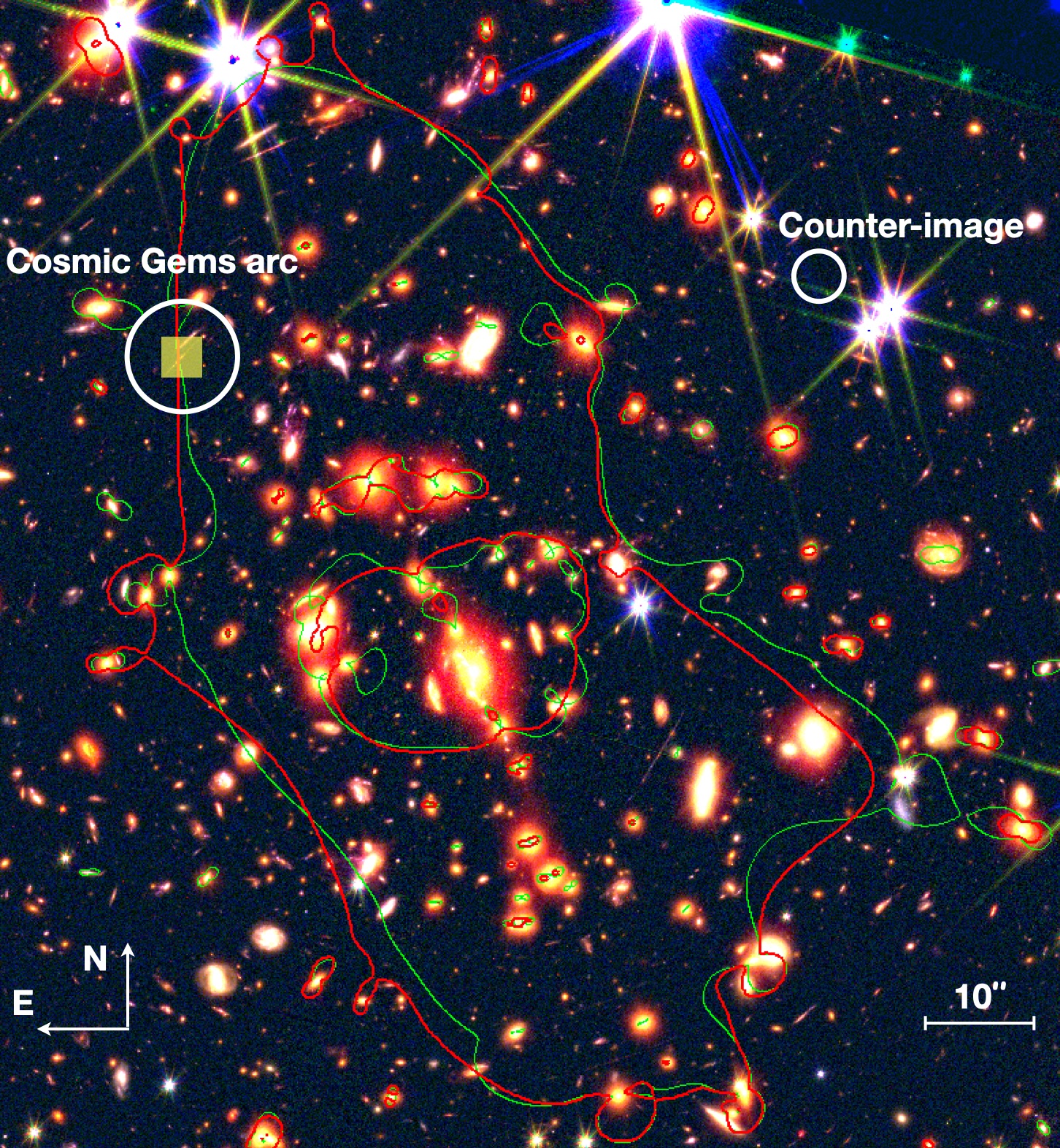}
 \caption{\JWST/ NIRCam color image of the galaxy cluster SPT0615 field with the locations of the Cosmic Gems arc and its counter-image marked, along with the critical lines for the $z=9.625$ lens models from {\sc Lenstool} (red line) and \glafic\ (green line). The yellow-shaded square marks the field of view of the \JWST/NIRSpec IFU observations.} 
 \label{fig:panoramic}
\end{figure}

We present updated mass models (with respect to what presented in \citealp{adamo2024a} and \citealp{Bradley25}) constructed with both \glafic\ \citep{oguri2010,oguri2021} and \lenstool\ \citep{jullo2007}. A panoramic view of the lensing cluster SPT0615 is given in Fig.~\ref{fig:panoramic}, along with the critical lines predicted by the two lens models at the redshift of the target. In this work, we refer to the magnification values obtained by the \glafic\ model as the reference ones, but we provide a one-to-one comparison of the two models in Appendix~\ref{sec:app:mul_image}. This direct comparison between the two models -- revealing differences of up to a factor of $\sim1.5$ -- provides insight into the systematic uncertainties that are typically not captured by the statistical uncertainties of the individual models.
Overall, the updated magnification values along the Cosmic Gems arc remain similar to the ones used in \citet{adamo2024a} and \citet{Bradley25}. 

\subsection{Multiple image catalog}
The updated models adopt 54 multiple images from 18 sources. Among the 18 sources, 12 have spectroscopic redshifts. In this paper, we identify 12 new multiple images from four sources. In addition to the new spectroscopic redshift for the Cosmic Gems arc from JWST, we add spectroscopic redshifts for four sources from our MUSE spectroscopy (see Section~\ref{sec:data:muse}). Photometric redshift estimates were used for four of the sources without spectroscopic redshift. Finally, for two sources (five multiple images in total) the redshift was left as a free parameter.
The details of the multiple images are summarized in Appendix~\ref{sec:app:mul_image}. 
The inclusion of newly identified multiple systems at $\rm z\gtrsim6$, located near the observed position of the Cosmic Gems, strengthens the constraints and improves the robustness of the updated lens models in the region of the arc. 

\subsection{Glafic model}
The updated \glafic\ mass model assumes the positional error of $0\farcs4$ for the multiple images (including the counter image of the Cosmic Gems arc), except for the Cosmic Gems arc for which the smaller position error of $0\farcs04$ is assumed to accurately reproduce the positions of the image clusters. The complex dark matter distribution of this cluster is modeled by four elliptical Navarro–Frenk–White (hereafter NFW, \citealp{navarro1997}) components. For the first three NFW components, their centers were fixed at ($93.9656010$, $-57.7801990$), ($93.9705078$, $-57.7753866$), and ($93.9613773$, $-57.7810248$) where bright cluster member galaxies are located, while the center of the remaining NFW component was fit to the data. The best-fitting position of this NFW component is ($93.9519126$, $-57.7804012$). In addition to the NFW components, we included the external shear, the third-order multipole perturbation, and cluster member galaxies. The latter are modeled by elliptical pseudo-Jaffe ellipsoids with ellipticities and position angles fixed to observed shapes of galaxies. The three parameters modeling the scaling relation of pseudo-Jaffe ellipsoids and galaxy luminosities were treated as free parameters \citep[see, e.g.,][]{kawamata2016}. The mass model was optimized by minimizing the chi-squared statistic evaluated in the source plane, taking account of the full magnification tensor \citep[see Appendix 2 of][]{oguri2010}. The best-fitting mass model has a root-mean-square between observed and model-predicted multiple image positions of $0\farcs32$, and $\chi^2=37.2$ for $45$ degrees of freedom. We ran a Markov chain Monte Carlo (MCMC) to derive errors on model-predicted quantities such as magnification.

To check the robustness of the identification of the counter-image of the Cosmic Gems arc at ($93.95000$, $-57.77023$) used as a constraint in our default setup, we also derived another \glafic\ mass model excluding the counter-image. This model predicts the position of the counter-image at ($93.94978$, $-57.77004$), i.e., $0\farcs8$ away from the observed position, with a $1\sigma$ error of $0\farcs8$. Therefore, this model correctly predicts the observed counter-image position within about $1\sigma$ level, which confirms the robustness of the identification of the counter-image (see also \citetalias{Vanzella2025}). From the magnification map at the redshift of the Cosmic Gems ($\rm z=9.625$, see Section~\ref{sec:map_redshift}), we extracted amplification values for the subregions of interest using small apertures at their observed positions (see Appendix~\ref{sec:app:mul_image} for more details).

\subsection{Lenstool model}
\lenstool\ uses a parametric approach to model the foreground lens, and MCMC to explore the parameter space, identify the best-fit parameters, and estimate uncertainties. 
Our lens modeling strategy follows \cite{Sharon2020}. We optimized the parameters of three halos, one associated with the mass of the cluster, one fixed to the position of the brightest cluster galaxy (BCG), and a third halo to allow contribution to the lensing potential from a foreground group. Cluster-member galaxies were selected using the red sequence technique \citep{gladders2000}. We used the dual pseudo-isothermal elliptical (dPIE) mass distribution with seven parameters, $x$, $y$, $e$, $\theta$, $r_{cut}$, $r_{core}$, and $\sigma_0$. The cut radius was fixed at 1500 kpc for the two cluster-scale halos. The mass associated with galaxies was linked to their magnitudes through scaling relations \citep{jullo2007}, and their positional parameters were fixed to the observed light distribution. 
As constraints, we used the positions of multiple images of emission clumps, and the spectroscopic redshifts of sources (Appendix~\ref{sec:app:mul_image}). Where a spectroscopic redshift was not available, the redshift of the source was left as a free parameter with broad priors. The positional uncertainty of all multiple image constraints was assumed to be $0\farcs3$. Finally, we used a positional constraint on the critical curve (CC) crossing of the arc, with a positional uncertainty set to $0\farcs03$.
Similar to the \glafic\ analysis, we produced two lens models: model A used the counter image of the Cosmic Gems arc as a constraint, and model B did not. 

We used 21 free parameters and 70 (68) constraints for the model with (without) the counter image constraint. Model A resulted in an image-plane RMS of $0\farcs65$, reducing to $0\farcs58$ for model B. The image-plane RMS of the four clumps along the arc ranges from $0\farcs01$ to $0\farcs09$, implying a high precision in this region. 
The model-A predicted position of the counter-image is $\sim1\farcs0$ from the observed counter image, at [$93.950439\pm0\farcs5$, $-57.770424\pm0\farcs2$].
Uncertainties on model predictions were derived by sampling 250-280 steps from the MCMC and computing lensing outputs (e.g., magnification) from these sets of parameters.  

\section{NIRSpec analysis of the arc} \label{sec:nirspec_analysis}
To extract 1D spectra from the IFU cube and investigate the sub-galactic regions of the Cosmic Gems, we defined six mask apertures (Fig.~\ref{fig:nircam_nirspec}, bottom-right panel), taking into account the identification of unique morphological features and NIRSpec resolution. The two images of cluster A were covered with individual masks ({\it A1} and {\it A2}), as they are bright isolated sources. On the other hand, the rest of the clusters were unresolved with the NIRSpec PSF and were observed as a single elongated source. We defined a mask that covers the entire region containing clusters B to E ({\it BCDE1} and {\it BCDE2} for the two images).
The narrow region between clusters E.1 and E.2, where the CC crosses the arc, was covered with a further mask ({\it CC}). The faint tail of the arc was covered by the {\it TAIL} mask. In the following analyses, we sometimes consider that the flux from two mirrored regions combined (e.g., {\it BCDE1,2} contains the flux from both {\it BCDE1} and {\it BCDE2}). 
Finally, a mask aperture covering the entire image 2 of the arc ({\it ARC2}) is used for the analysis in Section~\ref{sec:discussion:full_arc}, and 
an even larger one covering the full arc ({\it ARC1,2}) was used to extract the spectrum shown in Fig.~\ref{fig:nircam_nirspec}.
Spectra extracted from the aforementioned mask apertures are shown in Appendix~\ref{sec:app:all_spectra}.

\subsection{Spectroscopic redshift and emission map}
\label{sec:map_redshift}
\begin{figure*}
    \centering
    \includegraphics[width=0.99\textwidth]{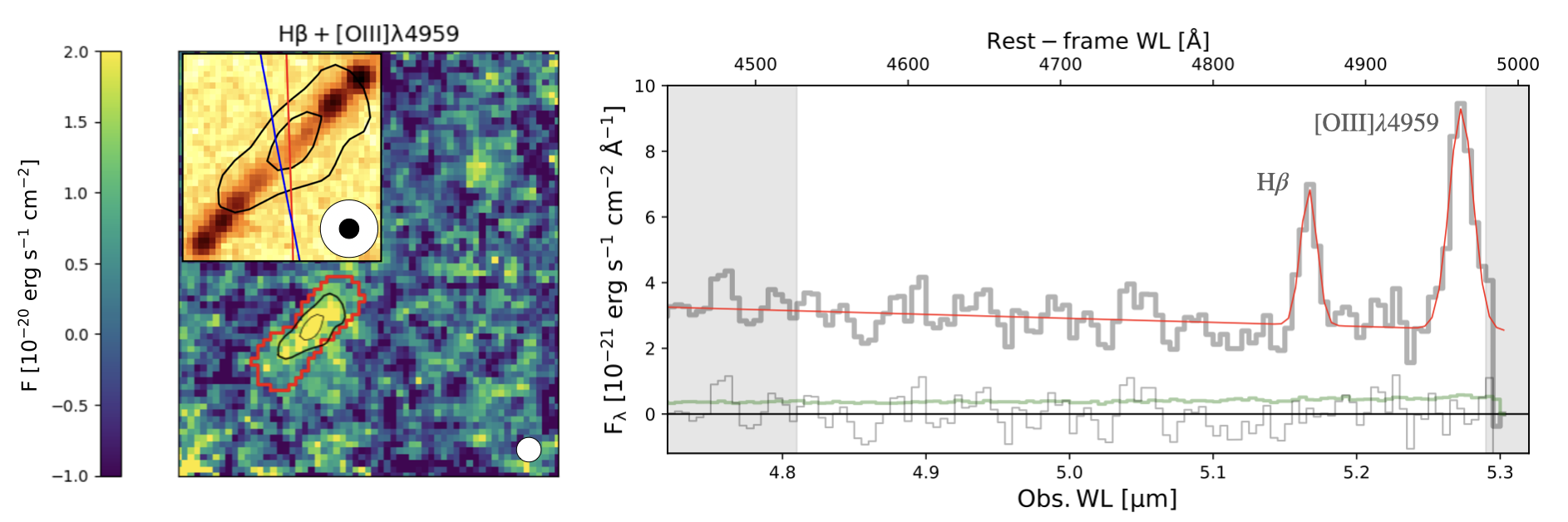}
    \caption{{\it Left:} 2D map of the stacked \Hb\ + \OIIIb\ in the wavelength range 5.15-5.3~$\mu$m (see text for more details). The red mask corresponds to the regions {\it BCDE1,2} and {\it CC} (defined in Fig.~\ref{fig:nircam_nirspec}), used to extract the spectrum on the right. The inset shows the NIRCam F150W+F200W data with overlaid black contours showing the peak of the \Hb\ + \OIIIb\ emission and the CCs from \textsc{glafic} (blue) and \lenstool\ (red). The white circle (in both the map and the inset) denote the FWHM of the NIRSpec-IFU PSF, while the NIRCam one is plotted as a black circle in the inset. {\it Right:} 1D spectrum (thick gray line) in the wavelength range of the detected emission lines. The continuum and line best-fitting result are shown as a red line. The error spectrum and spectral fit residuals are shown in green and light gray, respectively. The gray bands mark the portion of the spectrum not included in the fit.}
    \label{fig:map_redshift}
\end{figure*}

The 1D spectrum of the Cosmic Gems (Fig.~\ref{fig:nircam_nirspec}) is characterized by the clear \lya\ break, used in previous publications to constrain photometrically the redshift of the system ($\rm z_{phot}=10.2$, \citealp{Bradley25}) and by two faint emission lines, which we identify as \Hb\ and \OIIIb.
The spectroscopic redshifts derived from the two lines are $\rm z_{H\beta}=9.625\pm0.002$ and $\rm z_{[OIII]}=9.630\pm0.001$. Their redshift difference (consistent within $\rm 3\sigma$) is well within the spectral resolution at $\rm \sim5~\mu m$\footnote{The redshift difference $\rm \Delta z=0.005$ converts into $\rm \Delta v=140~km~s^{-1}$, while the resolving power of the prism at $\rm \sim5~\mu m$ ($\rm R\sim300$) corresponds to $\rm \Delta v(R)=1000~km~s^{-1}$.}.
We decided to keep $\rm z_{H\beta}=9.625$ as the reference redshift of the Cosmic Gems galaxy, due to the \OIIIb\ line being very close to the edge of the wavelength range detected (the red tail of the line was excluded from the fit, as shown in Fig.~\ref{fig:map_redshift}).

The spectroscopic redshift is lower than the photo-z previously published by \citet{Bradley25}.
The main reason behind this difference seems to be the part of the spectrum affected by strong \lya-damping wings. The shape of the observed spectrum is not accurately described by the sharp drop given by the intergalactic medium (IGM) absorption prescriptions by \citet{inoue2014}, adopted as standard prescription in the \texttt{EAZYPY}, \citep{brammer2008_eazy} photo-z fitting code used in \citet{Bradley25}\footnote{The same prescription is used also in the \texttt{Bagpipes} code used to fit the spectra in the current work. See Section~\ref{sec:analysis:bcde}.}. This effect has been widely discussed in the literature \citep[e.g.,][]{Fujimoto2023,Finkelstein2024,Hainline2024,Heintz2025}.
A deeper analysis of the \lya\ damping along the Cosmic Gems arc was performed in a separate study (i.e., Christensen et al., in prep.) and is briefly discussed also in the appendix of \citetalias{Vanzella2025}.
In brief, a very strong damping wing from hydrogen absorption is observed in the regions occupied by the clusters ({\it A} and {\it BCDE}), revealing high hydrogen column densities, $\rm log~N$(H\protect\scaleto{$I$}{1.2ex})$\rm /[cm^{-2}] \sim 22.4$. Densities an order of magnitude lower are found in the more diffuse regions of the arc (i.e., the {\it TAIL}). 

We created emission-line maps of \Hb\ and \OIIIb\ by collapsing the spectral channels including the lines, after continuum subtraction. For each pixel, the continuum level was estimated as the median of the slices in the (observed) wavelength range \hbox{5.20--5.24~$\mu$m}. The slices containing the lines are in ranges 5.16--5.18 and \hbox{5.27--5.29~$\mu$m} for \Hb\ and \OIIIb, respectively. 
The individual emission-line maps were combined in a \Hb$+$\OIIIb\ map shown in Fig.~\ref{fig:map_redshift} (left panel). The ionized gas emission is located in proximity of the critical line intersecting the arc, and it extends partly toward the {\it BCDE} region.
The spectral wavelengths containing the emission lines from this region of larger emission (combining the masks of {\it BCDE1,2} and {\it CC}, visualized as a red contour in the left panel), are shown in Fig.~\ref{fig:map_redshift} (right panel). The best fit of this spectrum (red line), modeled as the combination of a linear continuum and two Gaussian lines, was used to determine the spectroscopic redshift quoted above.

\subsection{BCDE regions}\label{sec:analysis:bcde}

\subsubsection{Main spectral features}
We extracted the spectrum for the region occupied by clusters B, C, D, and E from the {\it BCDE1} and {\it BCDE2} aperture masks shown in Fig.~\ref{fig:nircam_nirspec}. The stacked spectrum from both regions (dubbed {\it BCDE1,2}) is shown in Fig.~\ref{fig:bagpipes_bcde12}.
We extracted three main features of the spectrum, namely the EW of the emission lines, the Balmer break (Bb), and the UV slope ($\rm \beta_{UV}$). All are reported in Appendix~\ref{app:tab:spectroscopy}. 
The EWs of \Hb\ and \OIIIb\ were derived by simultaneously fitting a linear continuum and two Gaussians to the observed spectrum, as in Section~\ref{sec:map_redshift}, to derive the redshift of the galaxy. The resulting rest-frame EW values are EW(\OIIIb)$=47^{+6}_{-5}~\AA$ and EW(\Hb)$=18^{+3}_{-2}~\AA$. By assuming the intrinsic ratio of the [\OIII] lines ($\rm \lambda5007/\lambda4959=2.98$), we derived the expected flux in \OIIIa and consequently the $\rm R3\equiv[OIII]\lambda5007/H\beta$ ratio. The value $\rm R3=7.5\pm1.8$, suggests that the metallicity of the galaxy is in the range $\rm Z_{gas}\sim10-20\%~Z_\odot$ (assuming standard R3 metallicity calibrations, e.g., \citealp{nakajima2022,Sanders24,Sanders25arXiv}).

Following the prescription used in \citet{binggeli2019} and \citet{Endsley25}, we quantified the Balmer break amplitude as $\rm F_\nu(4200~\AA)/F_\nu(3500~\AA)$. In order to mitigate the variability of monochromatic values, we considered the fluxes averaged over the rest-frame ranges from $\rm 4050$ to $\rm 4330~\AA$ and from $3290$ to $\rm 3580~\AA$. A significant break was recovered, (Bb: $1.3\pm0.1$).
Finally, the rest-UV slope was fit in the range from $\rm 1600$ to $\rm 2600~\AA$, avoiding the wavelength region close to the \lya\ damping in the blue part. The best-fit value, $\rm \beta_{UV}=-2.53\pm0.03$, is steep but in line with average UV slopes found in $\rm z\gtrsim8$ galaxies \citep[e.g.,][]{cullen2023, Donnan2025, Tang2025}.

The values derived for the main spectral features (summarized in Appendix~\ref{app:tab:spectroscopy}) provide a powerful insight into the age of the system and of its SFH \citep[following e.g.,][]{Endsley25}. We can assume two extreme and opposite SFH cases, namely a constant star formation and an instantaneous burst (single stellar population, SSP), encompassing more complex SFHs evaluated below. In spite of the SFH assumption, the steep UV slope requires the system to be younger than $\rm \lesssim30~Myr$ in the two cases (assuming no extinction). On the other hand, the relatively low EWs (the total value for the sum of the $\rm H\beta+[\OIII]$ lines\footnote{The flux of the \OIIIa\ line is derived assuming again $\rm \lambda5007/\lambda4959=2.98$.} is $\rm 198~\AA$) imply an age $\rm \gtrsim7~Myr$ for an SSP, and much older ($\rm  \gtrsim500~Myr$) for constant SFHs. The small Balmer break observed is also consistent with a stellar age in the range $\rm \sim10-20~Myr$, in the SSP case. Long SFHs are disfavored, as the old ages required to explain the observed Bb and EWs under this assumption (up to $\rm \gtrsim500~Myr$ for a constant SF) are incompatible with the observed blue UV slope.

An absorption feature was detected in the spectrum near the expected wavelength of $\rm CIV~\lambda\lambda 1548,1550$~\AA. The absorption peak was observed at $1.628~\mu m$, where the prism mode provides its lowest spectral resolution ($\rm R \sim 40$), corresponding to $\sim 10,000$ \kms. Assuming this feature arises from $\rm CIV$ absorption, it would be blueshifted relative to the systemic redshift (as traced by optical emission lines) by $\Delta v \simeq -3200$~\kms (see Appendix~\ref{sec:app:all_spectra}). Given the large blueshift, the relatively low signal-to-noise ratio, and the limited spectral resolution at $1.6~\mu m$, we caution against a definitive interpretation of the presence of the $\rm CIV$ doublet in absorption. Higher spectral resolution spectroscopy is necessary to confirm this feature and, more generally, to investigate the potential presence of faint, narrow ultraviolet emission and/or absorption lines.

\subsubsection{Spectral fitting}
\begin{figure}
    \centering
\includegraphics[width=1.\columnwidth]{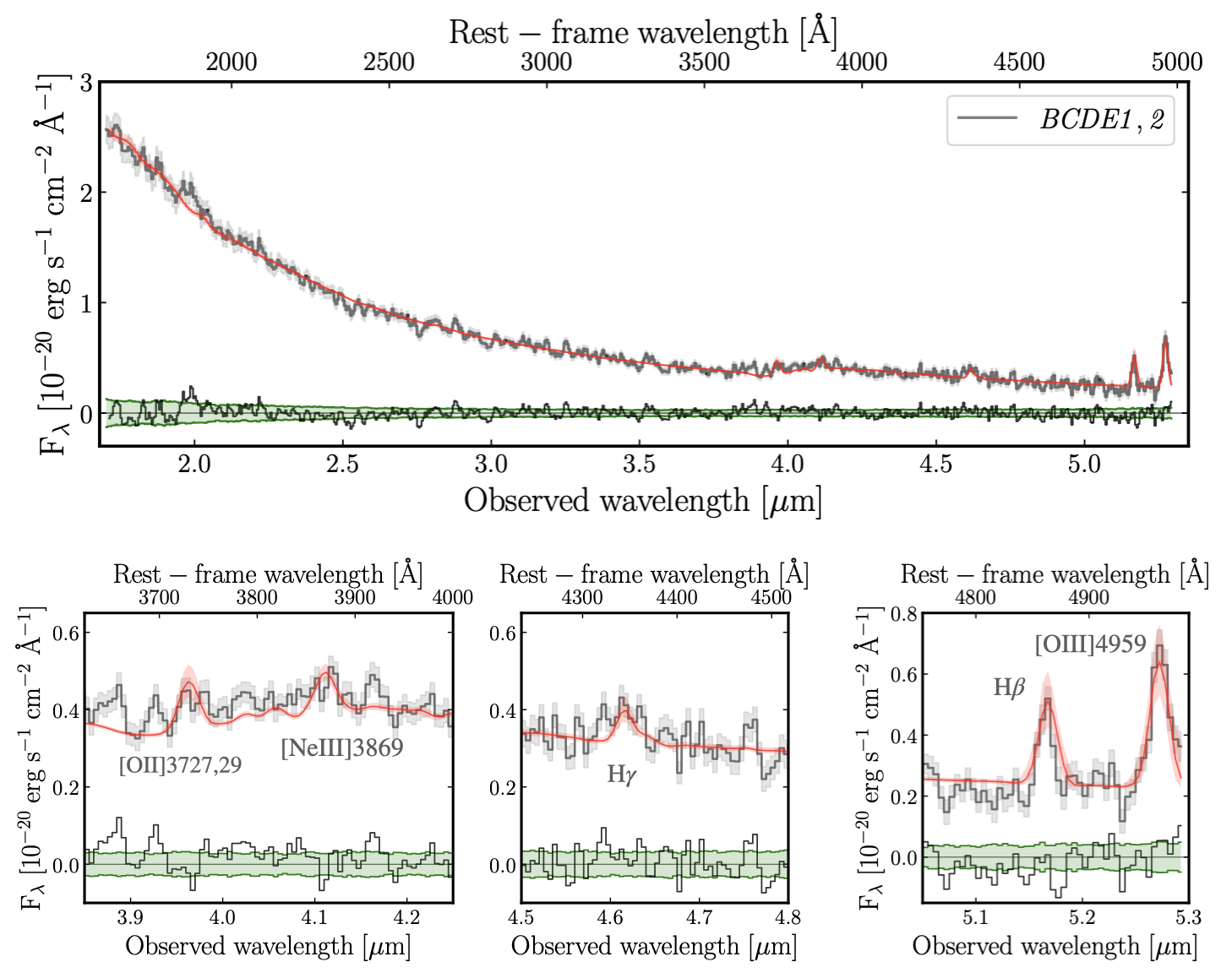}
    \caption{Spectrum of the combined BCDE12 regions (gray), along with the best-fit model (red), residuals (black), and $\pm1\sigma$ uncertainties (green). Bottom panels: Zoom-ins on the regions where the best-fit predicts the main emission lines.}
    \label{fig:bagpipes_bcde12}
\end{figure}

To further constrain the intrinsic properties of the region, we fit the spectrum with \texttt{Bagpipes} \citep{carnall2018_bagpipes,carnall2019_bagpipes}, using BPASS stellar models \citep{eldridge2017,stanway2018}, assuming a delayed exponentially decaying SFH and leaving the following free parameters: the age and the exponential index of the SFH ($\tau$); the stellar mass formed and its metallicity; the extinction $\rm A_V$ (with the \citealp{calzetti2000} attenuation law assumed); and the ionization parameter $\log(U)$. Redshift was allowed to vary within the range 9.625--9.630, to account for the slight redshift difference between the two observed lines (see Section~\ref{sec:map_redshift}). To account for possible underestimation, the spectral uncertainties were left free to be scaled by up to a factor of 50.
We used flat priors in the logarithmic space for all parameters excluding the extinction.
Finally, the velocity dispersion was left as a free parameter. We point out that the \bagpipes\ model spectra were adjusted to match the line spread function of the NIRSpec observations, before fitting.
We restricted the fit to wavelengths longer than $1600~\AA$ rest-frame ($1.7~\mu m$ observed) to avoid the part affected by \lya\ damping. 
The spectrum and best-fit model are shown in Fig.~\ref{fig:bagpipes_bcde12}. The model nicely reproduces the two emission lines, the small Balmer break and the UV slope. It also suggests the possible presence of other emission lines, namely $\rm H\gamma$, [O\protect\scaleto{$II$}{1.2ex}]$\rm \lambda\lambda3727,3729$ and [Ne\protect\scaleto{$III$}{1.2ex}]$\rm \lambda3869$. We zoom-in on their respective regions of the spectrum in the bottom-left panels of Fig.~\ref{fig:bagpipes_bcde12}. While these fainter lines are not robustly detected, their expected flux is consistent with the $\rm 1\sigma$ uncertainty given by the spectral fluctuations. 

The mass-weighted age of the region is constrained within a narrow range, $10-15$ Myr ($\rm16th$ to $\rm84th$ percentiles). The small value of the exponential index ($\rm \tau=~3$ Myr) robustly indicates that the region experienced a short burst of star formation. The SFH of the regions within the Cosmic Gems is further discussed in Section~\ref{sec:discussion:bcde12_a12}. The region is only mildly extincted ($\rm A_V=0.23\pm0.02$ mag) and has a best-fit metallicity $\rm Z=0.16\pm0.01~Z_\odot$, consistent with what is suggested by the R3 index previously discussed. Given the low spectral resolving power of the PRISM at the longest wavelengths ($\rm R\sim320$ at $\rm 5.3~\mu m$ corresponding to $\rm \sigma_v\approx1000~km/s$), the emission lines remain unresolved in the fit. 

In order to convert the best-fit observed stellar mass into an intrinsic value, we accounted for lensing. From the magnification map discussed in Section~\ref{sec:lens_model} we assumed a magnification of $\rm \mu_{tot}=137.7^{+20.6}_{-17.7}$ for the {\it BCDE1,2} region\footnote{This value is the average of the magnifications extracted for {\it BCDE1} and {\it BCDE2} in the Appendix~\ref{sec:app:mul_image}}. We took into account that our mask aperture included the flux from two mirrored images of the same region. The resulting intrinsic mass is $\rm log(M_\star/M_\odot)= 6.7\pm0.1$. This makes $\sim10\%$ of the mass budget of the arc, as reported in \citet{Bradley25} and further discussed in Section~\ref{sec:discussion:full_arc}.

Finally, according to the reference lens model, the {\it BCDE} region was characterized by a strong magnification gradient ranging approximately from $\rm \mu\sim80$ to $\rm \mu\gtrsim300$ (Appendix~\ref{sec:app:mul_image}).
To test the effect of this gradient on the results, we repeated the spectral analysis by dividing {\it BCDE1,2} into two subparts,
{\it BC1,2} and {\it DE1,2} (covering the positions of the respective star clusters).
The mask apertures used and the direct comparison between the two 1D spectra are discussed in Appendix~\ref{sec:app:bc_vs_de}. Overall, the main spectral features are very similar, with the {\it DE1,2} region having only slightly larger EW values than {\it BC1,2}. As a consequence, also the results of the spectral energy distribution (SED) fitting remain consistent in terms of age and duration of the burst. The fit results are listed in Appendix~\ref{app:tab:spectroscopy} and further discussed in Appendix~\ref{sec:app:bc_vs_de}.

\subsection{Region containing cluster A}
We repeated the analyses from the previous section for the spectrum extracted from the {\it A1,2} mask aperture, covering the two images of cluster A (the spectrum shown in Appendix~\ref{sec:app:all_spectra}). In this case, only a faint \Hb\ line ($3\sigma$ detection, resulting in $\rm EW(H\beta)=25^{+15}_{-7}~\AA$) emerges from the spectrum.
For \OIIIb\ we can only derive an upper limit $\rm EW([\OIII])<40~\AA$. We point out that, due to the lower continuum flux compared to the {\it BCDE1,2}, the limits derived in this case are almost as large as the detections of the previous section. The UV slope and the Balmer break also have values similar to the ones derived for {\it BCDE1,2}. The latter, however, has a large associated uncertainty ($\rm 1.2\pm0.4$), making it consistent with an undetectable Bb (i.e., a ratio of $1.0$). 

The best-fit \texttt{Bagpipes} model also returns similar properties as for {\it BCDE1,2}: the (mass-weighted) age is slightly older ($\rm 15^{+3}_{-2}~Myr$) but still with a SFH characterized by a short burst ($\rm \tau=2\pm1~Myr$); the attenuation remains low ($\rm 0.2\pm0.1$ mag). On the other hand, the metallicity of this region is lower (but within $2\sigma$ uncertainties). Given the absence of emission lines, we consider it to be poorly constrained. The best-fit mass for this region (after accounting for delensing, with $\rm \mu_{tot}=50.1^{+4.2}_{-3.6}$) is $\rm log(M_\star/M_\odot)=6.6\pm0.1$, which is very similar to that found for {\it BCDE}.

\subsection{The tail of the arc}
The northwest region of the arc\footnote{Only a fraction of the tail of the arc is observed with the IFU in the southeast region. In order to have a complete view of the region we only considered the northwest image.} (which we refer to as the {\it TAIL}) is characterized by a faint continuum without emission lines (see Appendix~\ref{sec:app:all_spectra}), for which we derived EW upper limits (see Appendix~\ref{app:tab:spectroscopy}). The UV slope of this region is the steepest observed within the arc, $\rm \beta=-2.7\pm0.1$, while no Balmer break is observed ($\rm Bb=1.0\pm0.5$). 
The best fit of the spectrum suggests an older region (with an age of $43^{+17}_{-14}$ Myr) described by a long SF episode ($\rm \tau=180^{+185}_{-107}$ Myr) compared to the ones populated by clusters ({\it BCDE} and {\it A}). The SFH is, however, much less constrained, with large uncertainties. 
Due to the lower magnification of the tail ($\rm \mu\sim20$), compared to the previous regions, its mass ($\rm log(M_\star/M_\odot)= 7.3\pm0.1$) comprises a considerable fraction of the mass of the entire arc, despite its faintness. In Section~\ref{sec:discussion:tail}, we further discuss the best-fit of the spectrum of the {\it TAIL} region and its derived SFH.

\subsection{The critical curve region}
The spectrum of the \textit{CC} region closely resembles that of the nearby \textit{BCDE} region but exhibits larger EWs: $\mathrm{EW}(\mathrm{H}\beta)=38$ \AA\ and $\mathrm{EW}([\mathrm{O,III}]\lambda4959)=75$ \AA, for a total EW(\Hb+[\OIII]$\rm )=338$ \AA.
These values, however, remain modest compared to the extreme emitters typically reported at high redshift \citep[e.g.,][]{Tang2025}. The Balmer break in this region is poorly constrained, with $\mathrm{Bb}=1.4\pm0.5$. 
An additional feature in emission corresponding to the doublet $\rm CIII]\lambda\lambda1907,1909$ (with $\rm EW=12^{+3}_{-2}~\AA$) is also possibly detected, see Appendix.~\ref{sec:app:all_spectra}.

One plausible scenario is that the observed nebular emission in the \textit{CC} region is ionized by one of the nearby stellar clusters. Given the extreme lensing magnification, the observed angular separation between \textit{CC} and the closest cluster (\textit{E}), approximately $0\farcs12$, corresponds to only $\sim1$ pc in the source plane\footnote{This calculation assumes a conservative magnification of $\mu=320$, corresponding to the magnification at the position of cluster \textit{E} (see Appendix~\ref{sec:app:mul_image}).}. 
Alternatively, the nebular emission may originate from in situ star formation of an \HII\ region very close to the caustic and appearing as an unresolved source on NIRSpec data and still stellar-continuum-undetected on NIRCam. Assuming \textit{CC} is unresolved in NIRSpec (FWHM~$\sim0\farcs2$), the intrinsic size of the emitting region would be $\lesssim0.9$ pc (for $\mu>500$).

In both cases, given the very large magnification, the contribution of \textit{CC} to the global arc emission (e.g., {\it BCDE1,2}) is intrinsically modest. As shown in Appendix~\ref{app:tab:spectroscopy}, the total spectrum from {\it BCDE1,2}$+${\it CC} is very similar to the {\it BCDE1,2} only, and returns similar physical intrinsic properties.

We refrain from inferring detailed physical properties for such a compact \ion{H}{ii} region, as this would require forward modeling under highly critical lensing conditions. Moreover, the spectral energy distribution and spectral fitting are limited by the absence of a clear counterpart in NIRCam imaging.

Despite these limitations, the \texttt{Bagpipes} fit to the \textit{CC} spectrum suggests older stellar populations compared to the other clumps, with a mass-weighted age of $95\pm14$ Myr and an extended star formation timescale ($\tau\sim300$ Myr). Such a prolonged episode of star formation appears implausible for a region of this compactness at $z\sim9.6$, where sizes are comparable to young massive star clusters (YMSCs). The mass upper limit for \textit{CC}, reported in Appendix~\ref{app:tab:spectroscopy}, combines the best-fit stellar mass with a conservative lower limit on magnification, resulting in $\rm M\lesssim5\times10^5~M_\odot$. A more refined lensing model and deeper imaging will be essential to improve these constraints and to securely identify any associated stellar counterpart.

\subsection{The intrinsic spectrum of the entire arc}\label{sec:discussion:full_arc}
In order to have a spectroscopic view of the Cosmic Gems arc as a whole, we analyze its spectrum extracted from one of its mirrored images (image 2, the northwest one, as it is better covered by the IFU pointing; see the mask aperture used in the bottom-central panel of Fig.~\ref{fig:nircam_nirspec}). Before the extraction, the flux of the cube was delensed spaxel-by-spaxel by assigning to each spaxel the corresponding magnification value as extracted from the reference lens model (see Appendix~\ref{sec:app:mul_image}). While still inaccurate\footnote{This methodology considers each pixel independently and therefore does not account for the PSF spreading the flux of compact regions over several pixels.}, this precaution allowed us to produce a spectrum, which is less biased toward the highly magnified regions close to the CC.
The delensed spectrum is overall similar to the {\it TAIL} region (i.e., the least magnified one). Its properties are provided in Appendix~\ref{app:tab:spectroscopy}, where the region is dubbed {\it ARC2}. Its best-fit (using the same assumptions as in Section~\ref{sec:nirspec_analysis}) returns a mass-weighted age of $\rm 68^{+17}_{-18}$ Myr (consistent with the age derived in the tail) and a stellar mass (already intrinsic by construction) $\rm log(M)=7.68^{+0.06}_{-0.06}$, slightly larger than that found in the tail region. The mass is within the range estimated from NIRCam SED in \citet{Bradley25}, $\rm M=(2.4-5.6)\times10^7~M_\odot$. The summed intrinsic masses from the individual mask apertures analyzed previously ({\it BCDE}, {\it A}, {\it TAIL} and {\it CC}) account for $75\%$ of the {\it ARC2} mass. 
The best-fit $\rm \tau=148^{+174}_{-98}$ Myr is in between the values found for the central regions of the arc and for the tail in the delayed-$\tau$ prescription, but remains unconstrained.

\section{The star clusters of the Cosmic Gems}\label{sec:ysc_main}

\subsection{Revising the properties of individual clusters}\label{sec:ysc}
A detailed analysis of the individual star clusters was already presented in \citet{adamo2024a}. We repeated the SED fitting analysis of the NIRCam photometry reported in \citet{adamo2024a} by fixing the redshift to the new $\rm z_{spec}=9.625$ and using the updated \glafic\ lens model to derive the intrinsic star cluster masses, densities and radii.
The fit was performed using \texttt{Bagpipes} with BPASS stellar populations models and the following two main assumptions. First, the SFHs of the clusters were assumed to be an exponential decay with a fixed $\rm \tau=1~Myr$ to simulate an SSP. Second, the dust attenuation was modeled with the \citet{calzetti2000} prescription. The output properties of the fit are summarized in Tab.~\ref{tab:ysc}.  
The (mass-weighted) ages of the clusters range between 7 and 27 Myr. They are, within uncertainties, consistent with the ages derived from the spectra of the corresponding regions. The one deviating the most from the "spectral" age is cluster E.1, with an age of $27^{+14}_{-9}$ Myr (compared to $12^{+3}_{-2}$ Myr for the \textit{BCDE} region, and $9^{+1}_{-1}$ Myr for the {\it DE} mask, discussed in Section~\ref{sec:app:bc_vs_de}). We point out that this cluster is also the faintest (resulting in large uncertainties in the SED fitting), implying that its overall contribution in the region, compared to clusters B, C, and D, is low and thus has minimal impact on the spectrum. 

\begin{figure}
    \centering
    \includegraphics[width=\columnwidth]{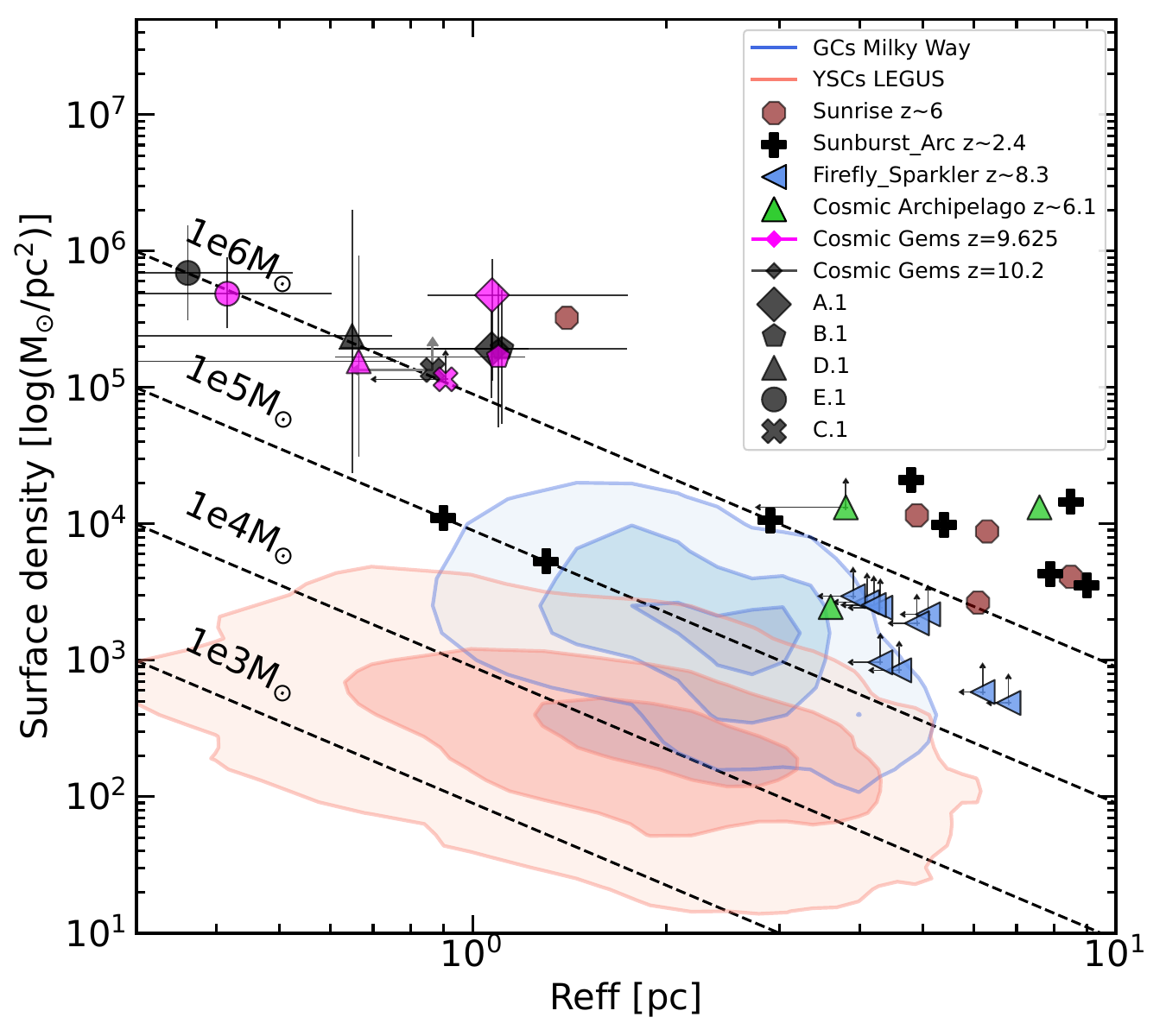}
    \caption{Stellar mass surface density ($\Sigma_{mass}$) and effective radii ($\rm R_{eff}$) of the star clusters discussed in this work (magenta symbols), along with the following collection of gravitationally lensed star clusters from literature: Sunburst arc \citep{vanzella2022}, Sunrise arc \citep{vanzella2023_sunrise}, Cosmic Archipelago \citep{Messa2025}, and Firefly Sparkle \citep{mowla2024}. A comparison with local YMCs from \citealp{brown2021} (red contours) and Milky Way globular clusters (\citealp{baumgardt2018}, blue contours) shows the high surface densities in high-z clusters when compared to local ones (see also the dotted lines marking the constant stellar mass tracks). For the Cosmic Gems clusters, the black symbols refer to the values published in \citet{adamo2024a} assuming $\rm z_{phot}=10.2$.}
    \label{fig:ysc_size_density}
\end{figure}

In Fig.~\ref{fig:ysc_size_density}, we plot the newly recovered radii and stellar densities as compared to the previously published values.  Overall, the main results of \citet{adamo2024a} remain valid. With the updated analysis, the clusters remain very compact ($\rm R_{eff}\lesssim1$ pc) and massive ($\rm M_\star\sim10^6~M_\odot$), leading to extreme densities ($\rm \Sigma_{M_\star}\sim10^5-10^6~M_\odot~pc^{-2}$) rarely observed in star clusters at any redshift. Given the similarity in the derived star cluster properties, the main conclusions of \citet{adamo2024a} still hold. 

\label{sec:ysc}
\begin{table*}
    \caption{\label{tab:ysc}Updated main properties of the individual star clusters in the Cosmic Gems.}
    \renewcommand{\arraystretch}{1.35}
    \centering
    \begin{tabular}{lrrrrrrrr}
    \hline
    \multicolumn{1}{c}{ID} & \multicolumn{1}{c}{$\rm \mu_{tot}$} & \multicolumn{1}{c}{$\rm \mu_{tan}$} & \multicolumn{1}{c}{$\rm R_{eff}$} & \multicolumn{1}{c}{Age} & \multicolumn{1}{c}{$\rm log(M/M_\odot)$} & \multicolumn{1}{c}{Av} & \multicolumn{1}{c}{$\rm Z/Z_\odot$} & \multicolumn{1}{c}{$\rm \Sigma_M$}  \\
    \multicolumn{1}{c}{} & \multicolumn{1}{c}{} & \multicolumn{1}{c}{} & \multicolumn{1}{c}{[pc]} & \multicolumn{1}{c}{[Myr]} & \multicolumn{1}{c}{} & \multicolumn{1}{c}{[mag]} & \multicolumn{1}{c}{[\%]} & \multicolumn{1}{c}{$\rm [10^5~M_\odot pc^{-2}]$}  \\
    \hline
    A.1     & $47.9^{+4.0}_{-4.0}$ & $46.5^{+3.1}_{-4.3}$ & $1.1^{+0.7}_{-0.2}$ & $15^{+7}_{-9}$  & $6.79^{+0.15}_{-0.59}$  & $0.28^{+0.10}_{-0.15}$  & $5.8^{+2.7}_{-3.4}$ & $4.8^{+4.1}_{-3.6}$	\\
    B.1     & $92.3^{+11.7}_{-10.6}$ & $89.3^{+10.2}_{-10.9}$ & $1.1^{+0.1}_{-0.5}$ & $10^{+6}_{-3}$  & $6.35^{+0.23}_{-0.34}$  & $0.20^{+0.13}_{-0.13}$  & $2.6^{+4.0}_{-1.9}$ & $1.7^{+4.1}_{-1.2}$ \\
    C.1     & $124.5^{+19.7}_{-16.8}$ & $120.0^{+17.7}_{-16.8}$ & $<0.9$ & $7^{+5}_{-2} $  & $6.02^{+0.35}_{-0.35}$  & $0.35^{+0.10}_{-0.16}$  & $4.4^{+3.4}_{-2.4}$ & $>1.1$ \\
    D.1     & $166.6^{+31.7}_{-26.5}$ & $162.4^{+31.0}_{-27.5}$ & $0.7^{+0.1}_{-0.6}$ & $7^{+10}_{-3}$  & $5.89^{+0.44}_{-0.43}$  & $0.33^{+0.11}_{-0.17}$  & $5.2^{+3.1}_{-2.6}$ & $1.6^{+7.7}_{-1.2}$ \\
    E.1     & $323.5^{+125.5}_{-82.4}$ & $308.3^{+120.1}_{-76.3}$ & $0.4^{+0.2}_{-0.1}$ & $27^{+14}_{-9}$ & $5.97^{+0.13}_{-0.15}$  & $0.26^{+0.16}_{-0.18}$  & $4.7^{+3.5}_{-3.2}$ & $4.9^{+4.1}_{-2.2}$ \\
    A.2     & $52.2^{+4.4}_{-3.2}$ & $50.0^{+4.2}_{-3.6}$ & $1.7^{+0.8}_{-0.4}$ & $15^{+7}_{-5} $ & $6.49^{+0.16}_{-0.24}$  & $0.17^{+0.14}_{-0.10}$  & $4.3^{+3.2}_{-3.0}$ & $1.0^{+1.4}_{-0.5}$ \\
    B.2     & $90.3^{+7.3}_{-7.4}$ & $87.1^{+7.0}_{-7.6}$ & $<1.3$ & $8^{+5}_{-3}$   & $6.47^{+0.29}_{-0.38}$  & $0.30^{+0.13}_{-0.15}$  & $4.3^{+3.1}_{-2.3}$ & $>1.7$ \\
    C.2+D.2 & $132.5^{+16.7}_{-14.4}$ & $126.5^{+17.2}_{-13.6}$ & $1.9^{+11.1}_{-1.8}$ & $17^{+7}_{-7}$  & $6.60^{+0.12}_{-0.23}$  & $0.20^{+0.15}_{-0.14}$  & $4.8^{+3.5}_{-3.1}$ & $1.0^{+6.5}_{-0.8}$ \\
    \hline
    \end{tabular}
    \tablefoot{Main updated properties of the five clusters (with their counterimages when detected) discussed in \citet{adamo2024a}. Age, mass, extinction, and metallicity of the clusters were derived from broadband SED fitting performed with \texttt{Bagpipes} (details in the main text).}
\end{table*}

\subsection{New cluster candidates in the tail of the Cosmic Gems}\label{sec:new_ysc}
\begin{figure}
    \centering
    \includegraphics[width=\columnwidth]{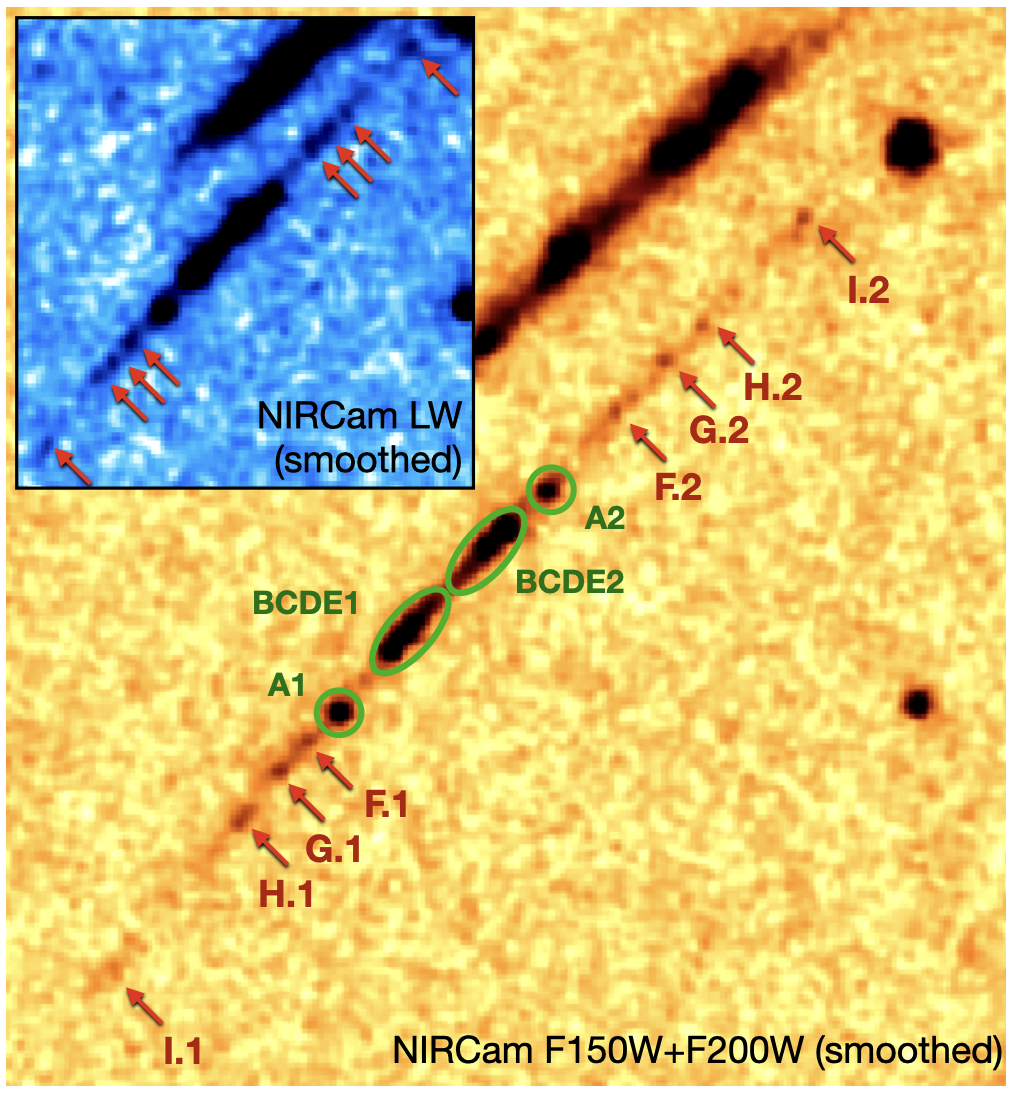}
    \caption{Sum of the F150W and F200W NIRCam observations, smoothed by a two-pixel Gaussian kernel to better visualize the inhomogeneities along the arc. The clusters discussed in \citet{adamo2024a} and Section~\ref{sec:ysc} of the current work are marked by green ellipses. The red arrows point to the position of the regions (detected at the $\rm 3-4\sigma$ level) discussed in the text. The same regions are also indicated in the inset panel, showing a smoothed version of the sum of the LW NIRCam filters available (F277W, F356W, F410M, and F444W).
    }
    \label{fig:new_ysc}
\end{figure}

The tail regions of the Cosmic Gems arc are characterized by unsmooth emission. Figure~\ref{fig:new_ysc} shows a rest-UV image obtained by stacking the F150W and F200W NIRCam bands and applying a two-pixel Gaussian smoothing to enhance the inhomogeneities. While some of these inhomogeneities can be due to noise fluctuation, we identify four compact regions, which we name F to I, detected at the 3-4$\rm \sigma$ level. These can be reliably considered compact regions of the Cosmic Gems, as they are observed in both mirrored images of the arc and are detected also in the long-wavelength (LW) filters (see the inset of Fig.~\ref{fig:new_ysc}).
These identified peaks are observationally faint, making it impossible to robustly derive their sizes. If we assume they are unresolved sources, considering the tangential magnifications in the tail (in the range $\rm \mu_{tan}\simeq 10-30$ according to the fiducial \glafic\ lens model), their compactness results in estimated upper limits on intrinsic radii ranging from 3 to 10 pc. A forward-modeling reconstruction of the arc (see next Section) also suggests that these regions should be smaller than $\rm \lesssim10$ pc in radii; otherwise, they would have a smoother morphology in NIRCam observations. 
Their measured observed magnitudes are in the range $\rm m_{AB}=28.3-29.3$ mag, corresponding to $\rm m_{AB}=32.0-33.0$ mag when delensed. Accounting for the redshift of the galaxy, we thus find that the magnitudes of these additional candidate clusters range from $\rm M_{AB}=-14.5$ to $\rm M_{AB}=-15.5$ mag. Assuming their ages are similar to the ones of the clusters discussed in Section~\ref{sec:ysc} ($\sim10-30$ Myr), their intrinsic masses should be $\rm M_\star\sim(0.5-8.0)\times10^6~M_\odot$. According to their mass and radius estimates, these faint sources could be further massive clusters formed in the Cosmic Gems. 
Deeper observations would be necessary to give more robust constraints on their properties.
The cluster mass function in the Cosmic Gems is further discussed in \citetalias{Vanzella2025}.

\subsection{Forward modeling reconstruction of the arc}
\begin{figure*}
    \centering
    \includegraphics[width=\textwidth]{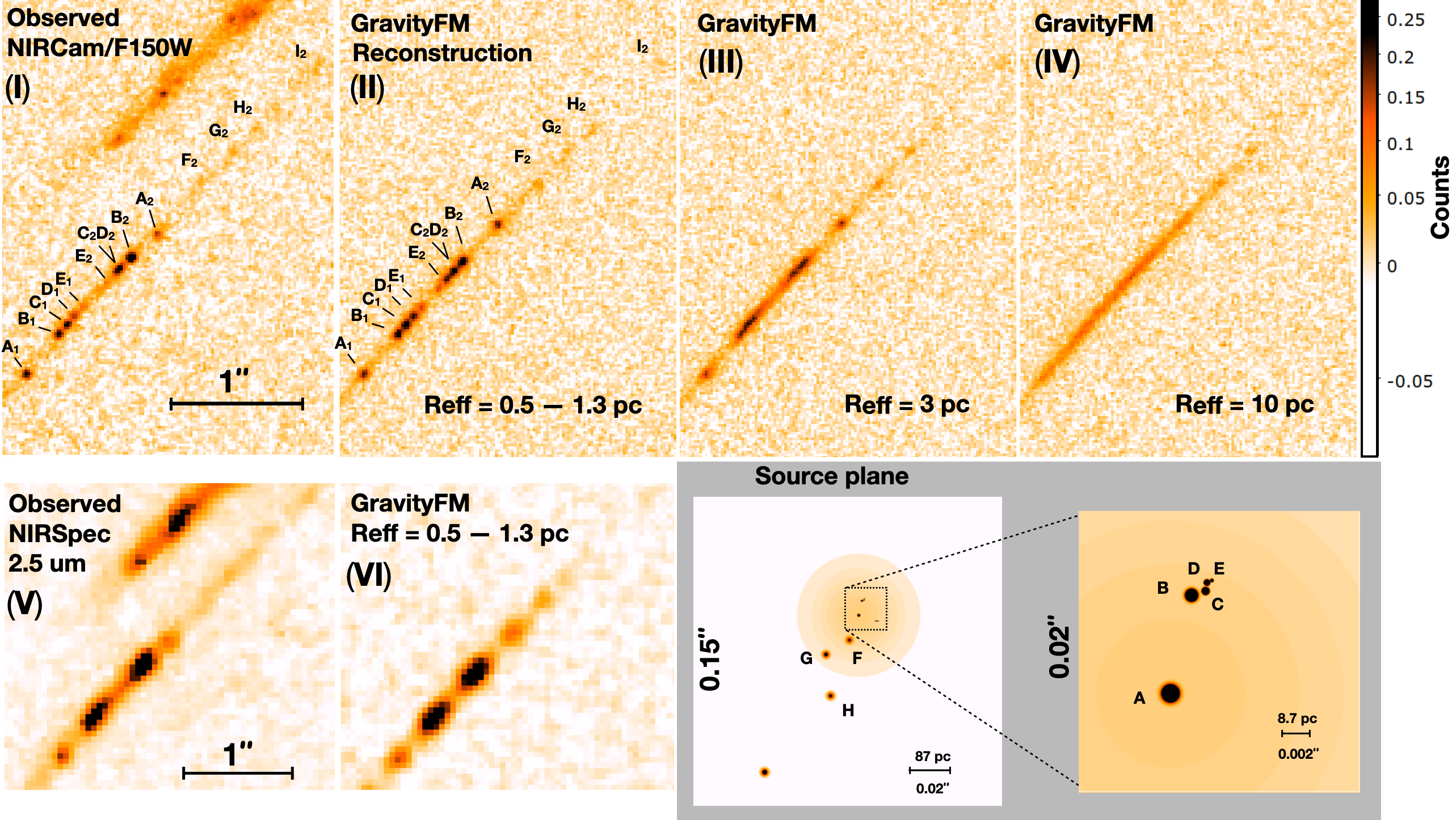}
    \caption{Reconstruction of the Cosmic Gems arc in the image and source planes with the GravityFM tool (Bergamini et al., in prep.). Top panels, from left to right: Observed F150W image (I); the arc reproduced by inserting radii of the clusters in the source plane spanning the range $0.5-1.3$ parsec (II, F150W PSF, with noise added); the same, with each assuming radii of 3 parsec (III); and the same, with each assuming radii of 10 parsec (IV). Bottom left panels: Same as in the top panels, but for NIRSpec data (V and VI, see text for more details). Bottom-right panels: Source plane reconstruction and the modeled sources inserted in the forward modeling reproducing the observed arc shown in panel II.}
    \label{fig:forward_mod}
\end{figure*}

The reference \glafic\ lens model was used to reconstruct the Cosmic Gems arc via a forward-modeling approach, based on the \texttt{GravityFM} software (Bergamini et al., in prep.). Sources with intrinsic sizes matching the values retrieved for the clusters (Tab.~\ref{tab:ysc}) were injected in the source plane, and their position and fluxes were chosen to match the observed NIRCam/F150W observations in the image plane (Fig.~\ref{fig:forward_mod}). In addition, an extended diffuse component was added to account for the diffuse light observed between the clusters (both in the case of NIRCam and of the NIRSpec). The exact origin of this diffuse light is unconstrained; here, we associated an emitting halo centered on cluster A with a radius of 65 parsec. Observational noise and the PSF effect were also accounted for, to ease the comparison with observations. Overall, this approach returns a good representation of the star clusters in the arc. We also notice that the forward modeling returns almost perfect mirrored images of the arc, while, in the case of observations, clusters C.2 and D.2 in the northwest image are blended, as already discussed by \citet{adamo2024a}. This asymmetry is probably due to a small perturbation in the lens model, currently not accounted for. Sources with radii of 4 pc were inserted to account for the clumps F to I discussed in the previous section. The source-plane position of all the sources is shown in the bottom-right panels of Fig.~\ref{fig:forward_mod}. Clusters A to E occupy a small region, $\sim50$ pc in diameter, and B to E are found within $\lesssim10$ pc from each other. If also the regions in the tail are considered, the arc extends to a diameter up to $\sim400$ pc. 
To test the reliability of our method of delensing the observed radii into intrinsic values, we repeated the modeling assigning constant radii of 3 and 10 pc to all the clusters (top-right panels in Fig.~\ref{fig:forward_mod}). Already in the 3 pc case, clusters B to E became indistinguishable in the observations, with only cluster A remaining distinct, even if slightly elongated. In the 10 pc reconstruction, the arc becomes completely smooth in the central part. Assuming even larger intrinsic radii, the faint emissions in the tails (sources F to I) become undistinguishable from a smooth emission. This test provides further confirmation that, given our fiducial lens model, the compact appearance of the sources in the Cosmic Gems arc implies small intrinsic sizes (with $\rm R_{eff}\lesssim1$ pc).

\section{Discussion} 
\label{sec:discussion}

\subsection{The post-burst phase in the star cluster regions}\label{sec:discussion:bcde12_a12}
\begin{figure}
    \centering
    \includegraphics[width=\columnwidth]{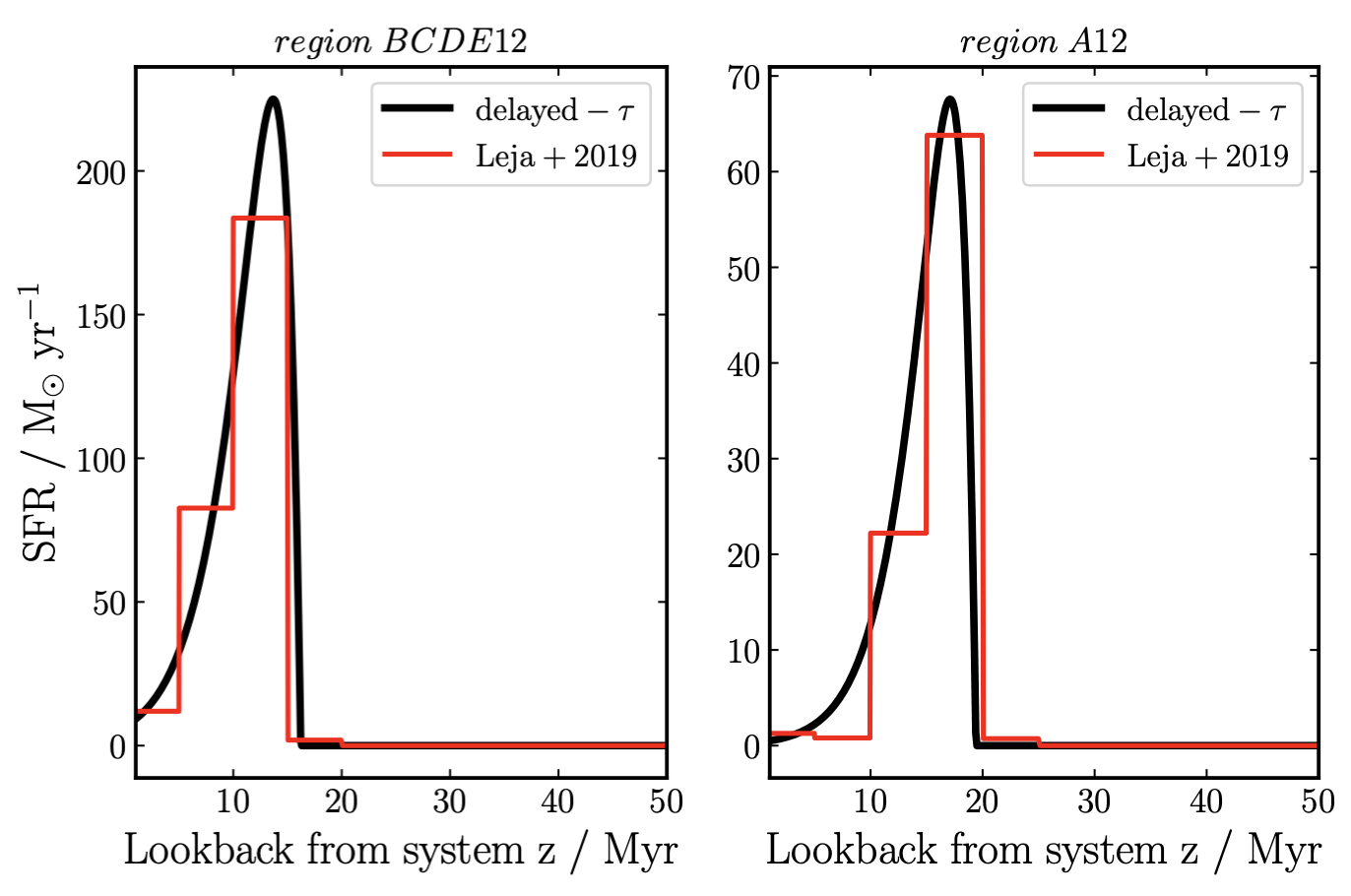}
    \caption{Comparison between the SFHs derived for the regions {\it BCDE} (left) and {\it A} (right) by assuming a parametric-delayed $\tau$-model (black lines) and the nonparametric continuity model by \citet{leja2019} (red curve). The SFR values are not rescaled to account for lensing.}
    \label{fig:SFH_BCDE_A}
\end{figure}

The analysis of the 1D spectra in Section~\ref{sec:nirspec_analysis} suggests that the regions corresponding to the position of the bright star clusters ({\it BCDE} and {\it A}) are in a post-burst phase, following a short ($\rm \tau\sim2~Myr$) star-formation episode around $10-20$ Myr ago during which the star clusters formed. 
In order to test that the post-burst solution is not driven by the shape of the SFH assumed in Section~\ref{sec:nirspec_analysis}, we refit the spectrum using the non-parametrized continuity SFH model by \citet{leja2019}, as implemented in \texttt{Bagpipes}. We considered age bins of 5 Myr in width in the range 0-30 Myr, and larger bins for older ages. We allowed sharp variations in star formation rate (SFR) between consecutive bins. The best-fit SFHs for the {\it BCDE12} and {\it A12} spectra (shown in Fig.~\ref{fig:SFH_BCDE_A}) confirm that the regions are currently in a "dormant" phase after a burst started $\sim15$ Myr earlier lasting $\sim5-10$ Myr. The fit of {\it A12} possibly suggests a slightly older burst (started $\sim20$ Myr ago) than in {\it BCDE}.

A possible alternative solution to the post-burst scenario is a younger SF region where the dearth of strong emission lines is due to a nonzero escape fraction (as also discussed in other studies of high-z quenched galaxy candidates, for example,~\citealp{looser2024,baker2025}). Steep ($\rm \beta\leq-2.5$), such as those observed in all regions of the Cosmic Gems, are considered indirect tracers of $\rm f_{esc}\gtrsim0.1$ \citep[e.g.,][]{chisholm2022,flury2022a,flury2022b,Mascia24,mascia2023a,topping2024}.
The main argument against this possibility is the presence of strong \lya\ damping wings in the observed spectrum. This feature indicates large column densities of \HI\ in the vicinity of the star clusters, as discussed in detail by Christensen et al., (in prep), and thus a high gas opacity that would prevent the escape of ionizing radiation along the line of sight. However, it is worth noting that simulations suggest Lyman continuum leakage (if any) could still occur along transverse directions, which are not probed by our observations \citep[][]{HeRG2020}. Another point weakening the optically thin scenario 
is the fact that the UV slope remains relatively steep along the entire arc. While this would imply an exceptionally large, spatially resolved, optically thin region of the interstellar medium (ISM) spanning dozens of parsecs, the more likely explanation is that the galaxy is in a post-burst phase, during which the stellar populations can naturally produce relatively steep UV slopes and weak optical emission lines without invoking substantial leakage of ionizing photons. On the other hand, it is also plausible that in the past, during its bursty phase, the Cosmic Gems galaxy may have experienced a period when copious ionizing photons were able to escape (e.g., \citetalias{Vanzella2025}; \citealp{Ferrara2025}).

The Cosmic Gems is thus the highest-redshift system observed in a post-starburst phase. Galaxies with similar properties (weak or absent emission lines and the presence of a Balmer break) have been recently observed up to $\rm z\sim7-9.5$ and are thought to represent the temporarily "dormant" (or mini-quenched) phase of galaxies characterized by a bursty SFH \citep[e.g.,][]{looser2024,baker2025, Tang2025}. The Cosmic Gems provides also the opportunity to look at the small sub-galactic scales of a post-burst system, revealing the presence of dense ($\rm \Sigma_M\sim10^5-10^6~M_\odot~pc^{-2}$) and massive ($M\sim10^6~M_\odot$) stellar clusters, as presented in Section~\ref{sec:ysc}. Those clusters are likely the main contributors to the observed spectra in this region.

We argue that their feedback (radiative or from supernovae, or both) may be the cause of the suppression of star formation in the region. This is expected from high-resolution cosmological simulations of systems that include young star clusters, which are formed in short bursts immediately followed by extended periods of quiescence or low-intensity activity \citep{calura2025,pascale2025}. Moreover, we know from local galaxies' studies that stellar feedback from clusters clears their natal clouds, creating cavities in the gas, and in general regulating the star-formation process in their surroundings \citep[see, e.g.,][for reviews]{krumholz2014,chevance2020b}. In starburst and in dwarf galaxies, stellar feedback can affect scales up to the entire host \citep[e.g.,][]{strickland2009,bik2015}. Strong radiation-driven feedback, temporarily pushing the gas and dust away from the galaxy, is one of the mechanisms that could explain the massive star-forming galaxies (sometimes referred to as “blue monsters”) observed at high-z, for example, in the “attenuation-free” scenario \citep{ferrara2025a,Ferrara2025}. 
As mentioned above, the study of Christensen et al. (in prep) suggests a large column density of gas aligned to the position of the clusters, thus implying that the galaxy is not devoid of gas and dust.
The possibility that the Cosmic Gems galaxy was a “blue monster” in its recent past is discussed in \citetalias{Vanzella2025}.
Finally, at high redshifts, the UV background could cause the suppression of star formation \citep{efstathiou1992}, especially in low-mass galaxies ($\rm 10^5-10^7~M_\odot$, e.g., \citealp{katz2020}). Given the mass of the Cosmic Gems ($\rm >10^7~M_\odot$, \citealp{Bradley25}; \citetalias{Vanzella2025}) and a redshift well before the end of the reionization (implying a limited UV background), we argue that it is not a main contribution to the SF suppression. Also, in this case, the large gas column density measured in Christensen et al., (in prep) favors an "internal" quenching scenario, also in agreement with high-resolution simulations of star cluster formation at $z>8$ \citep{calura2022, calura2025, GarciaR2023,Garcia25,SugimuraR2024}.

\subsection{Sub-galactic spatial variation of the SFH?}\label{sec:discussion:tail}
The best-fit results from the {\it CC} and {\it TAIL} regions return longer SF episodes than for the cluster regions, thus with large uncertainties on most of the fitted properties. We already discussed how the emission lines from the {\it CC} spectrum may be powered by the rest-UV sources in the {\it BCDE} region, given their proximity in the source plane. A long SFH also seems inappropriate for an intrinsically small region such as {\it CC} ($\rm \lesssim1~pc$ using the reference magnification from Appendix~\ref{app:tab:spectroscopy}). 

\begin{figure}
    \centering
    \includegraphics[width=\columnwidth]{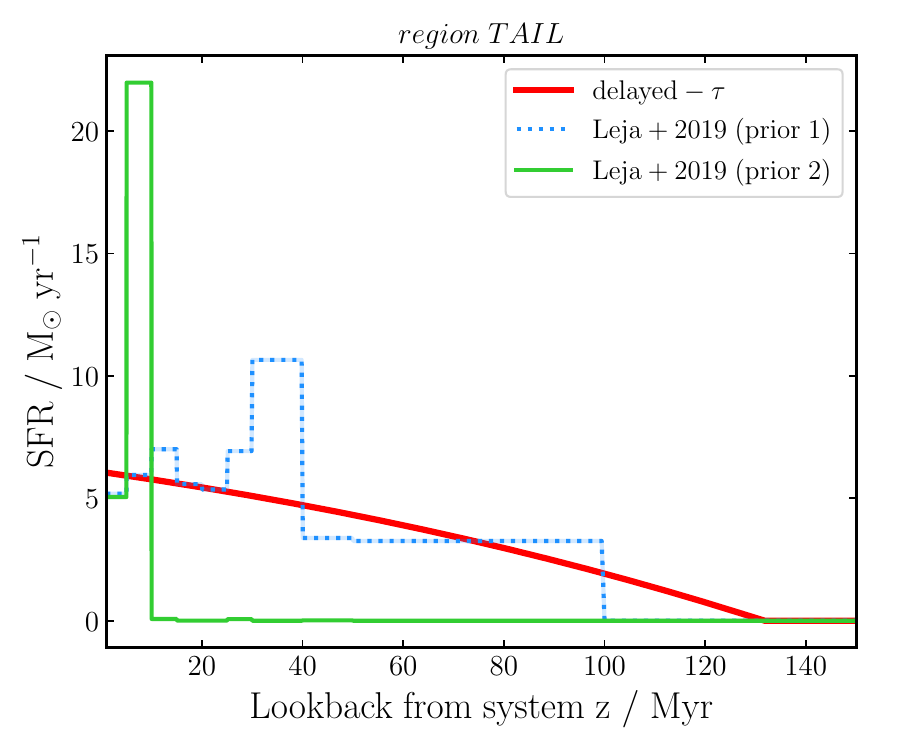}
    \caption{Comparison between the SFHs obtained via the \texttt{Bagpipes} fitting of the spectrum in the {\it TAIL}, in the case of a parametric-delayed $\tau$ model (red curve) and of two realizations of the continuity \citet{leja2019} model, assuming different priors (dotted-blue and solid green curves). The only difference between the priors used in the continuity model is the width of the distribution of the $\rm \Delta_{SFR}$ parameter, which regulates the difference in SFR between consecutive age bins; a small width produces the blue curve, while a large width, allowing bursty SF episodes, produces the green one.}
    \label{fig:SFH_TAIL}
\end{figure}

On the other hand, the tail region includes a large but observationally UV faint area of the galaxy. 
The main spectroscopic features of its spectrum, discussed in the previous sections (see the top half of Appendix~\ref{app:tab:spectroscopy}) are poorly constrained yet consistent with those found in the other regions; i.e., neither the emission lines nor the Balmer break is detected, but their upper limits are larger than the detections in BCDE and hence do not exclude their presence. 
The main difference among regions is the extremely steep slope seen in the tail ($\rm \beta=-2.74\pm0.06$).
We argue that the differences in the spectral fitting results may be due to the low signal and lack of spectral features in the tail. To further test the robustness of the results, we ran a fit with the continuity SFH assumption, as for the {\it BCDE} and {\it A} regions (Fig.~\ref{fig:SFH_TAIL}). The fit returns unconstrained SFHs. We find that the spectrum is well fit by a young burst, as in the cluster regions, and by a longer SF event\footnote{The difference in the two recovered SFHs is caused by different prior assumptions, namely the width of the {\it student-t} distribution used as prior for $\rm \Delta_{SFR}$, which describes the variation of SFR between consecutive time bins.} (as found with the delayed SFH parametrization in Section~\ref{sec:nirspec_analysis}). 

\section{Conclusions}
\label{sec:conclusions}

We reported the observation and analysis of JWST/NIRSpec prism IFU data (Program 5917, PI Vanzella) targeting the currently highest-redshift, strongly magnified ($\mu > 50$) Cosmic Gems arc at $z\simeq 10$. The data cube covers the spectral range 0.8–5.3 $\mu$m, with a spectral resolution varying from $R\sim30$ to $R\sim320$. Additionally, VLT/MUSE spectroscopic redshifts of newly identified multiple images (Program 0112.A-2069(A), PI Bauer) were presented and incorporated to update the lens models. The improved models, developed using both \texttt{Lenstool} and \texttt{Glafic}, account for the CC intersecting the arc and are constrained by 54 multiple images from 18 sources spanning the redshift interval $1<z<10$. The derived magnifications along the Cosmic Gems arc remain consistent with previous estimates, showing a steep gradient from $\mu\sim50$ to $\mu\sim320$ across the star clusters (A to E). These refined lens models enable a robust inference of the intrinsic properties of the subcomponents of the Cosmic Gems galaxy.

The key region of the arc is fully covered and spatially resolved in the NIRSpec IFU pointing. Our main results are summarized as follows:

\noindent (1) The redshift of the Cosmic Gems is confirmed at $z=9.625\pm0.002$ from a high signal-to-noise integrated spectrum (S/N~$\simeq$40–100), which exhibits a prominent \lya\ (damping wing) break, a very blue ultraviolet slope ($\beta_{UV}$ ranging from $-2.5$ to $-2.7$), a tentative Balmer break, and clear but relatively weak \Hb\ and \OIIIb\ emission lines (with EWs of $\leq38$ \AA\ and $\leq75$ \AA, respectively, in the subregions where the lines are detected).

\noindent (2) The spectral fitting of the subregions, encompassing both individual star clusters (e.g., A) and unresolved associations of clusters (BCDE), confirms the physical properties previously derived from imaging alone \citep{adamo2024a}. NIRCam-based SED fitting was updated to reflect the spectroscopic redshift (Section~\ref{sec:ysc}). In combination with spectral fitting (Section~\ref{sec:nirspec_analysis}) and revised magnification maps (Section~\ref{sec:lens_model}), the five clusters hosted within the Cosmic Gems arc (A, B, C, D, and E) exhibit radii in the range of 0.4–1.7 pc and stellar surface densities of $\sim10^5$ M$_\odot$ pc$^{-2}$. Four additional compact regions are identified along the extended tail of the arc, each with sizes $\lesssim10$ pc. Forward modeling reproduces the observed image-plane morphology and reveals that in the source plane the five star clusters are confined to a region $\sim50$ pc in extent, with four of them (B, C, D, and E) likely residing within a common $\sim$10 pc region. The entire arc spans an intrinsic area of radius $\sim400$ pc.

\noindent (3) The nature of the Cosmic Gems galaxy suggests it is in a post-burst or dormant phase, representing an example of a $z\sim10$ weak emitter, a class of galaxies requiring further investigation through deep spectroscopy and/or lensing studies. Stellar feedback from the observed dense star clusters could be the main reason for the (temporary) quenching of the SF in the Cosmic Gems.

\noindent (4) A compact and spatially unresolved \Hb\ and \OIIIb\ emission component is detected near the CC, indicating that the emitting region is intrinsically small, with a size $\lesssim0.9$ pc (for $\mu>500$). This feature may arise from photoionization by a nearby star cluster (within a few parsecs) or from in situ star formation in a tiny \ion{H}{ii} region. Although the emitting region is extremely compact, its inferred physical properties remain preliminary, limited by a low S/N and the absence of a detectable NIRCam counterpart. Given the large magnification, the contribution of the CC region to the total arc emission is negligible.

\begin{acknowledgements}
This work is based on observations made with the NASA/ESA/CSA \textit{James Webb} Space Telescope (\JWST) and \textit{Hubble} Space Telescope (HST).
These observations are associated with \JWST\ GO program n.4212 (PI L. Bradley) and n.5917 (PI E. Vanzella). The data can be obtained from the Mikulski Archive for Space Telescopes at the Space Telescope Science Institute, which is operated by the Association of Universities for Research in Astronomy, Inc., under NASA contract NAS 5-03127 for JWST.
We thank the anonymous referee for helping improving the manuscript.
MM, EV and FC acknowledge financial support through grants PRIN-MIUR 2020SKSTHZ, the INAF GO Grant 2022 ``The revolution is around the corner: JWST will probe globular cluster precursors and Population III stellar clusters at cosmic dawn'' and INAF GO Grant 2024 ``Mapping Star Cluster Feedback in a Galaxy 450 Myr after the Big Bang'', and by the European Union – NextGenerationEU within PRIN 2022 project n.20229YBSAN - Globular clusters in cosmological simulations and lensed fields: from their birth to the present epoch. This work was supported by JSPS KAKENHI Grant Numbers JP25H00662, JP22K21349.
AA acknowledges support by the Swedish research council Vetenskapsr{\aa}det (VR 2021-05559, and VR consolidator grant 2024-02061).
FEB acknowledges support from ANID-Chile BASAL CATA FB210003, FONDECYT Regular 1241005, and Millennium Science Initiative, AIM23-0001. MB acknowledges support from the Slovenian national research agency ARRS through grant N1-0238. EZ acknowledges project grant 2022-03804 from the Swedish Research Council.
R.A.W., S.H.C., and R.A.J. acknowledge support from NASA JWST Interdisciplinary Scientist grants NAG5-12460, NNX14AN10G and 80NSSC18K0200 from GSFC. 
AZ acknowledges support by grant 2020750 from the United States-Israel Binational Science Foundation (BSF) and grant 2109066 from the United States National Science Foundation (NSF), and by the Israel Science Foundation Grant No. 864/23.
\end{acknowledgements}

%
\bibliographystyle{aa} 
\bibliography{references} 

\begin{thebibliography}{111}
\expandafter\ifx\csname natexlab\endcsname\relax\def\natexlab#1{#1}\fi

\bibitem[{{Adamo} {et~al.}(2024){Adamo}, {Bradley}, {Vanzella}, {Claeyssens}, {Welch}, {Diego}, {Mahler}, {Oguri}, {Sharon}, {Abdurro'uf}, {Hsiao}, {Xu}, {Messa}, {Lassen}, {Zackrisson}, {Brammer}, {Coe}, {Kokorev}, {Ricotti}, {Zitrin}, {Fujimoto}, {Inoue}, {Resseguier}, {Rigby}, {Jim{\'e}nez-Teja}, {Windhorst}, {Hashimoto}, \& {Tamura}}]{adamo2024a}
{Adamo}, A., {Bradley}, L.~D., {Vanzella}, E., {et~al.} 2024, \nat, 632, 513

\bibitem[{{Arrabal Haro} {et~al.}(2023){Arrabal Haro}, {Dickinson}, {Finkelstein}, {Fujimoto}, {Fern{\'a}ndez}, {Kartaltepe}, {Jung}, {Cole}, {Burgarella}, {Chworowsky}, {Hutchison}, {Morales}, {Papovich}, {Simons}, {Amor{\'\i}n}, {Backhaus}, {Bagley}, {Bisigello}, {Calabr{\`o}}, {Castellano}, {Cleri}, {Dav{\'e}}, {Dekel}, {Ferguson}, {Fontana}, {Gawiser}, {Giavalisco}, {Harish}, {Hathi}, {Hirschmann}, {Holwerda}, {Huertas-Company}, {Koekemoer}, {Larson}, {Lucas}, {Mobasher}, {P{\'e}rez-Gonz{\'a}lez}, {Pirzkal}, {Rose}, {Santini}, {Trump}, {de la Vega}, {Wang}, {Weiner}, {Wilkins}, {Yang}, {Yung}, \& {Zavala}}]{Arrabal23}
{Arrabal Haro}, P., {Dickinson}, M., {Finkelstein}, S.~L., {et~al.} 2023, \apjl, 951, L22

\bibitem[{{Atek} {et~al.}(2024){Atek}, {Labb{\'e}}, {Furtak}, {Chemerynska}, {Fujimoto}, {Setton}, {Miller}, {Oesch}, {Bezanson}, {Price}, {Dayal}, {Zitrin}, {Kokorev}, {Weaver}, {Brammer}, {Dokkum}, {Williams}, {Cutler}, {Feldmann}, {Fudamoto}, {Greene}, {Leja}, {Maseda}, {Muzzin}, {Pan}, {Papovich}, {Nelson}, {Nanayakkara}, {Stark}, {Stefanon}, {Suess}, {Wang}, \& {Whitaker}}]{atek2024}
{Atek}, H., {Labb{\'e}}, I., {Furtak}, L.~J., {et~al.} 2024, \nat, 626, 975

\bibitem[{{Atek} {et~al.}(2023){Atek}, {Shuntov}, {Furtak}, {Richard}, {Kneib}, {Mahler}, {Zitrin}, {McCracken}, {Charlot}, {Chevallard}, \& {Chemerynska}}]{Atek23}
{Atek}, H., {Shuntov}, M., {Furtak}, L.~J., {et~al.} 2023, \mnras, 519, 1201

\bibitem[{{Baker} {et~al.}(2025){Baker}, {D'Eugenio}, {Maiolino}, {Bunker}, {Simmonds}, {Tacchella}, {Witstok}, {Arribas}, {Carniani}, {Charlot}, {Chevallard}, {Curti}, {Curtis-Lake}, {Jones}, {Kumari}, {Rinaldi}, {Robertson}, {Williams}, {Willott}, \& {Zhu}}]{baker2025}
{Baker}, W.~M., {D'Eugenio}, F., {Maiolino}, R., {et~al.} 2025, \aap, 697, A90

\bibitem[{{Baumgardt} \& {Hilker}(2018)}]{baumgardt2018}
{Baumgardt}, H. \& {Hilker}, M. 2018, \mnras, 478, 1520

\bibitem[{{Bik} {et~al.}(2015){Bik}, {{\"O}stlin}, {Hayes}, {Adamo}, {Melinder}, \& {Amram}}]{bik2015}
{Bik}, A., {{\"O}stlin}, G., {Hayes}, M., {et~al.} 2015, \aap, 576, L13

\bibitem[{{Binggeli} {et~al.}(2019){Binggeli}, {Zackrisson}, {Ma}, {Inoue}, {Vikaeus}, {Hashimoto}, {Mawatari}, {Shimizu}, \& {Ceverino}}]{binggeli2019}
{Binggeli}, C., {Zackrisson}, E., {Ma}, X., {et~al.} 2019, \mnras, 489, 3827

\bibitem[{{Bradley} {et~al.}(2025){Bradley}, {Adamo}, {Vanzella}, {Sharon}, {Brammer}, {Coe}, {Diego}, {Kokorev}, {Mahler}, {Oguri}, {Abdurro'uf}, {Bhatawdekar}, {Christensen}, {Fujimoto}, {Hashimoto}, {Hsiao}, {Inoue}, {Jim{\'e}nez-Teja}, {Messa}, {Norman}, {Ricotti}, {Tamura}, {Windhorst}, {Xu}, \& {Zitrin}}]{Bradley25}
{Bradley}, L.~D., {Adamo}, A., {Vanzella}, E., {et~al.} 2025, \apj, 991, 32

\bibitem[{{Brammer} {et~al.}(2008){Brammer}, {van Dokkum}, \& {Coppi}}]{brammer2008_eazy}
{Brammer}, G.~B., {van Dokkum}, P.~G., \& {Coppi}, P. 2008, \apj, 686, 1503

\bibitem[{{Brown} \& {Gnedin}(2021)}]{brown2021}
{Brown}, G. \& {Gnedin}, O.~Y. 2021, \mnras, 508, 5935

\bibitem[{{Bunker} {et~al.}(2023){Bunker}, {Saxena}, {Cameron}, {Willott}, {Curtis-Lake}, {Jakobsen}, {Carniani}, {Smit}, {Maiolino}, {Witstok}, {Curti}, {D'Eugenio}, {Jones}, {Ferruit}, {Arribas}, {Charlot}, {Chevallard}, {Giardino}, {de Graaff}, {Looser}, {L{\"u}tzgendorf}, {Maseda}, {Rawle}, {Rix}, {Del Pino}, {Alberts}, {Egami}, {Eisenstein}, {Endsley}, {Hainline}, {Hausen}, {Johnson}, {Rieke}, {Rieke}, {Robertson}, {Shivaei}, {Stark}, {Sun}, {Tacchella}, {Tang}, {Williams}, {Willmer}, {Baker}, {Baum}, {Bhatawdekar}, {Bowler}, {Boyett}, {Chen}, {Circosta}, {Helton}, {Ji}, {Kumari}, {Lyu}, {Nelson}, {Parlanti}, {Perna}, {Sandles}, {Scholtz}, {Suess}, {Topping}, {{\"U}bler}, {Wallace}, \& {Whitler}}]{Bunker23}
{Bunker}, A.~J., {Saxena}, A., {Cameron}, A.~J., {et~al.} 2023, \aap, 677, A88

\bibitem[{{Bushouse} {et~al.}(2023){Bushouse}, {Eisenhamer}, {Dencheva}, {Davies}, {Greenfield}, {Morrison}, {Hodge}, {Simon}, {Grumm}, {Droettboom}, {Slavich}, {Sosey}, {Pauly}, {Miller}, {Jedrzejewski}, {Hack}, {Davis}, {Crawford}, {Law}, {Gordon}, {Regan}, {Cara}, {MacDonald}, {Bradley}, {Shanahan}, {Jamieson}, {Teodoro}, \& {Williams}}]{bushouse23}
{Bushouse}, H., {Eisenhamer}, J., {Dencheva}, N., {et~al.} 2023, {JWST Calibration Pipeline}

\bibitem[{{Calura} {et~al.}(2022){Calura}, {Lupi}, {Rosdahl}, {Vanzella}, {Meneghetti}, {Rosati}, {Vesperini}, {Lacchin}, {Pascale}, \& {Gilli}}]{calura2022}
{Calura}, F., {Lupi}, A., {Rosdahl}, J., {et~al.} 2022, \mnras, 516, 5914

\bibitem[{{Calura} {et~al.}(2025){Calura}, {Pascale}, {Agertz}, {Andersson}, {Lacchin}, {Lupi}, {Meneghetti}, {Nipoti}, {Ragagnin}, {Rosdahl}, {Vanzella}, {Vesperini}, \& {Zanella}}]{calura2025}
{Calura}, F., {Pascale}, R., {Agertz}, O., {et~al.} 2025, \aap, 698, A207

\bibitem[{{Calzetti} {et~al.}(2000){Calzetti}, {Armus}, {Bohlin}, {Kinney}, {Koornneef}, \& {Storchi-Bergmann}}]{calzetti2000}
{Calzetti}, D., {Armus}, L., {Bohlin}, R.~C., {et~al.} 2000, \apj, 533, 682

\bibitem[{{Carnall} {et~al.}(2019){Carnall}, {McLure}, {Dunlop}, {Cullen}, {McLeod}, {Wild}, {Johnson}, {Appleby}, {Dav{\'e}}, {Amorin}, {Bolzonella}, {Castellano}, {Cimatti}, {Cucciati}, {Gargiulo}, {Garilli}, {Marchi}, {Pentericci}, {Pozzetti}, {Schreiber}, {Talia}, \& {Zamorani}}]{carnall2019_bagpipes}
{Carnall}, A.~C., {McLure}, R.~J., {Dunlop}, J.~S., {et~al.} 2019, \mnras, 490, 417

\bibitem[{{Carnall} {et~al.}(2018){Carnall}, {McLure}, {Dunlop}, \& {Dav{\'e}}}]{carnall2018_bagpipes}
{Carnall}, A.~C., {McLure}, R.~J., {Dunlop}, J.~S., \& {Dav{\'e}}, R. 2018, \mnras, 480, 4379

\bibitem[{{Carniani} {et~al.}(2024){Carniani}, {Hainline}, {D'Eugenio}, {Eisenstein}, {Jakobsen}, {Witstok}, {Johnson}, {Chevallard}, {Maiolino}, {Helton}, {Willott}, {Robertson}, {Alberts}, {Arribas}, {Baker}, {Bhatawdekar}, {Boyett}, {Bunker}, {Cameron}, {Cargile}, {Charlot}, {Curti}, {Curtis-Lake}, {Egami}, {Giardino}, {Isaak}, {Ji}, {Jones}, {Kumari}, {Maseda}, {Parlanti}, {P{\'e}rez-Gonz{\'a}lez}, {Rawle}, {Rieke}, {Rieke}, {Del Pino}, {Saxena}, {Scholtz}, {Smit}, {Sun}, {Tacchella}, {{\"U}bler}, {Venturi}, {Williams}, \& {Willmer}}]{Carniani24a}
{Carniani}, S., {Hainline}, K., {D'Eugenio}, F., {et~al.} 2024, \nat, 633, 318

\bibitem[{{Carvajal-Bohorquez} {et~al.}(2025){Carvajal-Bohorquez}, {Ciesla}, {Laporte}, {Boquien}, {Buat}, {Ilbert}, {Aufort}, {Shuntov}, {Witten}, {Oesch}, \& {Covelo-Paz}}]{Carvajal_Bohorquez2025}
{Carvajal-Bohorquez}, C., {Ciesla}, L., {Laporte}, N., {et~al.} 2025, arXiv e-prints, arXiv:2507.13160

\bibitem[{{Castellano} {et~al.}(2024){Castellano}, {Napolitano}, {Fontana}, {Roberts-Borsani}, {Treu}, {Vanzella}, {Zavala}, {Arrabal Haro}, {Calabr{\`o}}, {Llerena}, {Mascia}, {Merlin}, {Paris}, {Pentericci}, {Santini}, {Bakx}, {Bergamini}, {Cupani}, {Dickinson}, {Filippenko}, {Glazebrook}, {Grillo}, {Kelly}, {Malkan}, {Mason}, {Morishita}, {Nanayakkara}, {Rosati}, {Sani}, {Wang}, \& {Yoon}}]{Castellano24}
{Castellano}, M., {Napolitano}, L., {Fontana}, A., {et~al.} 2024, \apj, 972, 143

\bibitem[{{Chevance} {et~al.}(2020){Chevance}, {Kruijssen}, {Vazquez-Semadeni}, {Nakamura}, {Klessen}, {Ballesteros-Paredes}, {Inutsuka}, {Adamo}, \& {Hennebelle}}]{chevance2020b}
{Chevance}, M., {Kruijssen}, J.~M.~D., {Vazquez-Semadeni}, E., {et~al.} 2020, \ssr, 216, 50

\bibitem[{{Chisholm} {et~al.}(2022){Chisholm}, {Saldana-Lopez}, {Flury}, {Schaerer}, {Jaskot}, {Amor{\'\i}n}, {Atek}, {Finkelstein}, {Fleming}, {Ferguson}, {Fern{\'a}ndez}, {Giavalisco}, {Hayes}, {Heckman}, {Henry}, {Ji}, {Marques-Chaves}, {Mauerhofer}, {McCandliss}, {Oey}, {{\"O}stlin}, {Rutkowski}, {Scarlata}, {Thuan}, {Trebitsch}, {Wang}, {Worseck}, \& {Xu}}]{chisholm2022}
{Chisholm}, J., {Saldana-Lopez}, A., {Flury}, S., {et~al.} 2022, \mnras, 517, 5104

\bibitem[{{Choi} {et~al.}(2017){Choi}, {Conroy}, \& {Byler}}]{choi2017}
{Choi}, J., {Conroy}, C., \& {Byler}, N. 2017, \apj, 838, 159

\bibitem[{{Covelo-Paz} {et~al.}(2025){Covelo-Paz}, {Meuwly}, {Oesch}, {Witten}, {Weibel}, {Carvajal-Bohorquez}, {Ciesla}, {Giovinazzo}, \& {Brammer}}]{CoveloPaz2025}
{Covelo-Paz}, A., {Meuwly}, C., {Oesch}, P.~A., {et~al.} 2025, arXiv e-prints, arXiv:2506.22540

\bibitem[{{Cullen} {et~al.}(2023){Cullen}, {McLure}, {McLeod}, {Dunlop}, {Donnan}, {Carnall}, {Bowler}, {Begley}, {Hamadouche}, \& {Stanton}}]{cullen2023}
{Cullen}, F., {McLure}, R.~J., {McLeod}, D.~J., {et~al.} 2023, \mnras, 520, 14

\bibitem[{{Curtis-Lake} {et~al.}(2023){Curtis-Lake}, {Carniani}, {Cameron}, {Charlot}, {Jakobsen}, {Maiolino}, {Bunker}, {Witstok}, {Smit}, {Chevallard}, {Willott}, {Ferruit}, {Arribas}, {Bonaventura}, {Curti}, {D'Eugenio}, {Franx}, {Giardino}, {Looser}, {L{\"u}tzgendorf}, {Maseda}, {Rawle}, {Rix}, {Rodr{\'\i}guez del Pino}, {{\"U}bler}, {Sirianni}, {Dressler}, {Egami}, {Eisenstein}, {Endsley}, {Hainline}, {Hausen}, {Johnson}, {Rieke}, {Robertson}, {Shivaei}, {Stark}, {Tacchella}, {Williams}, {Willmer}, {Bhatawdekar}, {Bowler}, {Boyett}, {Chen}, {de Graaff}, {Helton}, {Hviding}, {Jones}, {Kumari}, {Lyu}, {Nelson}, {Perna}, {Sandles}, {Saxena}, {Suess}, {Sun}, {Topping}, {Wallace}, \& {Whitler}}]{Curtis23}
{Curtis-Lake}, E., {Carniani}, S., {Cameron}, A., {et~al.} 2023, Nature Astronomy, 7, 622

\bibitem[{{Dekel} {et~al.}(2023){Dekel}, {Sarkar}, {Birnboim}, {Mandelker}, \& {Li}}]{Dekel23}
{Dekel}, A., {Sarkar}, K.~C., {Birnboim}, Y., {Mandelker}, N., \& {Li}, Z. 2023, \mnras, 523, 3201

\bibitem[{{Dome} {et~al.}(2024){Dome}, {Tacchella}, {Fialkov}, {Ceverino}, {Dekel}, {Ginzburg}, {Lapiner}, \& {Looser}}]{dome2024}
{Dome}, T., {Tacchella}, S., {Fialkov}, A., {et~al.} 2024, \mnras, 527, 2139

\bibitem[{{Donnan} {et~al.}(2025){Donnan}, {Dickinson}, {Taylor}, {Arrabal Haro}, {Finkelstein}, {Stanton}, {Jung}, {Papovich}, {Akins}, {Koekemoer}, {McLeod}, {Napolitano}, {Amor{\'\i}n}, {Begley}, {Burgarella}, {Carnall}, {Casey}, {Calabr{\`o}}, {Cullen}, {Dunlop}, {Ellis}, {Fern{\'a}ndez}, {Giavalisco}, {Hirschmann}, {Hu}, {Illingworth}, {Kartaltepe}, {Kocevski}, {Kokorev}, {Leung}, {Lucas}, {Morales}, {McLure}, {Pentericci}, {P{\'e}rez-Gonz{\'a}lez}, {Somerville}, {Stevenson}, {Trump}, {Yung}, \& {Zavala}}]{Donnan2025}
{Donnan}, C.~T., {Dickinson}, M., {Taylor}, A.~J., {et~al.} 2025, arXiv e-prints, arXiv:2507.10518

\bibitem[{{Efstathiou}(1992)}]{efstathiou1992}
{Efstathiou}, G. 1992, \mnras, 256, 43P

\bibitem[{{Eldridge} {et~al.}(2017){Eldridge}, {Stanway}, {Xiao}, {McClelland}, {Taylor}, {Ng}, {Greis}, \& {Bray}}]{eldridge2017}
{Eldridge}, J.~J., {Stanway}, E.~R., {Xiao}, L., {et~al.} 2017, \pasa, 34, e058

\bibitem[{{Endsley} {et~al.}(2025){Endsley}, {Chisholm}, {Stark}, {Topping}, \& {Whitler}}]{Endsley25}
{Endsley}, R., {Chisholm}, J., {Stark}, D.~P., {Topping}, M.~W., \& {Whitler}, L. 2025, \apj, 987, 189

\bibitem[{{Endsley} {et~al.}(2024){Endsley}, {Stark}, {Whitler}, {Topping}, {Johnson}, {Robertson}, {Tacchella}, {Alberts}, {Baker}, {Bhatawdekar}, {Boyett}, {Bunker}, {Cameron}, {Carniani}, {Charlot}, {Chen}, {Chevallard}, {Curtis-Lake}, {Danhaive}, {Egami}, {Eisenstein}, {Hainline}, {Helton}, {Ji}, {Looser}, {Maiolino}, {Nelson}, {Pusk{\'a}s}, {Rieke}, {Rieke}, {Rix}, {Sandles}, {Saxena}, {Simmonds}, {Smit}, {Sun}, {Williams}, {Willmer}, {Willott}, \& {Witstok}}]{endsley2024}
{Endsley}, R., {Stark}, D.~P., {Whitler}, L., {et~al.} 2024, \mnras, 533, 1111

\bibitem[{{Faucher-Gigu{\`e}re}(2018)}]{fauchergiguere2018}
{Faucher-Gigu{\`e}re}, C.-A. 2018, \mnras, 473, 3717

\bibitem[{{Ferrara} {et~al.}(2025{\natexlab{a}}){Ferrara}, {Carniani}, {di Mascia}, {Bouwens}, {Oesch}, \& {Schouws}}]{ferrara2025a}
{Ferrara}, A., {Carniani}, S., {di Mascia}, F., {et~al.} 2025{\natexlab{a}}, \aap, 694, A215

\bibitem[{{Ferrara} {et~al.}(2023){Ferrara}, {Pallottini}, \& {Dayal}}]{Ferrara_2023_monsters}
{Ferrara}, A., {Pallottini}, A., \& {Dayal}, P. 2023, \mnras, 522, 3986

\bibitem[{{Ferrara} {et~al.}(2025{\natexlab{b}}){Ferrara}, {Pallottini}, \& {Sommovigo}}]{Ferrara2025}
{Ferrara}, A., {Pallottini}, A., \& {Sommovigo}, L. 2025{\natexlab{b}}, \aap, 694, A286

\bibitem[{{Finkelstein} {et~al.}(2024){Finkelstein}, {Leung}, {Bagley}, {Dickinson}, {Ferguson}, {Papovich}, {Akins}, {Arrabal Haro}, {Dav{\'e}}, {Dekel}, {Kartaltepe}, {Kocevski}, {Koekemoer}, {Pirzkal}, {Somerville}, {Yung}, {Amor{\'\i}n}, {Backhaus}, {Behroozi}, {Bisigello}, {Bromm}, {Casey}, {Ch{\'a}vez Ortiz}, {Cheng}, {Chworowsky}, {Cleri}, {Cooper}, {Davis}, {de la Vega}, {Elbaz}, {Franco}, {Fontana}, {Fujimoto}, {Giavalisco}, {Grogin}, {Holwerda}, {Huertas-Company}, {Hirschmann}, {Iyer}, {Jogee}, {Jung}, {Larson}, {Lucas}, {Mobasher}, {Morales}, {Morley}, {Mukherjee}, {P{\'e}rez-Gonz{\'a}lez}, {Ravindranath}, {Rodighiero}, {Rowland}, {Tacchella}, {Taylor}, {Trump}, \& {Wilkins}}]{Finkelstein2024}
{Finkelstein}, S.~L., {Leung}, G. C.~K., {Bagley}, M.~B., {et~al.} 2024, \apjl, 969, L2

\bibitem[{{Flury} {et~al.}(2022{\natexlab{a}}){Flury}, {Jaskot}, {Ferguson}, {Worseck}, {Makan}, {Chisholm}, {Saldana-Lopez}, {Schaerer}, {McCandliss}, {Wang}, {Ford}, {Heckman}, {Ji}, {Giavalisco}, {Amorin}, {Atek}, {Blaizot}, {Borthakur}, {Carr}, {Castellano}, {Cristiani}, {De Barros}, {Dickinson}, {Finkelstein}, {Fleming}, {Fontanot}, {Garel}, {Grazian}, {Hayes}, {Henry}, {Mauerhofer}, {Micheva}, {Oey}, {Ostlin}, {Papovich}, {Pentericci}, {Ravindranath}, {Rosdahl}, {Rutkowski}, {Santini}, {Scarlata}, {Teplitz}, {Thuan}, {Trebitsch}, {Vanzella}, {Verhamme}, \& {Xu}}]{flury2022a}
{Flury}, S.~R., {Jaskot}, A.~E., {Ferguson}, H.~C., {et~al.} 2022{\natexlab{a}}, \apjs, 260, 1

\bibitem[{{Flury} {et~al.}(2022{\natexlab{b}}){Flury}, {Jaskot}, {Ferguson}, {Worseck}, {Makan}, {Chisholm}, {Saldana-Lopez}, {Schaerer}, {McCandliss}, {Xu}, {Wang}, {Oey}, {Ford}, {Heckman}, {Ji}, {Giavalisco}, {Amor{\'\i}n}, {Atek}, {Blaizot}, {Borthakur}, {Carr}, {Castellano}, {De Barros}, {Dickinson}, {Finkelstein}, {Fleming}, {Fontanot}, {Garel}, {Grazian}, {Hayes}, {Henry}, {Mauerhofer}, {Micheva}, {Ostlin}, {Papovich}, {Pentericci}, {Ravindranath}, {Rosdahl}, {Rutkowski}, {Santini}, {Scarlata}, {Teplitz}, {Thuan}, {Trebitsch}, {Vanzella}, \& {Verhamme}}]{flury2022b}
{Flury}, S.~R., {Jaskot}, A.~E., {Ferguson}, H.~C., {et~al.} 2022{\natexlab{b}}, \apj, 930, 126

\bibitem[{{Fujimoto} {et~al.}(2023){Fujimoto}, {Arrabal Haro}, {Dickinson}, {Finkelstein}, {Kartaltepe}, {Larson}, {Burgarella}, {Bagley}, {Behroozi}, {Chworowsky}, {Hirschmann}, {Trump}, {Wilkins}, {Yung}, {Koekemoer}, {Papovich}, {Pirzkal}, {Ferguson}, {Fontana}, {Grogin}, {Grazian}, {Kewley}, {Kocevski}, {Lotz}, {Pentericci}, {Ravindranath}, {Somerville}, {Wilkins}, {Amor{\'\i}n}, {Backhaus}, {Calabr{\`o}}, {Casey}, {Cooper}, {Fern{\'a}ndez}, {Franco}, {Giavalisco}, {Hathi}, {Harish}, {Hutchison}, {Iyer}, {Jung}, {Lucas}, \& {Zavala}}]{Fujimoto2023}
{Fujimoto}, S., {Arrabal Haro}, P., {Dickinson}, M., {et~al.} 2023, \apjl, 949, L25

\bibitem[{{Fujimoto} {et~al.}(2024){Fujimoto}, {Wang}, {Weaver}, {Kokorev}, {Atek}, {Bezanson}, {Labbe}, {Brammer}, {Greene}, {Chemerynska}, {Dayal}, {de Graaff}, {Furtak}, {Oesch}, {Setton}, {Price}, {Miller}, {Williams}, {Whitaker}, {Zitrin}, {Cutler}, {Leja}, {Pan}, {Coe}, {van Dokkum}, {Feldmann}, {Fudamoto}, {Goulding}, {Khullar}, {Marchesini}, {Maseda}, {Nanayakkara}, {Nelson}, {Smit}, {Stefanon}, \& {Weibel}}]{Fujimoto24}
{Fujimoto}, S., {Wang}, B., {Weaver}, J.~R., {et~al.} 2024, \apj, 977, 250

\bibitem[{{Gaia Collaboration} {et~al.}(2018){Gaia Collaboration}, {Brown}, {Vallenari}, {Prusti}, {de Bruijne}, {Babusiaux}, {Bailer-Jones}, {Biermann}, {Evans}, {Eyer}, {Jansen}, {Jordi}, {Klioner}, {Lammers}, {Lindegren}, {Luri}, {Mignard}, {Panem}, {Pourbaix}, {Randich}, {Sartoretti}, {Siddiqui}, {Soubiran}, {van Leeuwen}, {Walton}, {Arenou}, {Bastian}, {Cropper}, {Drimmel}, {Katz}, {Lattanzi}, {Bakker}, {Cacciari}, {Casta{\~n}eda}, {Chaoul}, {Cheek}, {De Angeli}, {Fabricius}, {Guerra}, {Holl}, {Masana}, {Messineo}, {Mowlavi}, {Nienartowicz}, {Panuzzo}, {Portell}, {Riello}, {Seabroke}, {Tanga}, {Th{\'e}venin}, {Gracia-Abril}, {Comoretto}, {Garcia-Reinaldos}, {Teyssier}, {Altmann}, {Andrae}, {Audard}, {Bellas-Velidis}, {Benson}, {Berthier}, {Blomme}, {Burgess}, {Busso}, {Carry}, {Cellino}, {Clementini}, {Clotet}, {Creevey}, {Davidson}, {De Ridder}, {Delchambre}, {Dell'Oro}, {Ducourant}, {Fern{\'a}ndez-Hern{\'a}ndez}, {Fouesneau}, {Fr{\'e}mat}, {Galluccio}, {Garc{\'\i}a-Torres},
  {Gonz{\'a}lez-N{\'u}{\~n}ez}, {Gonz{\'a}lez-Vidal}, {Gosset}, {Guy}, {Halbwachs}, {Hambly}, {Harrison}, {Hern{\'a}ndez}, {Hestroffer}, {Hodgkin}, {Hutton}, {Jasniewicz}, {Jean-Antoine-Piccolo}, {Jordan}, {Korn}, {Krone-Martins}, {Lanzafame}, {Lebzelter}, {L{\"o}ffler}, {Manteiga}, {Marrese}, {Mart{\'\i}n-Fleitas}, {Moitinho}, {Mora}, {Muinonen}, {Osinde}, {Pancino}, {Pauwels}, {Petit}, {Recio-Blanco}, {Richards}, {Rimoldini}, {Robin}, {Sarro}, {Siopis}, {Smith}, {Sozzetti}, {S{\"u}veges}, {Torra}, {van Reeven}, {Abbas}, {Abreu Aramburu}, {Accart}, {Aerts}, {Altavilla}, {{\'A}lvarez}, {Alvarez}, {Alves}, {Anderson}, {Andrei}, {Anglada Varela}, {Antiche}, {Antoja}, {Arcay}, {Astraatmadja}, {Bach}, {Baker}, {Balaguer-N{\'u}{\~n}ez}, {Balm}, {Barache}, {Barata}, {Barbato}, {Barblan}, {Barklem}, {Barrado}, {Barros}, {Barstow}, {Bartholom{\'e} Mu{\~n}oz}, {Bassilana}, {Becciani}, {Bellazzini}, {Berihuete}, {Bertone}, {Bianchi}, {Bienaym{\'e}}, {Blanco-Cuaresma}, {Boch}, {Boeche}, {Bombrun}, {Borrachero},
  {Bossini}, {Bouquillon}, {Bourda}, {Bragaglia}, {Bramante}, {Breddels}, {Bressan}, {Brouillet}, {Br{\"u}semeister}, {Brugaletta}, {Bucciarelli}, {Burlacu}, {Busonero}, {Butkevich}, {Buzzi}, {Caffau}, {Cancelliere}, {Cannizzaro}, {Cantat-Gaudin}, {Carballo}, {Carlucci}, {Carrasco}, {Casamiquela}, {Castellani}, {Castro-Ginard}, {Charlot}, {Chemin}, {Chiavassa}, {Cocozza}, {Costigan}, {Cowell}, {Crifo}, {Crosta}, {Crowley}, {Cuypers}, {Dafonte}, {Damerdji}, {Dapergolas}, {David}, {David}, {de Laverny}, {De Luise}, {De March}, {de Martino}, {de Souza}, {de Torres}, {Debosscher}, {del Pozo}, {Delbo}, {Delgado}, {Delgado}, {Di Matteo}, {Diakite}, {Diener}, {Distefano}, {Dolding}, {Drazinos}, {Dur{\'a}n}, {Edvardsson}, {Enke}, {Eriksson}, {Esquej}, {Eynard Bontemps}, {Fabre}, {Fabrizio}, {Faigler}, {Falc{\~a}o}, {Farr{\`a}s Casas}, {Federici}, {Fedorets}, {Fernique}, {Figueras}, {Filippi}, {Findeisen}, {Fonti}, {Fraile}, {Fraser}, {Fr{\'e}zouls}, {Gai}, {Galleti}, {Garabato}, {Garc{\'\i}a-Sedano}, {Garofalo},
  {Garralda}, {Gavel}, {Gavras}, {Gerssen}, {Geyer}, {Giacobbe}, {Gilmore}, {Girona}, {Giuffrida}, {Glass}, {Gomes}, {Granvik}, {Gueguen}, {Guerrier}, {Guiraud}, {Guti{\'e}rrez-S{\'a}nchez}, {Haigron}, {Hatzidimitriou}, {Hauser}, {Haywood}, {Heiter}, {Helmi}, {Heu}, {Hilger}, {Hobbs}, {Hofmann}, {Holland}, {Huckle}, {Hypki}, {Icardi}, {Jan{\ss}en}, {Jevardat de Fombelle}, {Jonker}, {Juh{\'a}sz}, {Julbe}, {Karampelas}, {Kewley}, {Klar}, {Kochoska}, {Kohley}, {Kolenberg}, {Kontizas}, {Kontizas}, {Koposov}, {Kordopatis}, {Kostrzewa-Rutkowska}, {Koubsky}, {Lambert}, {Lanza}, {Lasne}, {Lavigne}, {Le Fustec}, {Le Poncin-Lafitte}, {Lebreton}, {Leccia}, {Leclerc}, {Lecoeur-Taibi}, {Lenhardt}, {Leroux}, {Liao}, {Licata}, {Lindstr{\o}m}, {Lister}, {Livanou}, {Lobel}, {L{\'o}pez}, {Managau}, {Mann}, {Mantelet}, {Marchal}, {Marchant}, {Marconi}, {Marinoni}, {Marschalk{\'o}}, {Marshall}, {Martino}, {Marton}, {Mary}, {Massari}, {Matijevi{\v{c}}}, {Mazeh}, {McMillan}, {Messina}, {Michalik}, {Millar}, {Molina}, {Molinaro},
  {Moln{\'a}r}, {Montegriffo}, {Mor}, {Morbidelli}, {Morel}, {Morris}, {Mulone}, {Muraveva}, {Musella}, {Nelemans}, {Nicastro}, {Noval}, {O'Mullane}, {Ord{\'e}novic}, {Ord{\'o}{\~n}ez-Blanco}, {Osborne}, {Pagani}, {Pagano}, {Pailler}, {Palacin}, {Palaversa}, {Panahi}, {Pawlak}, {Piersimoni}, {Pineau}, {Plachy}, {Plum}, {Poggio}, {Poujoulet}, {Pr{\v{s}}a}, {Pulone}, {Racero}, {Ragaini}, {Rambaux}, {Ramos-Lerate}, {Regibo}, {Reyl{\'e}}, {Riclet}, {Ripepi}, {Riva}, {Rivard}, {Rixon}, {Roegiers}, {Roelens}, {Romero-G{\'o}mez}, {Rowell}, {Royer}, {Ruiz-Dern}, {Sadowski}, {Sagrist{\`a} Sell{\'e}s}, {Sahlmann}, {Salgado}, {Salguero}, {Sanna}, {Santana-Ros}, {Sarasso}, {Savietto}, {Schultheis}, {Sciacca}, {Segol}, {Segovia}, {S{\'e}gransan}, {Shih}, {Siltala}, {Silva}, {Smart}, {Smith}, {Solano}, {Solitro}, {Sordo}, {Soria Nieto}, {Souchay}, {Spagna}, {Spoto}, {Stampa}, {Steele}, {Steidelm{\"u}ller}, {Stephenson}, {Stoev}, {Suess}, {Surdej}, {Szabados}, {Szegedi-Elek}, {Tapiador}, {Taris}, {Tauran}, {Taylor},
  {Teixeira}, {Terrett}, {Teyssand ier}, {Thuillot}, {Titarenko}, {Torra Clotet}, {Turon}, {Ulla}, {Utrilla}, {Uzzi}, {Vaillant}, {Valentini}, {Valette}, {van Elteren}, {Van Hemelryck}, {van Leeuwen}, {Vaschetto}, {Vecchiato}, {Veljanoski}, {Viala}, {Vicente}, {Vogt}, {von Essen}, {Voss}, {Votruba}, {Voutsinas}, {Walmsley}, {Weiler}, {Wertz}, {Wevers}, {Wyrzykowski}, {Yoldas}, {{\v{Z}}erjal}, {Ziaeepour}, {Zorec}, {Zschocke}, {Zucker}, {Zurbach}, \& {Zwitter}}]{gaia2018}
{Gaia Collaboration}, {Brown}, A.~G.~A., {Vallenari}, A., {et~al.} 2018, \aap, 616, A1

\bibitem[{{Garcia} {et~al.}(2025){Garcia}, {Ricotti}, \& {Sugimura}}]{Garcia25}
{Garcia}, F. A.~B., {Ricotti}, M., \& {Sugimura}, K. 2025, The Open Journal of Astrophysics, 8, 146

\bibitem[{{Garcia} {et~al.}(2023){Garcia}, {Ricotti}, {Sugimura}, \& {Park}}]{GarciaR2023}
{Garcia}, F. A.~B., {Ricotti}, M., {Sugimura}, K., \& {Park}, J. 2023, \mnras, 522, 2495

\bibitem[{{Gelli} {et~al.}(2023){Gelli}, {Salvadori}, {Ferrara}, {Pallottini}, \& {Carniani}}]{gelli2023}
{Gelli}, V., {Salvadori}, S., {Ferrara}, A., {Pallottini}, A., \& {Carniani}, S. 2023, \apjl, 954, L11

\bibitem[{{Gladders} \& {Yee}(2000)}]{gladders2000}
{Gladders}, M.~D. \& {Yee}, H.~K.~C. 2000, \aj, 120, 2148

\bibitem[{{Hainline} {et~al.}(2024){Hainline}, {D'Eugenio}, {Jakobsen}, {Chevallard}, {Carniani}, {Witstok}, {Ji}, {Curtis-Lake}, {Johnson}, {Robertson}, {Tacchella}, {Curti}, {Charlot}, {Helton}, {Arribas}, {Bhatawdekar}, {Bunker}, {Cameron}, {Egami}, {Eisenstein}, {Hausen}, {Kumari}, {Maiolino}, {P{\'e}rez-Gonz{\'a}lez}, {Rieke}, {Saxena}, {Scholtz}, {Smit}, {Sun}, {Williams}, {Willmer}, \& {Willott}}]{Hainline2024}
{Hainline}, K.~N., {D'Eugenio}, F., {Jakobsen}, P., {et~al.} 2024, \apj, 976, 160

\bibitem[{{He} {et~al.}(2020){He}, {Ricotti}, \& {Geen}}]{HeRG2020}
{He}, C.-C., {Ricotti}, M., \& {Geen}, S. 2020, \mnras, 492, 4858

\bibitem[{{Heintz} {et~al.}(2025){Heintz}, {Brammer}, {Watson}, {Oesch}, {Keating}, {Hayes}, {Abdurro'uf}, {Arellano-C{\'o}rdova}, {Carnall}, {Christiansen}, {Cullen}, {Dav{\'e}}, {Dayal}, {Ferrara}, {Finlator}, {Fynbo}, {Flury}, {Gelli}, {Gillman}, {Gottumukkala}, {Gould}, {Greve}, {Hardin}, {Hsiao}, {Hutter}, {Jakobsson}, {Killi}, {Khosravaninezhad}, {Laursen}, {Lee}, {Magdis}, {Matthee}, {Naidu}, {Narayanan}, {Pollock}, {Prescott}, {Rusakov}, {Shuntov}, {Sneppen}, {Smit}, {Tanvir}, {Terp}, {Toft}, {Valentino}, {Vijayan}, {Weaver}, {Wise}, \& {Witstok}}]{Heintz2025}
{Heintz}, K.~E., {Brammer}, G.~B., {Watson}, D., {et~al.} 2025, \aap, 693, A60

\bibitem[{{Helton} {et~al.}(2025){Helton}, {Rieke}, {Alberts}, {Wu}, {Eisenstein}, {Hainline}, {Carniani}, {Ji}, {Baker}, {Bhatawdekar}, {Bunker}, {Cargile}, {Charlot}, {Chevallard}, {D'Eugenio}, {Egami}, {Johnson}, {Jones}, {Lyu}, {Maiolino}, {P{\'e}rez-Gonz{\'a}lez}, {Rieke}, {Robertson}, {Saxena}, {Scholtz}, {Shivaei}, {Sun}, {Tacchella}, {Whitler}, {Williams}, {Willmer}, {Willott}, {Witstok}, \& {Zhu}}]{Helton25}
{Helton}, J.~M., {Rieke}, G.~H., {Alberts}, S., {et~al.} 2025, Nature Astronomy, 9, 729

\bibitem[{{Hsiao} {et~al.}(2024){Hsiao}, {Abdurro'uf}, {Coe}, {Larson}, {Jung}, {Mingozzi}, {Dayal}, {Kumari}, {Kokorev}, {Vikaeus}, {Brammer}, {Furtak}, {Adamo}, {Andrade-Santos}, {Antwi-Danso}, {Brada{\v{c}}}, {Bradley}, {Broadhurst}, {Carnall}, {Conselice}, {Diego}, {Donahue}, {Eldridge}, {Fujimoto}, {Henry}, {Hernandez}, {Hutchison}, {James}, {Norman}, {Park}, {Pirzkal}, {Postman}, {Ricotti}, {Rigby}, {Vanzella}, {Welch}, {Wilkins}, {Windhorst}, {Xu}, {Zackrisson}, \& {Zitrin}}]{Hsiao24}
{Hsiao}, T. Y.-Y., {Abdurro'uf}, {Coe}, D., {et~al.} 2024, \apj, 973, 8

\bibitem[{{Hutter} {et~al.}(2025){Hutter}, {Cueto}, {Dayal}, {Gottl{\"o}ber}, {Trebitsch}, \& {Yepes}}]{Hutter_2025}
{Hutter}, A., {Cueto}, E.~R., {Dayal}, P., {et~al.} 2025, \aap, 694, A254

\bibitem[{{Inoue} {et~al.}(2014){Inoue}, {Shimizu}, {Iwata}, \& {Tanaka}}]{inoue2014}
{Inoue}, A.~K., {Shimizu}, I., {Iwata}, I., \& {Tanaka}, M. 2014, \mnras, 442, 1805

\bibitem[{{Jullo} {et~al.}(2007){Jullo}, {Kneib}, {Limousin}, {El{\'\i}asd{\'o}ttir}, {Marshall}, \& {Verdugo}}]{jullo2007}
{Jullo}, E., {Kneib}, J.~P., {Limousin}, M., {et~al.} 2007, New Journal of Physics, 9, 447

\bibitem[{{Katz} {et~al.}(2020){Katz}, {Ramsoy}, {Rosdahl}, {Kimm}, {Blaizot}, {Haehnelt}, {Michel-Dansac}, {Garel}, {Laigle}, {Devriendt}, \& {Slyz}}]{katz2020}
{Katz}, H., {Ramsoy}, M., {Rosdahl}, J., {et~al.} 2020, \mnras, 494, 2200

\bibitem[{{Kawamata} {et~al.}(2016){Kawamata}, {Oguri}, {Ishigaki}, {Shimasaku}, \& {Ouchi}}]{kawamata2016}
{Kawamata}, R., {Oguri}, M., {Ishigaki}, M., {Shimasaku}, K., \& {Ouchi}, M. 2016, \apj, 819, 114

\bibitem[{{Kroupa}(2001)}]{kroupa2001}
{Kroupa}, P. 2001, \mnras, 322, 231

\bibitem[{{Krumholz}(2014)}]{krumholz2014}
{Krumholz}, M.~R. 2014, \physrep, 539, 49

\bibitem[{{Leja} {et~al.}(2019){Leja}, {Carnall}, {Johnson}, {Conroy}, \& {Speagle}}]{leja2019}
{Leja}, J., {Carnall}, A.~C., {Johnson}, B.~D., {Conroy}, C., \& {Speagle}, J.~S. 2019, \apj, 876, 3

\bibitem[{{Li} {et~al.}(2024){Li}, {Dekel}, {Sarkar}, {Aung}, {Giavalisco}, {Mandelker}, \& {Tacchella}}]{Li24}
{Li}, Z., {Dekel}, A., {Sarkar}, K.~C., {et~al.} 2024, \aap, 690, A108

\bibitem[{{Looser} {et~al.}(2025){Looser}, {D'Eugenio}, {Maiolino}, {Tacchella}, {Curti}, {Arribas}, {Baker}, {Baum}, {Bonaventura}, {Boyett}, {Bunker}, {Carniani}, {Charlot}, {Chevallard}, {Curtis-Lake}, {Lola Danhaive}, {Eisenstein}, {de Graaff}, {Hainline}, {Ji}, {Johnson}, {Kumari}, {Nelson}, {Parlanti}, {Rix}, {Robertson}, {Del Pino}, {Sandles}, {Scholtz}, {Smit}, {Stark}, {{\"U}bler}, {Williams}, {Willott}, \& {Witstok}}]{looser2025}
{Looser}, T.~J., {D'Eugenio}, F., {Maiolino}, R., {et~al.} 2025, \aap, 697, A88

\bibitem[{{Looser} {et~al.}(2024){Looser}, {D'Eugenio}, {Maiolino}, {Witstok}, {Sandles}, {Curtis-Lake}, {Chevallard}, {Tacchella}, {Johnson}, {Baker}, {Suess}, {Carniani}, {Ferruit}, {Arribas}, {Bonaventura}, {Bunker}, {Cameron}, {Charlot}, {Curti}, {de Graaff}, {Maseda}, {Rawle}, {Rix}, {Del Pino}, {Smit}, {{\"U}bler}, {Willott}, {Alberts}, {Egami}, {Eisenstein}, {Endsley}, {Hausen}, {Rieke}, {Robertson}, {Shivaei}, {Williams}, {Boyett}, {Chen}, {Ji}, {Jones}, {Kumari}, {Nelson}, {Perna}, {Saxena}, \& {Scholtz}}]{looser2024}
{Looser}, T.~J., {D'Eugenio}, F., {Maiolino}, R., {et~al.} 2024, \nat, 629, 53

\bibitem[{{Mascia} {et~al.}(2024){Mascia}, {Pentericci}, {Calabr{\`o}}, {Santini}, {Napolitano}, {Arrabal Haro}, {Castellano}, {Dickinson}, {Ocvirk}, {Lewis}, {Amor{\'\i}n}, {Bagley}, {Bhatawdekar}, {Cleri}, {Costantin}, {Dekel}, {Finkelstein}, {Fontana}, {Giavalisco}, {Grogin}, {Hathi}, {Hirschmann}, {Holwerda}, {Jung}, {Kartaltepe}, {Koekemoer}, {Lucas}, {Papovich}, {P{\'e}rez-Gonz{\'a}lez}, {Pirzkal}, {Trump}, {Wilkins}, \& {Yung}}]{Mascia24}
{Mascia}, S., {Pentericci}, L., {Calabr{\`o}}, A., {et~al.} 2024, \aap, 685, A3

\bibitem[{{Mascia} {et~al.}(2023){Mascia}, {Pentericci}, {Calabr{\`o}}, {Treu}, {Santini}, {Yang}, {Napolitano}, {Roberts-Borsani}, {Bergamini}, {Grillo}, {Rosati}, {Vulcani}, {Castellano}, {Boyett}, {Fontana}, {Glazebrook}, {Henry}, {Mason}, {Merlin}, {Morishita}, {Nanayakkara}, {Paris}, {Roy}, {Williams}, {Wang}, {Brammer}, {Brada{\v{c}}}, {Chen}, {Kelly}, {Koekemoer}, {Trenti}, \& {Windhorst}}]{mascia2023a}
{Mascia}, S., {Pentericci}, L., {Calabr{\`o}}, A., {et~al.} 2023, \aap, 672, A155

\bibitem[{{Mason} {et~al.}(2023){Mason}, {Trenti}, \& {Treu}}]{Mason23}
{Mason}, C.~A., {Trenti}, M., \& {Treu}, T. 2023, \mnras, 521, 497

\bibitem[{{Messa} {et~al.}(2025){Messa}, {Vanzella}, {Loiacono}, {Bergamini}, {Castellano}, {Sun}, {Willott}, {Windhorst}, {Yan}, {Angora}, {Rosati}, {Adamo}, {Annibali}, {Bolamperti}, {Brada{\v{c}}}, {Bradley}, {Calura}, {Claeyssens}, {Comastri}, {Conselice}, {D'Silva}, {Dickinson}, {Frye}, {Grillo}, {Grogin}, {Gruppioni}, {Koekemoer}, {Meneghetti}, {Me{\v{s}}tri{\'c}}, {Pascale}, {Ravindranath}, {Ricotti}, {Summers}, \& {Zanella}}]{Messa2025}
{Messa}, M., {Vanzella}, E., {Loiacono}, F., {et~al.} 2025, \aap, 694, A59

\bibitem[{{Mowla} {et~al.}(2024){Mowla}, {Iyer}, {Asada}, {Desprez}, {Tan}, {Martis}, {Sarrouh}, {Strait}, {Abraham}, {Brada{\v{c}}}, {Brammer}, {Muzzin}, {Pacifici}, {Ravindranath}, {Sawicki}, {Willott}, {Estrada-Carpenter}, {Jahan}, {Noirot}, {Matharu}, {Rihtar{\v{s}}i{\v{c}}}, \& {Zabl}}]{mowla2024}
{Mowla}, L., {Iyer}, K., {Asada}, Y., {et~al.} 2024, \nat, 636, 332

\bibitem[{{Naidu} {et~al.}(2025){Naidu}, {Oesch}, {Brammer}, {Weibel}, {Li}, {Matthee}, {Chisholm}, {Pollock}, {Heintz}, {Johnson}, {Shen}, {Hviding}, {Leja}, {Tacchella}, {Ganguly}, {Witten}, {Atek}, {Belli}, {Bose}, {Bouwens}, {Dayal}, {Decarli}, {de Graaff}, {Fudamoto}, {Giovinazzo}, {Greene}, {Illingworth}, {Inoue}, {Kane}, {Labbe}, {Leonova}, {Marques-Chaves}, {Meyer}, {Nelson}, {Roberts-Borsani}, {Schaerer}, {Simcoe}, {Stefanon}, {Sugahara}, {Toft}, {van der Wel}, {van Dokkum}, {Walter}, {Watson}, {Weaver}, \& {Whitaker}}]{naidu2025_arxiv}
{Naidu}, R.~P., {Oesch}, P.~A., {Brammer}, G., {et~al.} 2025, arXiv e-prints, arXiv:2505.11263

\bibitem[{{Naidu} {et~al.}(2022){Naidu}, {Oesch}, {van Dokkum}, {Nelson}, {Suess}, {Brammer}, {Whitaker}, {Illingworth}, {Bouwens}, {Tacchella}, {Matthee}, {Allen}, {Bezanson}, {Conroy}, {Labbe}, {Leja}, {Leonova}, {Magee}, {Price}, {Setton}, {Strait}, {Stefanon}, {Toft}, {Weaver}, \& {Weibel}}]{Naidu2022}
{Naidu}, R.~P., {Oesch}, P.~A., {van Dokkum}, P., {et~al.} 2022, \apjl, 940, L14

\bibitem[{{Nakajima} {et~al.}(2022){Nakajima}, {Ouchi}, {Xu}, {Rauch}, {Harikane}, {Nishigaki}, {Isobe}, {Kusakabe}, {Nagao}, {Ono}, {Onodera}, {Sugahara}, {Kim}, {Komiyama}, {Lee}, \& {Zahedy}}]{nakajima2022}
{Nakajima}, K., {Ouchi}, M., {Xu}, Y., {et~al.} 2022, \apjs, 262, 3

\bibitem[{{Napolitano} {et~al.}(2025){Napolitano}, {Castellano}, {Pentericci}, {Arrabal Haro}, {Fontana}, {Treu}, {Bergamini}, {Calabr{\`o}}, {Mascia}, {Morishita}, {Roberts-Borsani}, {Santini}, {Vanzella}, {Vulcani}, {Zakharova}, {Bakx}, {Dickinson}, {Grillo}, {Leethochawalit}, {Llerena}, {Merlin}, {Paris}, {Rojas-Ruiz}, {Rosati}, {Wang}, {Yoon}, \& {Zavala}}]{Napolitano2025}
{Napolitano}, L., {Castellano}, M., {Pentericci}, L., {et~al.} 2025, \aap, 693, A50

\bibitem[{{Navarro} {et~al.}(1997){Navarro}, {Frenk}, \& {White}}]{navarro1997}
{Navarro}, J.~F., {Frenk}, C.~S., \& {White}, S. D.~M. 1997, \apj, 490, 493

\bibitem[{{Oguri}(2010)}]{oguri2010}
{Oguri}, M. 2010, \pasj, 62, 1017

\bibitem[{{Oguri}(2021)}]{oguri2021}
{Oguri}, M. 2021, \pasp, 133, 074504

\bibitem[{{Pallottini} \& {Ferrara}(2023)}]{Pallottini23}
{Pallottini}, A. \& {Ferrara}, A. 2023, \aap, 677, L4

\bibitem[{{Pascale} {et~al.}(2025){Pascale}, {Calura}, {Vesperini}, {Rosdahl}, {Nipoti}, {Giunchi}, {Lacchin}, {Lupi}, {Messa}, {Meneghetti}, {Ragagnin}, {Vanzella}, \& {Zanella}}]{pascale2025}
{Pascale}, R., {Calura}, F., {Vesperini}, E., {et~al.} 2025, \aap, 699, A31

\bibitem[{{Paterno-Mahler} {et~al.}(2018){Paterno-Mahler}, {Sharon}, {Coe}, {Mahler}, {Cerny}, {Johnson}, {Schrabback}, {Andrade-Santos}, {Avila}, {Brada{\v{c}}}, {Bradley}, {Carrasco}, {Czakon}, {Dawson}, {Frye}, {Hoag}, {Huang}, {Jones}, {Lam}, {Livermore}, {Lovisari}, {Mainali}, {Oesch}, {Ogaz}, {Past}, {Peterson}, {Ryan}, {Salmon}, {Sendra-Server}, {Stark}, {Umetsu}, {Vulcani}, \& {Zitrin}}]{paterno2018}
{Paterno-Mahler}, R., {Sharon}, K., {Coe}, D., {et~al.} 2018, \apj, 863, 154

\bibitem[{{Piqueras} {et~al.}(2019){Piqueras}, {Conseil}, {Shepherd}, {Bacon}, {Leclercq}, \& {Richard}}]{Piqueras2019}
{Piqueras}, L., {Conseil}, S., {Shepherd}, M., {et~al.} 2019, in Astronomical Society of the Pacific Conference Series, Vol. 521, Astronomical Data Analysis Software and Systems XXVI, ed. M.~{Molinaro}, K.~{Shortridge}, \& F.~{Pasian}, 545

\bibitem[{{Richard} {et~al.}(2021){Richard}, {Claeyssens}, {Lagattuta}, {Guaita}, {Bauer}, {Pello}, {Carton}, {Bacon}, {Soucail}, {Lyon}, {Kneib}, {Mahler}, {Cl{\'e}ment}, {Mercier}, {Variu}, {Tamone}, {Ebeling}, {Schmidt}, {Nanayakkara}, {Maseda}, {Weilbacher}, {Bouch{\'e}}, {Bouwens}, {Wisotzki}, {de la Vieuville}, {Martinez}, \& {Patr{\'\i}cio}}]{richard2021}
{Richard}, J., {Claeyssens}, A., {Lagattuta}, D., {et~al.} 2021, \aap, 646, A83

\bibitem[{{Roberts-Borsani} {et~al.}(2023){Roberts-Borsani}, {Treu}, {Chen}, {Morishita}, {Vanzella}, {Zitrin}, {Bergamini}, {Castellano}, {Fontana}, {Glazebrook}, {Grillo}, {Kelly}, {Merlin}, {Nanayakkara}, {Paris}, {Rosati}, {Yang}, {Acebron}, {Bonchi}, {Boyett}, {Brada{\v{c}}}, {Brammer}, {Broadhurst}, {Calabr{\'o}}, {Diego}, {Dressler}, {Furtak}, {Filippenko}, {Henry}, {Koekemoer}, {Leethochawalit}, {Malkan}, {Mason}, {Mercurio}, {Metha}, {Pentericci}, {Pierel}, {Rieck}, {Roy}, {Santini}, {Strait}, {Strausbaugh}, {Trenti}, {Vulcani}, {Wang}, {Wang}, \& {Windhorst}}]{robertsborsani2023}
{Roberts-Borsani}, G., {Treu}, T., {Chen}, W., {et~al.} 2023, \nat, 618, 480

\bibitem[{{Robertson} {et~al.}(2024){Robertson}, {Johnson}, {Tacchella}, {Eisenstein}, {Hainline}, {Arribas}, {Baker}, {Bunker}, {Carniani}, {Cargile}, {Carreira}, {Charlot}, {Chevallard}, {Curti}, {Curtis-Lake}, {D'Eugenio}, {Egami}, {Hausen}, {Helton}, {Jakobsen}, {Ji}, {Jones}, {Maiolino}, {Maseda}, {Nelson}, {P{\'e}rez-Gonz{\'a}lez}, {Pusk{\'a}s}, {Rieke}, {Smit}, {Sun}, {{\"U}bler}, {Whitler}, {Williams}, {Willmer}, {Willott}, \& {Witstok}}]{Robertson24}
{Robertson}, B., {Johnson}, B.~D., {Tacchella}, S., {et~al.} 2024, \apj, 970, 31

\bibitem[{{Robertson} {et~al.}(2023){Robertson}, {Tacchella}, {Johnson}, {Hainline}, {Whitler}, {Eisenstein}, {Endsley}, {Rieke}, {Stark}, {Alberts}, {Dressler}, {Egami}, {Hausen}, {Rieke}, {Shivaei}, {Williams}, {Willmer}, {Arribas}, {Bonaventura}, {Bunker}, {Cameron}, {Carniani}, {Charlot}, {Chevallard}, {Curti}, {Curtis-Lake}, {D'Eugenio}, {Jakobsen}, {Looser}, {L{\"u}tzgendorf}, {Maiolino}, {Maseda}, {Rawle}, {Rix}, {Smit}, {{\"U}bler}, {Willott}, {Witstok}, {Baum}, {Bhatawdekar}, {Boyett}, {Chen}, {de Graaff}, {Florian}, {Helton}, {Hviding}, {Ji}, {Kumari}, {Lyu}, {Nelson}, {Sandles}, {Saxena}, {Suess}, {Sun}, {Topping}, \& {Wallace}}]{Robertson23}
{Robertson}, B.~E., {Tacchella}, S., {Johnson}, B.~D., {et~al.} 2023, Nature Astronomy, 7, 611

\bibitem[{{Salmon} {et~al.}(2018){Salmon}, {Coe}, {Bradley}, {Brada{\v{c}}}, {Strait}, {Paterno-Mahler}, {Huang}, {Oesch}, {Zitrin}, {Acebron}, {Cibirka}, {Kikuchihara}, {Oguri}, {Brammer}, {Sharon}, {Trenti}, {Avila}, {Ogaz}, {Andrade-Santos}, {Carrasco}, {Cerny}, {Dawson}, {Frye}, {Hoag}, {Jones}, {Mainali}, {Ouchi}, {Rodney}, {Stark}, \& {Umetsu}}]{salmon2018}
{Salmon}, B., {Coe}, D., {Bradley}, L., {et~al.} 2018, \apjl, 864, L22

\bibitem[{{Sanders} {et~al.}(2025){Sanders}, {Shapley}, {Topping}, {Reddy}, {Berg}, {Khostovan}, {Bouwens}, {Brammer}, {Carnall}, {Cullen}, {Dav{\'e}}, {Dunlop}, {Ellis}, {F{\"o}rster Schreiber}, {Furlanetto}, {Glazebrook}, {Illingworth}, {Jones}, {Kriek}, {McLeod}, {McLure}, {Narayanan}, {Oesch}, {Pahl}, {Pettini}, {Schaerer}, {Stark}, {Steidel}, {Tang}, {Clarke}, {Donnan}, \& {Kehoe}}]{Sanders25arXiv}
{Sanders}, R.~L., {Shapley}, A.~E., {Topping}, M.~W., {et~al.} 2025, arXiv e-prints, arXiv:2508.10099

\bibitem[{{Sanders} {et~al.}(2024){Sanders}, {Shapley}, {Topping}, {Reddy}, \& {Brammer}}]{Sanders24}
{Sanders}, R.~L., {Shapley}, A.~E., {Topping}, M.~W., {Reddy}, N.~A., \& {Brammer}, G.~B. 2024, \apj, 962, 24

\bibitem[{{Sharon} {et~al.}(2020){Sharon}, {Bayliss}, {Dahle}, {Dunham}, {Florian}, {Gladders}, {Johnson}, {Mahler}, {Paterno-Mahler}, {Rigby}, {Whitaker}, {Akhshik}, {Koester}, {Murray}, {Remolina Gonz{\'a}lez}, \& {Wuyts}}]{Sharon2020}
{Sharon}, K., {Bayliss}, M.~B., {Dahle}, H., {et~al.} 2020, \apjs, 247, 12

\bibitem[{{Somerville} {et~al.}(2025){Somerville}, {Yung}, {Lancaster}, {Menon}, {Sommovigo}, \& {Finkelstein}}]{Somerville25}
{Somerville}, R.~S., {Yung}, L.~Y.~A., {Lancaster}, L., {et~al.} 2025, \mnras [\eprint[arXiv]{2505.05442}]

\bibitem[{{Stanway} \& {Eldridge}(2018)}]{stanway2018}
{Stanway}, E.~R. \& {Eldridge}, J.~J. 2018, \mnras, 479, 75

\bibitem[{{Stasi{\'n}ska} \& {Leitherer}(1996)}]{stasinska1996}
{Stasi{\'n}ska}, G. \& {Leitherer}, C. 1996, \apjs, 107, 661

\bibitem[{{Strait} {et~al.}(2020){Strait}, {Brada{\v{c}}}, {Coe}, {Bradley}, {Salmon}, {Lemaux}, {Huang}, {Zitrin}, {Sharon}, {Acebron}, {Andrade-Santos}, {Avila}, {Frye}, {Hoag}, {Mahler}, {Nonino}, {Ogaz}, {Oguri}, {Ouchi}, {Paterno-Mahler}, \& {Pelliccia}}]{strait2020}
{Strait}, V., {Brada{\v{c}}}, M., {Coe}, D., {et~al.} 2020, \apj, 888, 124

\bibitem[{{Strickland} \& {Heckman}(2009)}]{strickland2009}
{Strickland}, D.~K. \& {Heckman}, T.~M. 2009, \apj, 697, 2030

\bibitem[{{Sugimura} {et~al.}(2024){Sugimura}, {Ricotti}, {Park}, {Garcia}, \& {Yajima}}]{SugimuraR2024}
{Sugimura}, K., {Ricotti}, M., {Park}, J., {Garcia}, F. A.~B., \& {Yajima}, H. 2024, \apj, 970, 14

\bibitem[{{Sun} {et~al.}(2023{\natexlab{a}}){Sun}, {Faucher-Gigu{\`e}re}, {Hayward}, \& {Shen}}]{sun2023b}
{Sun}, G., {Faucher-Gigu{\`e}re}, C.-A., {Hayward}, C.~C., \& {Shen}, X. 2023{\natexlab{a}}, \mnras, 526, 2665

\bibitem[{{Sun} {et~al.}(2023{\natexlab{b}}){Sun}, {Faucher-Gigu{\`e}re}, {Hayward}, {Shen}, {Wetzel}, \& {Cochrane}}]{sun2023a}
{Sun}, G., {Faucher-Gigu{\`e}re}, C.-A., {Hayward}, C.~C., {et~al.} 2023{\natexlab{b}}, \apjl, 955, L35

\bibitem[{{Tacchella} {et~al.}(2016){Tacchella}, {Dekel}, {Carollo}, {Ceverino}, {DeGraf}, {Lapiner}, {Mandelker}, \& {Primack Joel}}]{tacchella2016}
{Tacchella}, S., {Dekel}, A., {Carollo}, C.~M., {et~al.} 2016, \mnras, 457, 2790

\bibitem[{{Tacchella} {et~al.}(2023){Tacchella}, {Eisenstein}, {Hainline}, {Johnson}, {Baker}, {Helton}, {Robertson}, {Suess}, {Chen}, {Nelson}, {Pusk{\'a}s}, {Sun}, {Alberts}, {Egami}, {Hausen}, {Rieke}, {Rieke}, {Shivaei}, {Williams}, {Willmer}, {Bunker}, {Cameron}, {Carniani}, {Charlot}, {Curti}, {Curtis-Lake}, {Looser}, {Maiolino}, {Maseda}, {Rawle}, {Rix}, {Smit}, {{\"U}bler}, {Willott}, {Witstok}, {Baum}, {Bhatawdekar}, {Boyett}, {Danhaive}, {de Graaff}, {Endsley}, {Ji}, {Lyu}, {Sandles}, {Saxena}, {Scholtz}, {Topping}, \& {Whitler}}]{Tacchella23}
{Tacchella}, S., {Eisenstein}, D.~J., {Hainline}, K., {et~al.} 2023, \apj, 952, 74

\bibitem[{{Tacchella} {et~al.}(2020){Tacchella}, {Forbes}, \& {Caplar}}]{tacchella2020}
{Tacchella}, S., {Forbes}, J.~C., \& {Caplar}, N. 2020, \mnras, 497, 698

\bibitem[{{Tang} {et~al.}(2025){Tang}, {Stark}, {Mason}, {Gelli}, {Chen}, \& {Topping}}]{Tang2025}
{Tang}, M., {Stark}, D.~P., {Mason}, C.~A., {et~al.} 2025, arXiv e-prints, arXiv:2507.08245

\bibitem[{{Topping} {et~al.}(2024){Topping}, {Stark}, {Endsley}, {Whitler}, {Hainline}, {Johnson}, {Robertson}, {Tacchella}, {Chen}, {Alberts}, {Baker}, {Bunker}, {Carniani}, {Charlot}, {Chevallard}, {Curtis-Lake}, {DeCoursey}, {Egami}, {Eisenstein}, {Ji}, {Maiolino}, {Williams}, {Willmer}, {Willott}, \& {Witstok}}]{topping2024}
{Topping}, M.~W., {Stark}, D.~P., {Endsley}, R., {et~al.} 2024, \mnras, 529, 4087

\bibitem[{{Trinca} {et~al.}(2024){Trinca}, {Schneider}, {Valiante}, {Graziani}, {Ferrotti}, {Omukai}, \& {Chon}}]{Trinca24}
{Trinca}, A., {Schneider}, R., {Valiante}, R., {et~al.} 2024, \mnras, 529, 3563

\bibitem[{{Vanzella} {et~al.}(2022){Vanzella}, {Castellano}, {Bergamini}, {Meneghetti}, {Zanella}, {Calura}, {Caminha}, {Rosati}, {Cupani}, {Me{\v{s}}tri{\'c}}, {Brammer}, {Tozzi}, {Mercurio}, {Grillo}, {Sani}, {Cristiani}, {Nonino}, {Merlin}, \& {Pignataro}}]{vanzella2022}
{Vanzella}, E., {Castellano}, M., {Bergamini}, P., {et~al.} 2022, \aap, 659, A2

\bibitem[{{Vanzella} {et~al.}(2023){Vanzella}, {Claeyssens}, {Welch}, {Adamo}, {Coe}, {Diego}, {Mahler}, {Khullar}, {Kokorev}, {Oguri}, {Ravindranath}, {Furtak}, {Hsiao}, {Abdurro'uf}, {Mandelker}, {Brammer}, {Bradley}, {Brada{\v{c}}}, {Conselice}, {Dayal}, {Nonino}, {Andrade-Santos}, {Windhorst}, {Pirzkal}, {Sharon}, {de Mink}, {Fujimoto}, {Zitrin}, {Eldridge}, \& {Norman}}]{vanzella2023_sunrise}
{Vanzella}, E., {Claeyssens}, A., {Welch}, B., {et~al.} 2023, \apj, 945, 53

\bibitem[{{Vanzella} {et~al.}(2024){Vanzella}, {Loiacono}, {Messa}, {Castellano}, {Bergamini}, {Zanella}, {Annibali}, {Sun}, {Dickinson}, {Adamo}, {Calura}, {Ricotti}, {Rosati}, {Meneghetti}, {Grillo}, {Brada{\v{c}}}, {Conselice}, {Yan}, {Bolamperti}, {Me{\v{s}}tri{\'c}}, {Gilli}, {Gronke}, {Willott}, {Sani}, {Acebron}, {Comastri}, {Mignoli}, {Gruppioni}, {Mercurio}, {Strait}, {Pascale}, {Annunziatella}, {Frye}, {Bradley}, {Grogin}, {Koekemoer}, {Ravindranath}, {D'Silva}, {Summers}, {Rihtar{\v{s}}i{\v{c}}}, \& {Windhorst}}]{Vanzella2024}
{Vanzella}, E., {Loiacono}, F., {Messa}, M., {et~al.} 2024, \aap, 691, A251

\bibitem[{{Vanzella} {et~al.}(2025){Vanzella}, {Messa}, {Adamo}, {Loiacono}, {Oguri}, {Sharon}, {Bradley}, {Bergamini}, {Meneghetti}, {Claeyssens}, {Welch}, {Bradac}, {Zanella}, {Bolamperti}, {Calura}, {Y-Y. Hsiao}, {Zackrisson}, {Ricotti}, {Christensen}, {Diego}, {Bauer}, {Xu}, {Fujimoto}, {Grillo}, {Lombardi}, {Rosati}, {Resseguier}, {Zitrin}, {Bik}, {Richard}, {Abdurro'uf}, {Bhatawdekar}, {Coe}, {Frye}, {Jimenez-Teja}, {Norman}, {Rigby}, {Trenti}, \& {Hashimoto}}]{Vanzella2025}
{Vanzella}, E., {Messa}, M., {Adamo}, A., {et~al.} 2025, arXiv e-prints, arXiv:2507.18699

\bibitem[{{Wang} {et~al.}(2023){Wang}, {Fujimoto}, {Labb{\'e}}, {Furtak}, {Miller}, {Setton}, {Zitrin}, {Atek}, {Bezanson}, {Brammer}, {Leja}, {Oesch}, {Price}, {Chemerynska}, {Cutler}, {Dayal}, {van Dokkum}, {Goulding}, {Greene}, {Fudamoto}, {Khullar}, {Kokorev}, {Marchesini}, {Pan}, {Weaver}, {Whitaker}, \& {Williams}}]{Wang23}
{Wang}, B., {Fujimoto}, S., {Labb{\'e}}, I., {et~al.} 2023, \apjl, 957, L34

\bibitem[{{Weilbacher} {et~al.}(2020){Weilbacher}, {Palsa}, {Streicher}, {Bacon}, {Urrutia}, {Wisotzki}, {Conseil}, {Husemann}, {Jarno}, {Kelz}, {P{\'e}contal-Rousset}, {Richard}, {Roth}, {Selman}, \& {Vernet}}]{Weilbacher2020}
{Weilbacher}, P.~M., {Palsa}, R., {Streicher}, O., {et~al.} 2020, \aap, 641, A28

\bibitem[{{Welch} {et~al.}(2023){Welch}, {Coe}, {Zitrin}, {Diego}, {Windhorst}, {Mandelker}, {Vanzella}, {Ravindranath}, {Zackrisson}, {Florian}, {Bradley}, {Sharon}, {Brada{\v{c}}}, {Rigby}, {Frye}, \& {Fujimoto}}]{welch2023}
{Welch}, B., {Coe}, D., {Zitrin}, A., {et~al.} 2023, \apj, 943, 2

\bibitem[{{Whitler} {et~al.}(2025){Whitler}, {Stark}, {Topping}, {Robertson}, {Rieke}, {Hainline}, {Endsley}, {Chen}, {Baker}, {Bhatawdekar}, {Bunker}, {Carniani}, {Charlot}, {Chevallard}, {Curtis-Lake}, {Egami}, {Eisenstein}, {Helton}, {Ji}, {Johnson}, {P{\'e}rez-Gonz{\'a}lez}, {Rinaldi}, {Tacchella}, {Williams}, {Willmer}, {Willott}, \& {Witstok}}]{Whitler25}
{Whitler}, L., {Stark}, D.~P., {Topping}, M.~W., {et~al.} 2025, \apj, 992, 63

\bibitem[{{Zavala} {et~al.}(2025){Zavala}, {Castellano}, {Akins}, {Bakx}, {Burgarella}, {Casey}, {Ch{\'a}vez Ortiz}, {Dickinson}, {Finkelstein}, {Mitsuhashi}, {Nakajima}, {P{\'e}rez-Gonz{\'a}lez}, {Arrabal Haro}, {Bergamini}, {Buat}, {Backhaus}, {Calabr{\`o}}, {Cleri}, {Fern{\'a}ndez-Arenas}, {Fontana}, {Franco}, {Grillo}, {Giavalisco}, {Grogin}, {Hathi}, {Hirschmann}, {Ikeda}, {Jung}, {Kartaltepe}, {Koekemoer}, {Larson}, {McKinney}, {Papovich}, {Rosati}, {Saito}, {Santini}, {Terlevich}, {Terlevich}, {Treu}, \& {Yung}}]{Zavala25}
{Zavala}, J.~A., {Castellano}, M., {Akins}, H.~B., {et~al.} 2025, Nature Astronomy, 9, 155

\end{thebibliography}
%
%

\begin{appendix}
\onecolumn
\section{Additional plots on data reduction and PSF-matching of the cube}\label{sec:app:reduction}
\begin{figure}[!h]
    \centering \includegraphics[width=0.40\textwidth]{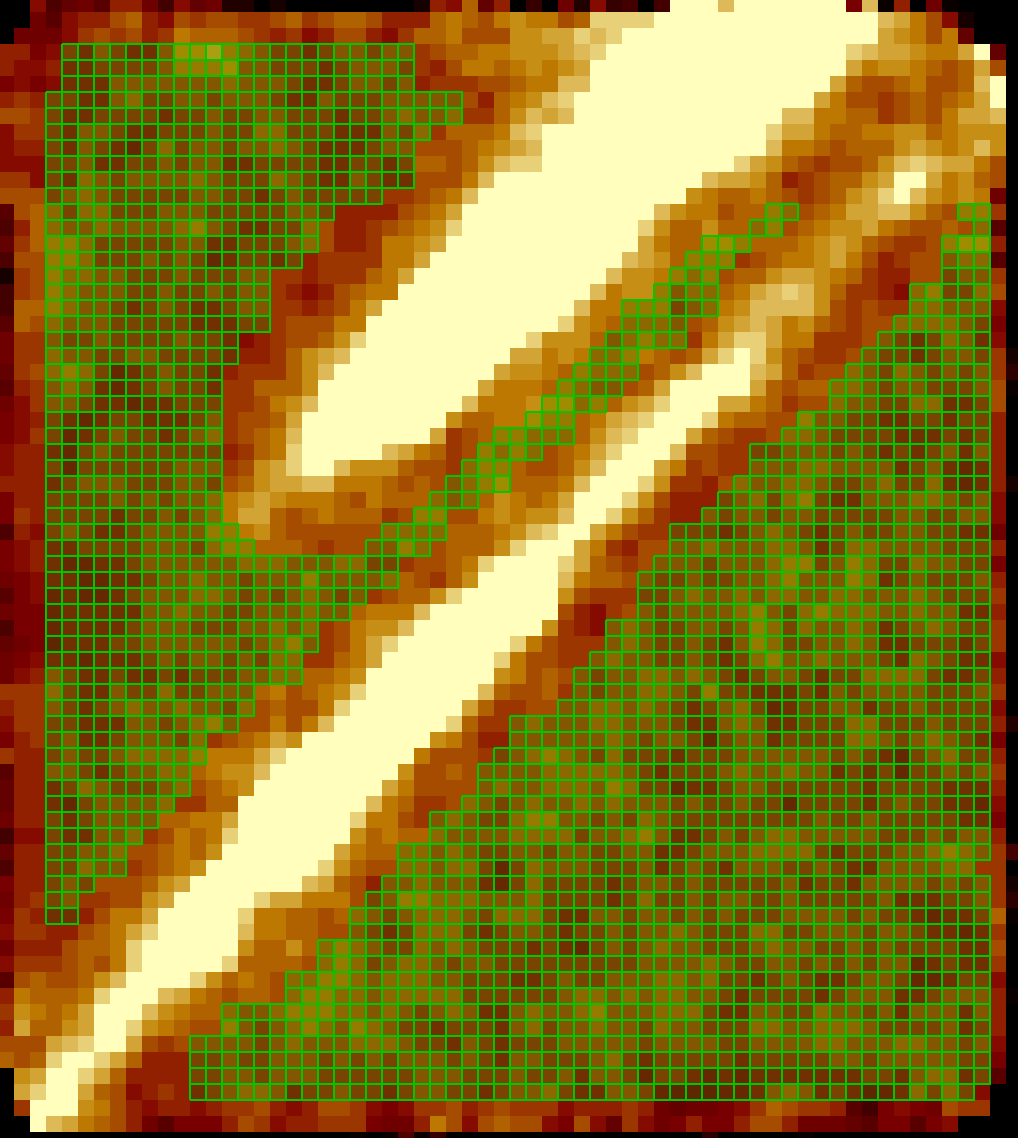}
    \caption{Green spaxels used for background estimation. The underlying map, produced by collapsing all the slices, is used to show the position of the target arc at $z=9.625$ and the position of a lower-redshift galaxy (z=2.531) within the field of view.}
    \label{fig:app:background}
\end{figure}

In this appendix we provide additional material about the data reduction process. Fig.~\ref{fig:app:background} shows the cube spaxels used to estimate the background, by fitting a 2D polynomial in each wavelength slice. As mentioned in the main text, the background-subtracted cubes produced with polynomial fits of zeroth to second order are almost indistinguishable. 

\begin{figure*}
    \centering
\includegraphics[width=0.99\textwidth]{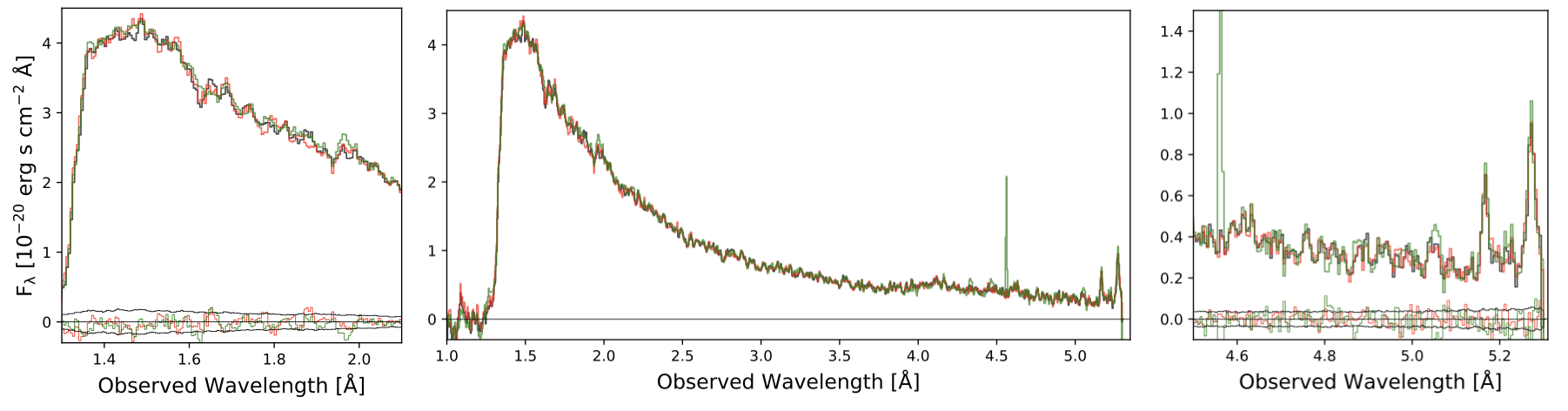}
    \caption{Comparison of the reference reduction (customized, at 50 mas, black curve) with the reduction at 100 mas (red curve) and the standard pipeline reduction (green curve). All spectra were extracted from the same region (BCDE12\_CC, shown in the main text in Fig.~\ref{fig:nircam_nirspec}). Central panel: Full spectrum. Lateral panels: Insets in the range of the \lya\ drop (left) and of the emission lines (right). The difference between the reference and the other reductions is compared to the uncertainties.}
\label{fig:app:comparison_reductions}
\end{figure*}

The comparison between one spectrum coming from stage 3 of the official pipeline, reduced at 100 mas scale, and two spectra from our customized pipeline (at 50 and 100 mas), all taken from the same region, is presented in Fig.~\ref{fig:app:comparison_reductions}. 

\begin{figure*}
    \centering
\includegraphics[width=0.95\textwidth]{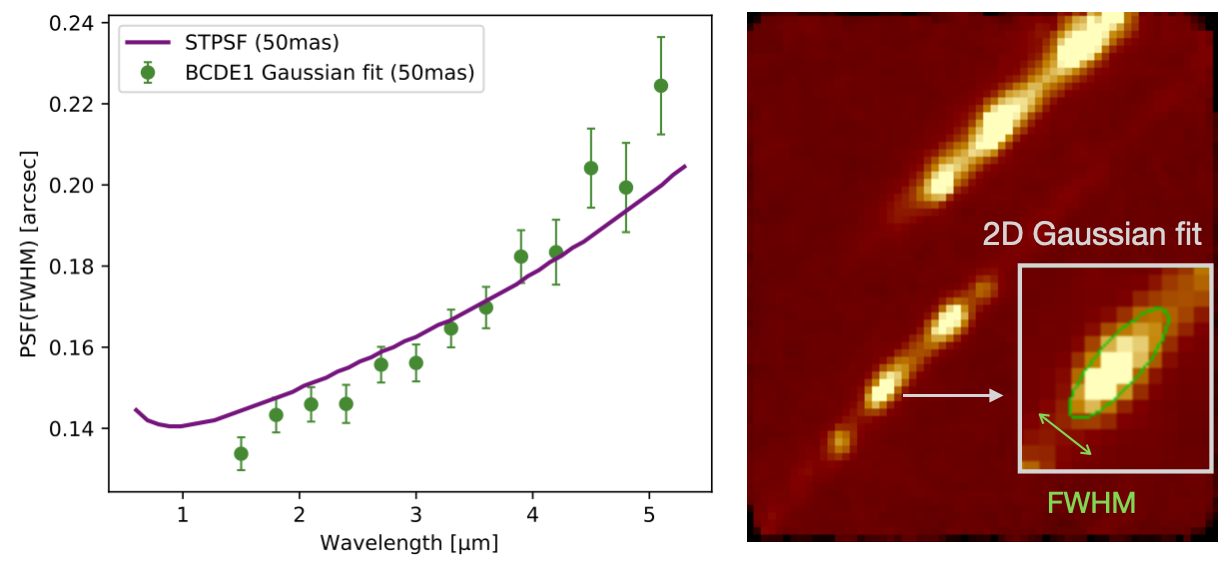}
    \caption{Variation of the PSF across wavelengths in the cube (purple line) used to create a PSF-matched final cube. The line was estimated from the FWHM of a series of PSFs created by \texttt{STPSF}. The green points, used for comparison, show estimates of the FWHM at various wavelengths on the minor (unresolved) axis of source BCDE1 (as shown in the right panel).}
    \label{fig:app:psfmatch}
\end{figure*}

Finally, Fig.~\ref{fig:app:psfmatch} shows the relation used to PSF-match the cube across wavelengths. This reference curve (purple line in the figure) comes from the FWHM of the PSFs produced by the \texttt{STPSF} tool. For comparison, we estimate the FWHM in the transversal direction of the arc of the BCDE1 clump in our data (green data in the figure). To increase the signal of this source, stackings of 30 slices are created along the wavelength range; all fits are performed assuming a 2D ellipse Gaussian model, and fixing the position angle to the direction of the arc. The assumption behind this test is that the source is spatially unresolved in NIRSpec-IFU, as supported by its appearance in NIRCam imaging. 
The wavelength trend recovered in this way (shown in Fig.~\ref{fig:app:psfmatch}) is overall consistent with the reference one. 

\FloatBarrier


\section{Multiple images, lens model comparison, and magnification map}\label{sec:app:mul_image}

A comparison of the magnifications, derived at the position of the star clusters, by the \glafic\ and \lenstool\  lens models (see Section~\ref{sec:lens_model}) is given in Tab.\ref{tab:app:mu_comparison}. On average \lenstool\ predicts larger magnification values, by a factor of up to 1.5. 

\begin{figure*}
    \centering    \includegraphics[width=0.99\textwidth]{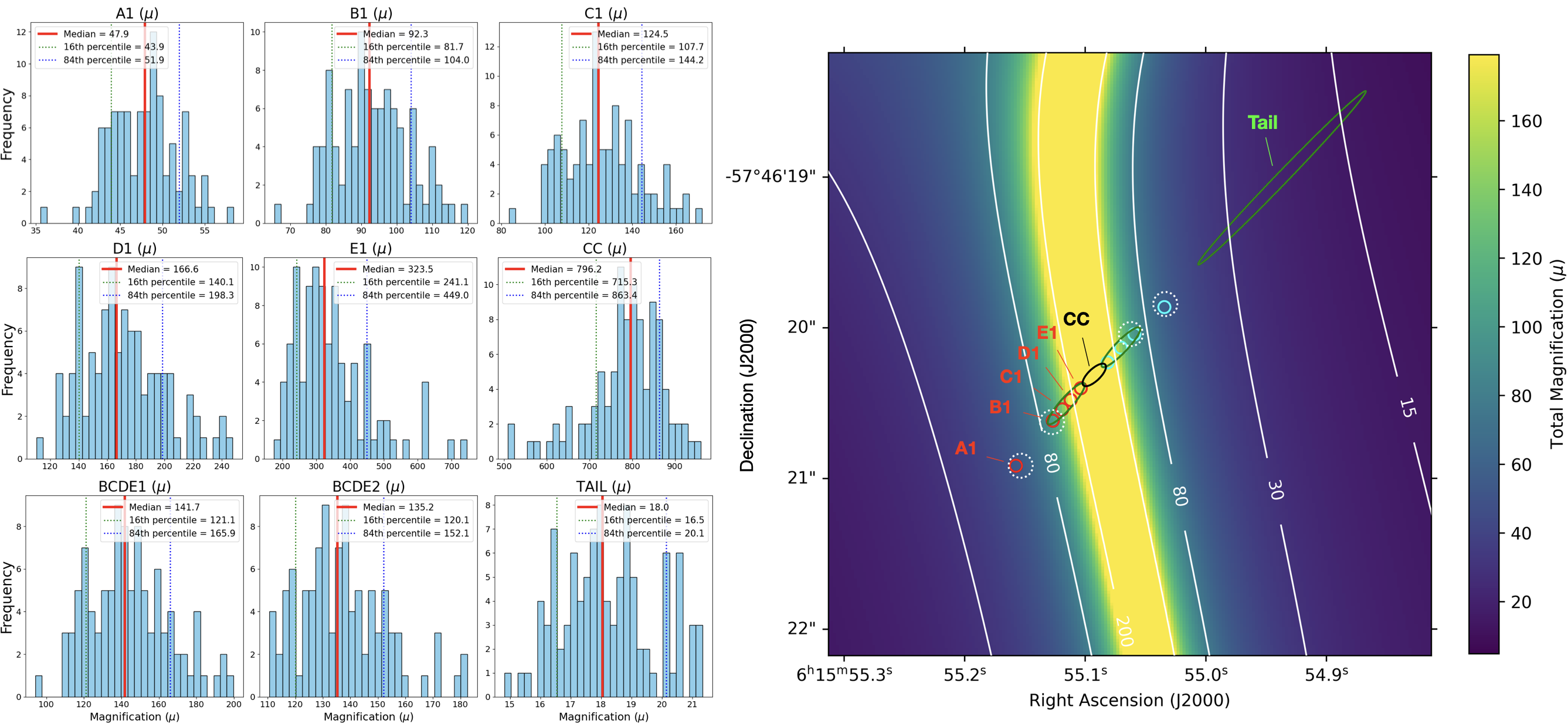}
    \caption{MCMC extractions from 100 magnification maps produced with \glafic\ as described in Section~\ref{sec:lens_model}. Left: Histograms, medians, and 16-84\% percentiles of the magnifications for the relevant regions of the Cosmic Gems arc discussed in this work. Right: Color-coded total magnification map indicating regions within which the magnifications were calculated (by computing the median of the values within the aperture). The regions correspond to the observed positions of the entries discussed in this work. As a comparison, the dotted white circles mark the model positions of the regions used among the constraints of the model. By construction, the observed and model positions are very close to each other.}
    \label{fig:map_glafic}
\end{figure*}

Multiple images used for the strong lens mass modeling are summarized in
Tab.~\ref{tab:mul_image}. Two systems lack reliable redshift (photometric or spectroscopic), which is thus left as a free parameter in the fit.
The magnification map at $\rm z=9.625$ in the region occupied by the Cosmic Gems arc is shown in Fig.~\ref{fig:map_glafic}, together with the apertures used to extract the amplification values assigned to the star clusters and to the NIRSpec masks defined in the main text. These apertures (solid circles and ellipses) are centered at the sources' observed positions; in the case of star clusters A.1, A.2, B.1 and B.2 we also show the predicted positions from the lens model (dotted white circles) that, by construction, are very close to the observed ones. Fig.~\ref{fig:map_glafic} also shows the distribution of amplification values obtained by considering the 100 MCMC extractions of the lens model (see Section.~\ref{sec:lens_model} for details); the median magnification is considered as "reference" for each region. When mirrored regions are considered together (i.e., {\it BCDE1,2} in Appendix~\ref{app:tab:spectroscopy}), the final magnification is the average of the two. 

\begin{table}[!h]
\caption{\label{tab:app:mu_comparison}Star cluster magnifications in different lens models.}
\renewcommand{\arraystretch}{1.35}
\centering
\begin{tabular}{llll}
\hline\hline
ID & \glafic\ & \lenstool & Ratio \\
\hline
A.1 & $47.9^{+4.0}_{-4.0}$ & $67.7^{+9.1}_{-0.3}$ & $1.4^{+0.2}_{-0.1}$ \\
B.1 & $92.3^{+11.7}_{-10.6}$ & $125.4^{+23.2}_{-3.2}$ & $1.4^{+0.3}_{-0.2}$   \\
C.1 & $124.5^{+19.7}_{-16.8}$ & $160.6^{+35.1}_{-6.7}$ & $1.3^{+0.3}_{-0.2}$   \\
D.1 & $166.6^{+31.7}_{-26.5}$ & $216.3^{+61.8}_{-13.4}$ & $1.3^{+0.4}_{-0.3}$   \\
E.1 & $323.5^{+125.5}_{-82.4}$ & $369.8^{+198.9}_{-48.9}$ & $1.1^{+0.7}_{-0.5}$   \\
A.2 & $52.2^{+4.4}_{-3.2}$ & $74.1^{+8.0}_{-3.1}$ & $1.4^{+0.2}_{-0.1}$   \\
B.2 & $90.3^{+7.3}_{-7.4}$ & $131.6^{+18.3}_{-10.2}$ & $1.5^{+0.2}_{-0.2}$   \\
C.2+D.2 & $132.5^{+16.7}_{-14.4}$ & $200.6^{+40.6}_{-22.8}$ & $1.5^{+0.3}_{-0.3}$   \\
\hline
\hline
\end{tabular}
\tablefoot{Shown is a comparison between the magnification values obtained for the five main clusters in the case of the two lens models studied (\glafic\ and {\sc Lenstool}). Their ratios, reported in the last column, represent the scaling factor used to convert the intrinsic masses, derived assuming one of the models, to the ones obtained by the other.}
\end{table}

\begin{table}
\caption{\label{tab:mul_image}Multiple images used for strong lens mass modeling.}
\centering
\begin{tabular}{lllll}
\hline\hline
ID & RA(J2000) & Dec(J2000) & Redshift & Ref.\tablefootmark{a}\\
\hline
 1.1 & $ 93.967707$ & $-57.780612$ &   $ 1.358$ & 1,2\\ 
 1.2 & $ 93.966248$ & $-57.779779$ &            & \\ 
 1.3 & $ 93.962831$ & $-57.779168$ &            & \\ 
\hline
 2.1 & $ 93.967415$ & $-57.780807$ &   $ 1.358$ & 1,2\\ 
 2.2 & $ 93.966082$ & $-57.780001$ &            & \\ 
 2.3 & $ 93.962582$ & $-57.779334$ &            & \\ 
 2.4 & $ 93.965665$ & $-57.780529$ &            & \\ 
\hline
 3.1 & $ 93.967498$ & $-57.780918$ &   $ 1.358$ & 1,2\\ 
 3.2 & $ 93.965915$ & $-57.780029$ &            & \\ 
 3.3 & $ 93.962624$ & $-57.779445$ &            & \\ 
 3.4 & $ 93.965623$ & $-57.780584$ &            & \\ 
\hline
 4.1 & $ 93.967248$ & $-57.781140$ &   $ 1.358$ & 1,2\\ 
 4.2 & $ 93.962528$ & $-57.779634$ &            & \\ 
 4.3 & $ 93.965373$ & $-57.780945$ &            & \\ 
\hline
 5.1 & $ 93.955832$ & $-57.781418$ &            & 1,2\\ 
 5.2 & $ 93.957207$ & $-57.782612$ &            & \\ 
\hline
 6.1 & $ 93.958303$ & $-57.781589$ &   $ 4.013$ & 1,2\\ 
 6.2 & $ 93.955874$ & $-57.776168$ &            & \\ 
 6.3 & $ 93.965873$ & $-57.776112$ &            & \\ 
 6.4 & $ 93.966582$ & $-57.774862$ &            & \\ 
 6.5 & $ 93.964284$ & $-57.779477$ &            & \\ 
\hline
 7.1 & $ 93.958837$ & $-57.779130$ & $1.7\pm0.5$ & 3,4\\ 
 7.2 & $ 93.964036$ & $-57.782170$ &            & \\ 
 7.3 & $ 93.970302$ & $-57.782715$ &            & \\ 
\hline
 8.1 & $ 93.965173$ & $-57.777030$ & $1.8\pm0.5$ & 3,4\\ 
 8.2 & $ 93.963320$ & $-57.776747$ &            & \\ 
\hline
 9.1 & $ 93.979824$ & $-57.772476$ &   $ 9.625$\tablefootmark{b} & 1,2,3,4,5\\ 
 9.2 & $ 93.979314$ & $-57.772182$ &            & \\ 
 9.3 & $ 93.950004$ & $-57.770225$ &            & \\ 
\hline
10.1 & $ 93.979695$ & $-57.772393$ &   $ 9.625$\tablefootmark{b} & 1,2,3,4,5\\ 
10.2 & $ 93.979412$ & $-57.772235$ &            & \\ 
10.3 & $ 93.950003$ & $-57.770225$ &            & \\ 
\hline
11.1 & $ 93.978702$ & $-57.771832$ &   $ 9.625$\tablefootmark{b} & 1,2,3,4,5\\ 
11.2 & $ 93.980386$ & $-57.772834$ &            & \\ 
\hline
12.1 & $ 93.973333$ & $-57.767401$ &   $ 6.102$\tablefootmark{b} & 3,4\\ 
12.2 & $ 93.981540$ & $-57.769895$ &            & \\ 
12.3 & $ 93.964613$ & $-57.766720$ &            & \\ 
\hline
13.1 & $ 93.973327$ & $-57.777517$ & $6.4\pm0.5$ & 3,4\\ 
13.2 & $ 93.944498$ & $-57.774728$ &            & \\ 
\hline
14.1 & $ 93.981011$ & $-57.776970$ & $3.2\pm0.5$ & 3,4\\ 
14.2 & $ 93.970332$ & $-57.772476$ &            & \\ 
14.3 & $ 93.957967$ & $-57.772538$ &            & \\ 
\hline
15.1 & $ 93.969522$ & $-57.783962$ &            & 5\\ 
15.2 & $ 93.964478$ & $-57.783726$ &            & \\ 
15.3 & $ 93.955549$ & $-57.778803$ &            & \\ 
\hline
16.1 & $ 93.950320$ & $-57.779373$ &   $ 3.569$\tablefootmark{b} & 5\\ 
16.2 & $ 93.960220$ & $-57.786883$ &            & \\ 
16.3 & $ 93.972837$ & $-57.787069$ &            & \\ 
\hline
17.1 & $ 93.950000$ & $-57.775647$ &   $ 4.880$\tablefootmark{b} & 5\\ 
17.2 & $ 93.960709$ & $-57.786008$ &            & \\ 
17.3 & $ 93.970043$ & $-57.774477$ &            & \\ 
17.4 & $ 93.969428$ & $-57.775505$ &            & \\ 
\hline
18.1 & $ 93.946965$ & $-57.781198$ &   $ 5.842$\tablefootmark{b} & 5\\ 
18.2 & $ 93.970939$ & $-57.789100$ &            & \\ 
\hline
\hline
\end{tabular}
\tablefoot{The redshifts without errors show spectroscopic redshifts, while those with errors correspond to photometric redshifts. Larger errors of $0.5$ in photometric redshifts, compared to measurement errors, are used as Gaussian priors for redshifts in mass modeling with \glafic. The \lenstool\ model assumed broad priors of [1.5--8] for all arcs. Arcs number 5.1, 5.2, 13.1, 13.2, and 17.4 were not used in the \lenstool\ model. Otherwise, the models employed the same constraints. \\
\tablefoottext{a}{References for multiple images. 1 -- \citet{paterno2018}; 2 -- \citet{salmon2018}; 3 -- \citet{adamo2024a}; 4 -- \citet{Bradley25}; 5 -- this paper.}
\tablefoottext{b}{New spectroscopic redshifts reported in this paper.}
}
\end{table}

\FloatBarrier

\twocolumn
\onecolumn

\section{Spectra from all mask apertures}\label{sec:app:all_spectra}
\begin{figure*}[!h]
    \centering
    \includegraphics[width=0.99\textwidth]{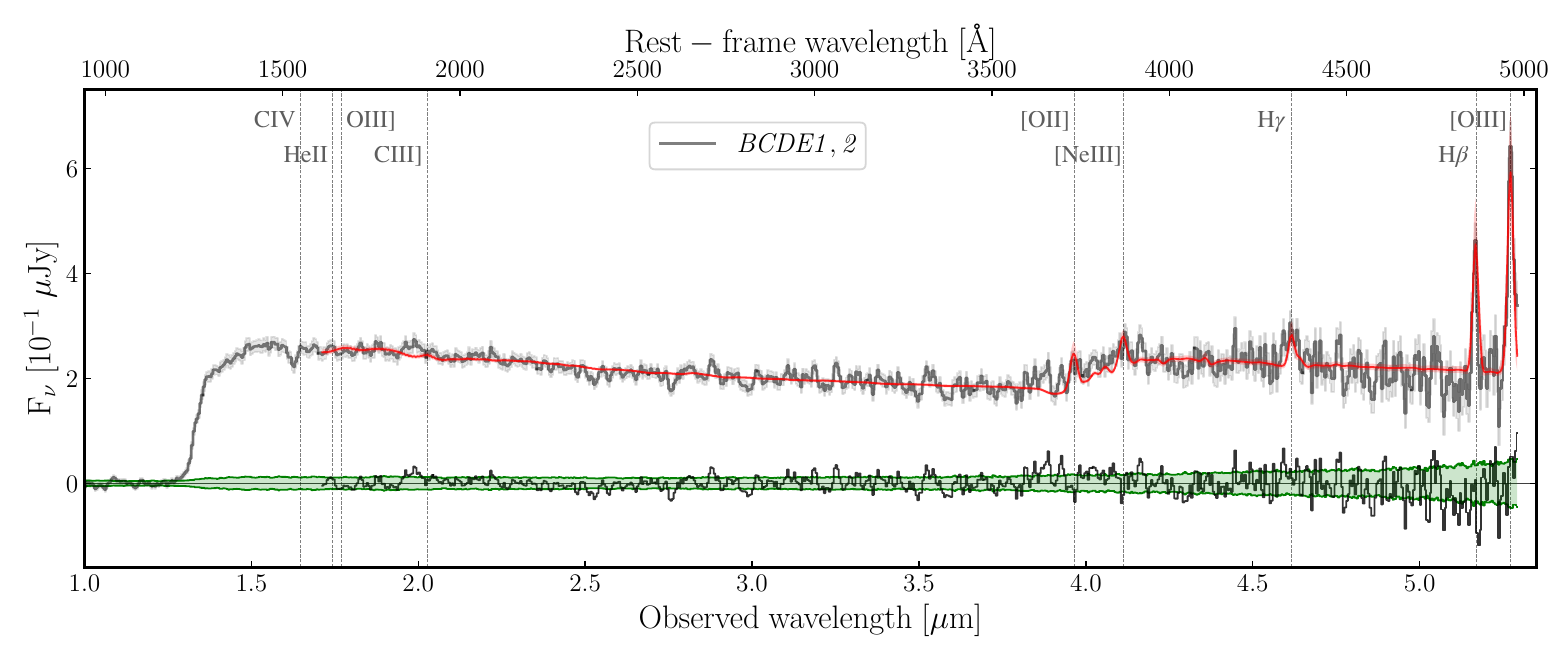}
    \includegraphics[width=0.99\textwidth]{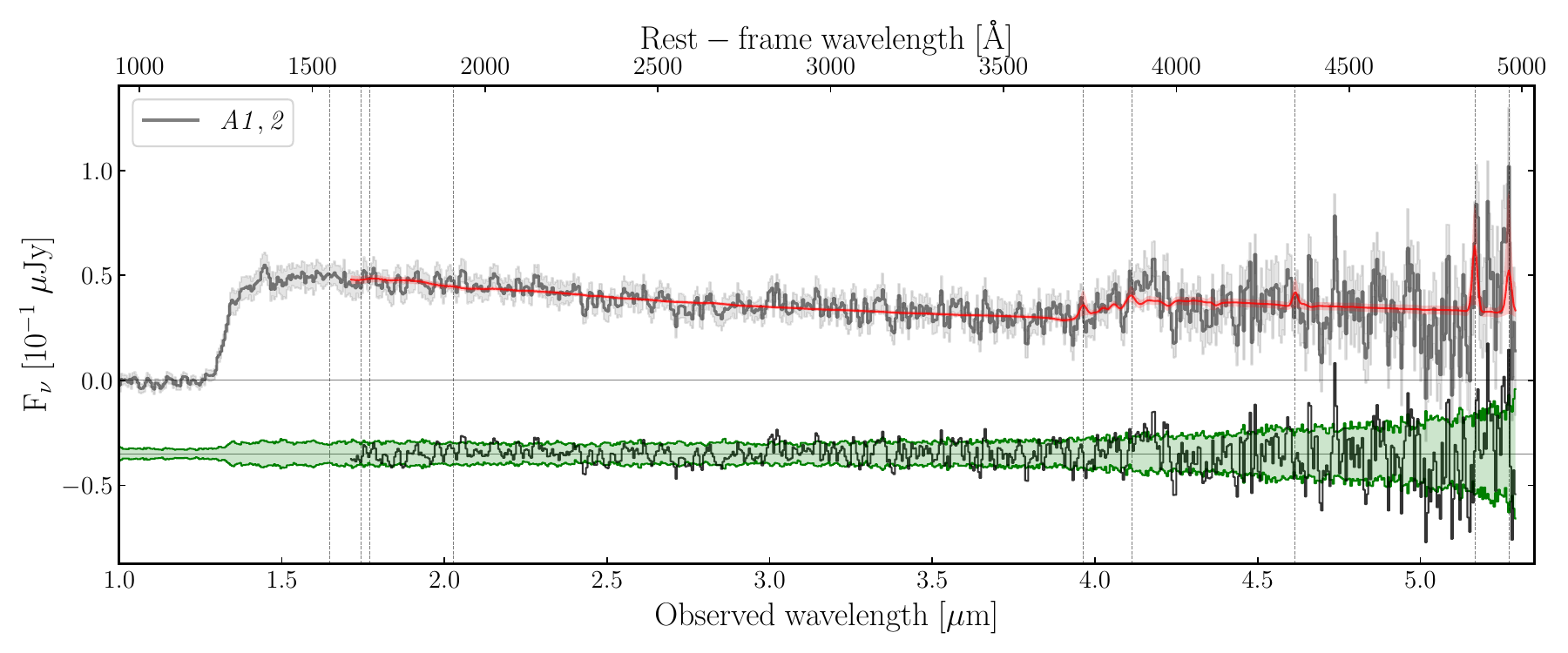}
    \caption{1D observed and best-fit spectra (gray and red curves, respectively) for the main mask apertures discussed in the main text (Section~\ref{sec:nirspec_analysis}). The residuals between the observed and best-model spectra are shown as a black curve, while the green band encloses the $\rm \pm1\sigma$ uncertainties from the observations. Both the residuals and uncertainties are, in some cases, shifted arbitrarily to avoid confusion with the observed spectra. We note that the fit is performed only for (observed) wavelengths longer than $1.7~\mu m$. The expected wavelengths of the main UV and optical lines are shown.}
    \label{fig:all_spectra}
\end{figure*}
\begin{figure*}
  \ContinuedFloat
  \centering
    \includegraphics[width=0.99\textwidth]{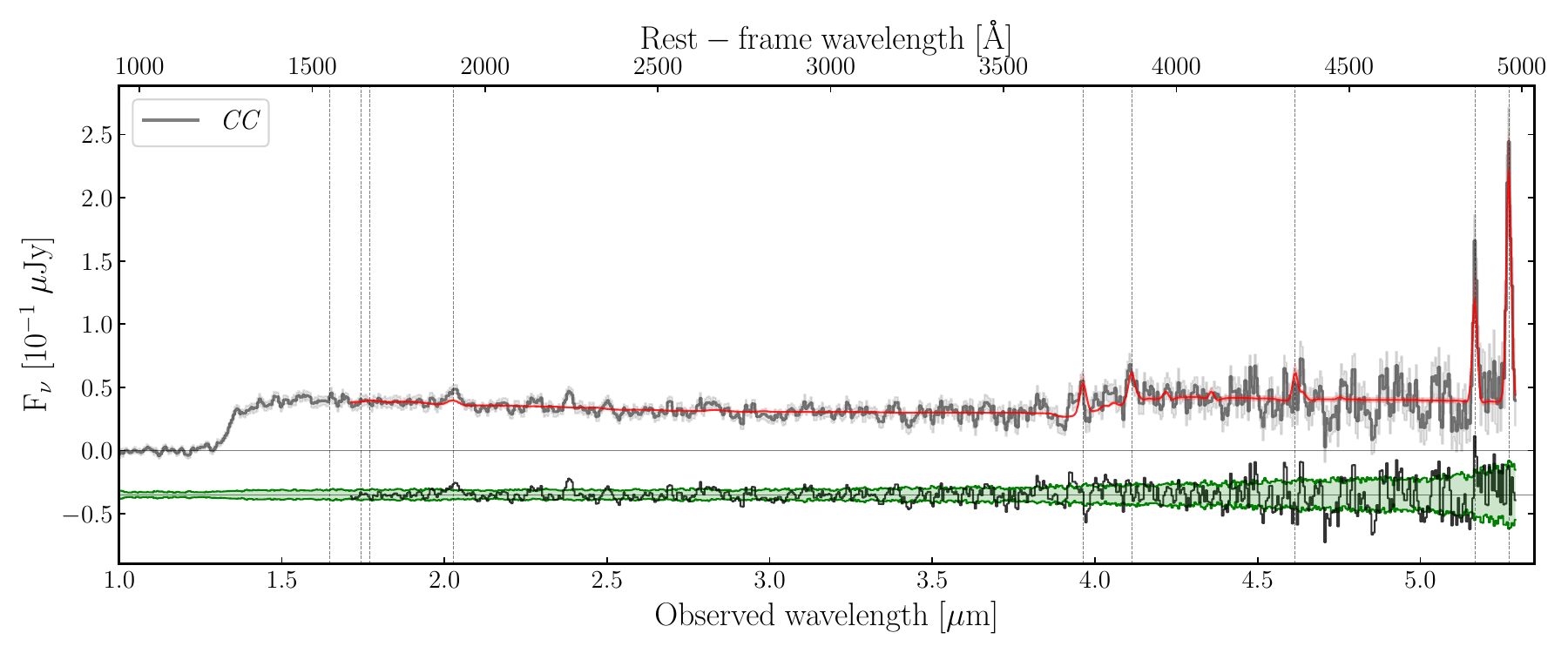}
    \includegraphics[width=0.99\textwidth]{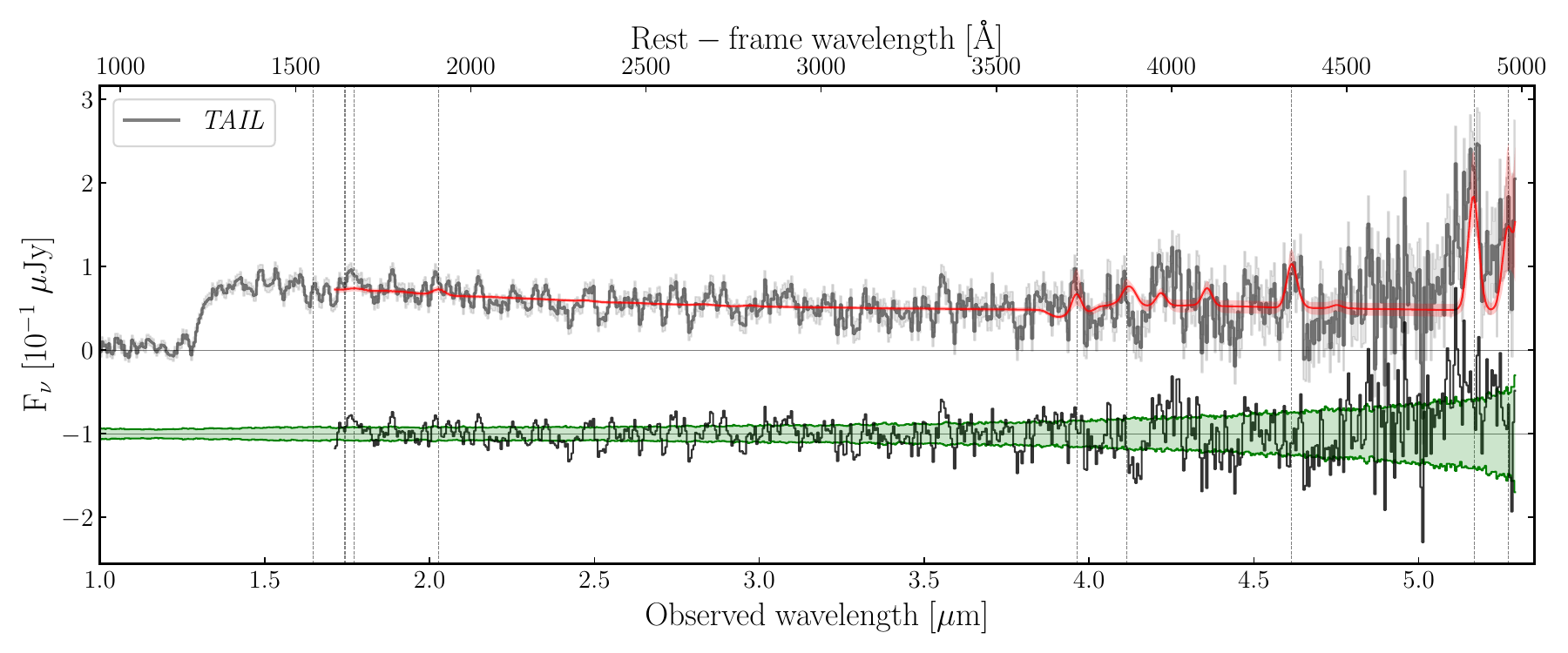}
    \includegraphics[width=0.99\textwidth]{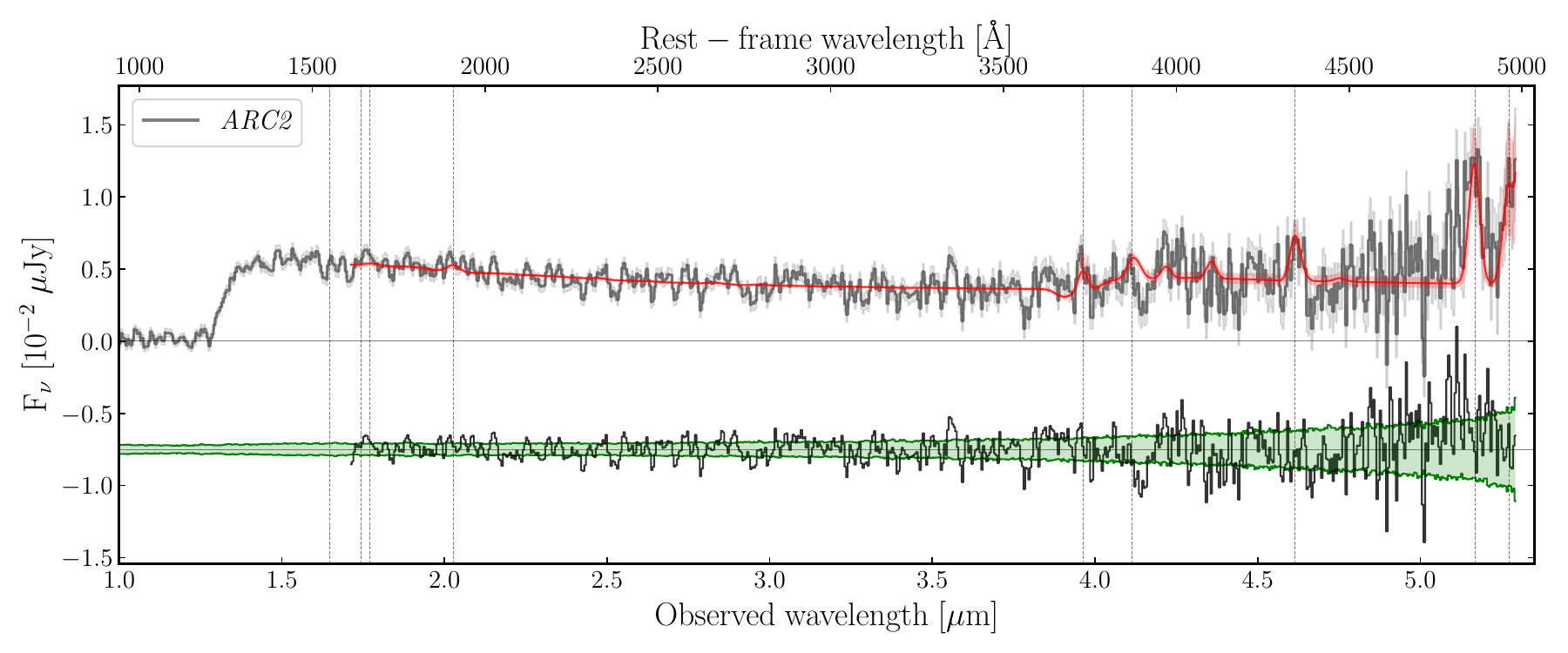}
  \caption{Continued.}
\end{figure*}
\FloatBarrier

\begin{sidewaystable*}
\section{Table of spectral properties}\label{app:tab:spectroscopy}
\caption{\label{tab:spectroscopy}Main spectroscopic properties of the Cosmic Gems arc and its subregions.}
\renewcommand{\arraystretch}{1.4}
\centering
    \begin{tabular}{lrrrrr|rrr}
    \hline
    \multicolumn{1}{c}{} & \multicolumn{1}{c}{\it BCDE1,2} & \multicolumn{1}{c}{\it A1,2} & \multicolumn{1}{c}{\it TAIL} & \multicolumn{1}{c}{\it CC} & \multicolumn{1}{c}{\it BCDE1,2+CC} & \multicolumn{1}{|c}{\it ARC2} & \multicolumn{1}{c}{\it BC1,2} & \multicolumn{1}{c}{\it DE1,2} \\
    \hline
\multicolumn{2}{l}{spectral features} &  & & & & & & \\
    EW([O\protect\scaleto{$III$}{1.2ex}]4959)~[\AA] & $47^{+6}_{-5}$ & $<40$ & $<20$ & $75^{+15}_{-12}$ & $51^{+6}_{-5}$ & $<23$ & $42^{+8}_{-7}$ & $54^{+10}_{-9}$\\
    EW($\rm H\beta$)~[\AA] & $18^{+3}_{-2}$ & $25^{+15}_{-7}$ & $<52$ & $38^{+8}_{-6}$ & $21^{+3}_{-2}$ & $<59$ & $13^{+3}_{-3}$ & $23^{+5}_{-4}$ \\
    R3 & $7.5\pm1.8$ & $-$ & $-$ & $5.9\pm1.3$ & $7.0\pm1.4$  & $-$ & $9.0\pm3.0$ & $6.5\pm1.3$ \\
    $\rm \beta_{UV}$ & $-2.53^{+0.03}_{-0.03}$ & $-2.60^{+0.06}_{-0.06}$ & $-2.74^{+0.06}_{-0.06}$ & $-2.52^{+0.06}_{-0.06}$ & $-2.53^{+0.02}_{-0.02}$ & $-2.66^{+0.05}_{-0.05}$ & $-2.55^{+0.03}_{-0.03}$ & $-2.52^{+0.04}_{-0.03}$ \\
    Balmer break & $1.3\pm0.1$ & $1.2\pm0.4$ & $1.0\pm0.5$ & $1.4\pm0.4$ & $1.3\pm0.1$ & $1.2\pm0.4$ & $1.2\pm0.2$ & $1.34\pm0.23$ \\
    \hline
\multicolumn{2}{l}{\texttt{Bagpipes} fit} &  & & & & \\
    (mw)age [Myr] & $12^{+3}_{-2}$ ($11$) & $15^{+3}_{-2}$ ($15$) & $49^{+18}_{-15}~(45)$ & $95^{+14}_{-14}~(108)$  & $10^{+2}_{-1}$ ($10$) & $68^{+17}_{-18}~(77)$ & $14^{+2}_{-2}~(13)$ & $9^{+1}_{-1}~(8)$ \\
    $\rm \tau~[Myr]$ & $3^{+1}_{-0}$ ($2$) & $2^{+1}_{-1}$ ($2$) & $206^{+181}_{-127}~(106)$ & $306^{+121}_{-131}~(449)$ & $2^{+1}_{-0}$ ($2$) & $148^{+174}_{-98}~(344)$ & $3^{+0}_{-0}~(3)$ & $2^{+0}_{-0}~(1)$ \\
    $\rm A_V~[mag]$ & $0.23^{+0.02}_{-0.02}$ ($0.22$) & $0.22^{+0.07}_{-0.08}$ ($0.24$) & $0.03^{+0.03}_{-0.02}~(0.01)$ & $0.05^{+0.04}_{-0.03}~(0.03)$ & $0.21^{+0.02}_{-0.02}$ ($0.22$) & $0.03^{+0.03}_{-0.02}~(0.00)$ & $0.24^{+0.03}_{-0.03}~(0.24)$ & $0.20^{+0.03}_{-0.03}~(0.19)$ \\
    $\rm Z_\star~[Z_\odot]$ & $0.16^{+0.01}_{-0.01}$ ($0.16$) & $0.06^{+0.05}_{-0.04} (0.04)$ & $0.06^{+0.02}_{-0.02}~(0.08)$ & $0.18^{+0.02}_{-0.02}~(0.19)$ & $0.17^{+0.01}_{-0.01}$ ($0.17$) & $0.07^{+0.02}_{-0.02}~(0.08)$ & $0.13^{+0.02}_{-0.02}~(0.14)$ & $0.17^{+0.01}_{-0.01}~(0.17)$ \\
    $\rm log(M_\star/M_\odot)^\dag$ & $6.72^{+0.10}_{-0.09}~(6.67)$ & $6.56^{+0.11}_{-0.13}~(6.56)$ &  $7.42^{+0.10}_{-0.12}~(7.39)$ & $<5.68^{+0.03}_{-0.03}~(5.70)^\ddag$ & $6.69^{+0.10}_{-0.08}~(6.68)$ & $7.68^{+0.06}_{-0.06}~(7.66)$ & $6.68^{+0.08}_{-0.09}~(6.63)$ & $5.95^{+0.11}_{-0.14}~(5.89)$ \\
    $\rm \mu_{tot}$ & $138^{+21}_{-18}~(153)$ & $50^{+4}_{-4}~(55)$ & $18^{+2}_{-2}~(21)$  & $>500^\ddag$ & $138^{+21}_{-18}~(153)$ & $-$ & $107^{+15}_{-14}~(120)$ & $226^{+60}_{-45}~(259)$ \\
    \hline
    \end{tabular}
    \tablefoot{Main spectral features from the 1D spectra extracted from the regions outlined in Fig.~\ref{fig:nircam_nirspec} (top lines) and best-fit properties of the same regions, as derived from the \texttt{Bagpipes} fit of the spectra (bottom lines). The equivalent widths represent the rest-frame values. The median values and $\pm1\sigma$ uncertainties are reported for the fit results. The values within the parenthesis refer to the maximum likelihood values.
    The magnifications reported in the bottom line are estimated from the \glafic\ model, described in Section~\ref{sec:lens_model}. $^\dag$The masses quoted are {\it intrinsic}, i.e., the mass resulting from the fit was delensed (using the magnifications quoted in the last line of the table) and, in the case of {\it BCDE1,2}, {\it A1,2}, and {\it CC}, was divided by half to account for the number of counterimages included in the masks. The lens modeling uncertainty is propagated to the final intrinsic mass. The systematic uncertainty on magnification (i.e., considering different lens models), ranges between 10--40\% (see Appendix~\ref{sec:app:mul_image}). $^\ddag$To conservatively estimate the intrinsic mass of the {\it CC} region, we considered a lower limit in magnification, as estimated from the MCMC extractions in Appendix~\ref{sec:app:mul_image}, resulting in an upper limit on the mass estimate.}
\end{sidewaystable*}
\FloatBarrier

\section{{\it BC} and {\it DE} subregions}\label{sec:app:bc_vs_de}
\begin{figure*}[!h]
    \centering
    \includegraphics[width=0.99\textwidth]{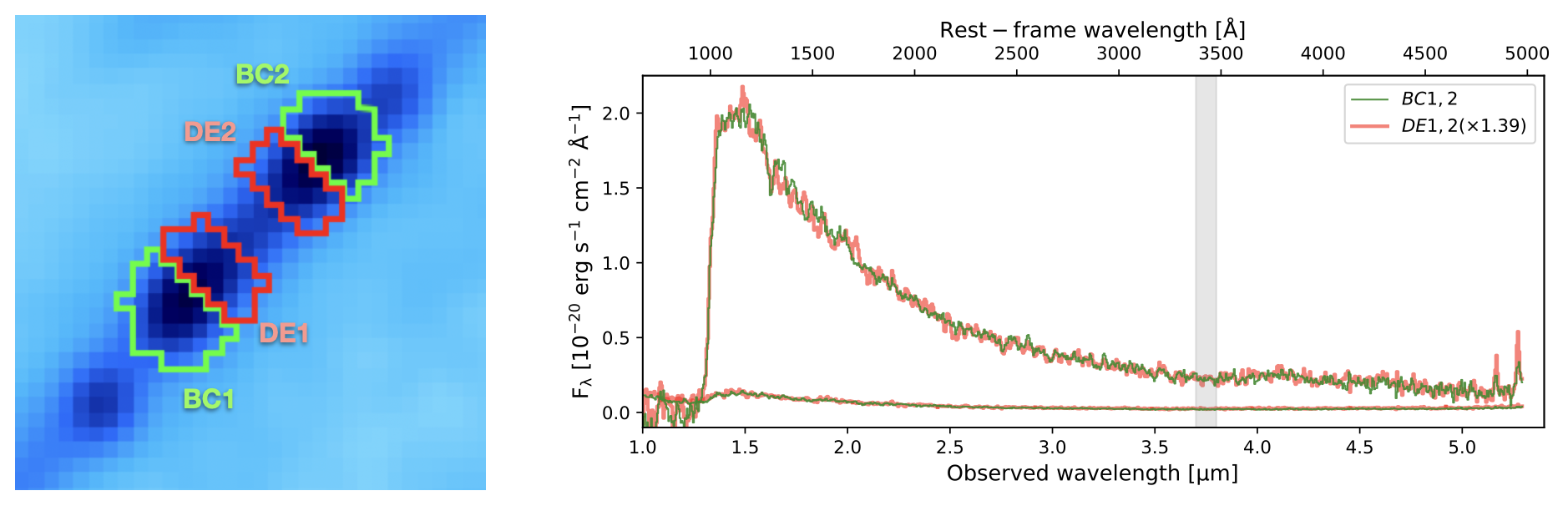}
    \caption{{\it (Left)}: Division of the {\it BCDE1} and {\it BCDE2} mask apertures into the sub-apertures {\it BC1}-{\it DE1} and {\it BC2}-{\it DE2}. {\it (Right)}: Comparison of the 1D spectra extracted from the {\it BC1,2} and {\it DE1,2} mask apertures. The latter were rescaled (by a factor of $1.39$) in order to match the average $\rm F_\lambda$ of {\it BC1,2} in the observed wavelength range $3.7$ to $\rm 3.8~\mu m$. This interval was chosen to highlight possible differences between the spectra in the UV slope (blueward of $\rm 3.7~\mu m$) or in the Balmer break amplitude (redward of $\rm 3.8~\mu m$).}
    \label{fig:bc_vs_de}
\end{figure*}

In order to test possible small-scale variations within the {\it BCDE1,2} region, we divide the mask aperture in two ({\it BC1,2} and {\it DE1,2}, see Fig.~\ref{fig:bc_vs_de}). We remind that given the large PSF of the NIRSpec/IFU, compared to NIRCam, the emission within {\it BCDE1,2} is spatially only marginally resolved (see Fig.~\ref{fig:nircam_nirspec}). The two spectra show very similar $\rm \beta_{UV}\approx-2.5$ and Balmer break ($\rm \approx1.2-1.3$), with {\it DE1,2} having slightly larger EWs than {\it BC1,2}. The direct comparison of the two spectra is shown in Fig.~\ref{fig:bc_vs_de}. The best-fit from \texttt{Bagpipes} are also overall consistent among the two subregions, returning young ages ($9-14$ Myr) and a short burst ($\rm \tau=2-3$ Myr). Given the large difference in magnification between the two subregions ($\rm \mu_{BC1,2}=107$ vs. $\rm \mu_{DE1,2}=225$), the intrinsic mass in the {\it BC} region, dominates the mass budget of the {\it BCDE} region. The total mass coming from the two fits remains consistent with the mass of {\it BCDE1,2} estimated in Section~\ref{sec:nirspec_analysis}. 

\end{appendix}

\end{document}